\documentclass[12pt]{article}

\usepackage{empheq}
\usepackage{hyperref}
\usepackage{scrextend}
\hypersetup{
    colorlinks = true,
    linkcolor = [rgb]{0 .4 .8},
    citecolor=[rgb]{1 .2 .2}
}
\usepackage{setspace}
\usepackage[dviwin]{graphicx}
\usepackage{amsfonts}
\usepackage{amssymb}
\usepackage{mathrsfs}
\usepackage{amsmath}
\usepackage{epsfig}
\usepackage{relsize}
\usepackage{graphicx}
\usepackage{mathabx}
\usepackage{floatrow}
\usepackage{textpos}

\usepackage{dsfont}
\usepackage{mathrsfs}
\usepackage{esint}
\usepackage{wrapfig}
\usepackage{lipsum}

\usepackage{array}

\usepackage{amsthm}

\newtheoremstyle{mytheoremstyle} 
    {0.3cm}                    
    {0.3cm}                    
{\bfseries}
    {}                           
    {\scshape}                   
    {.}                          
    {.5em}                       
    {}  
    \theoremstyle{mytheoremstyle}
\newtheorem{thm}{Proposition}

\newtheorem{lem}{Lemma}

\newcommand{\bea}{\begin{equation}}
\newcommand{\eea}{\end{equation}}
\newcommand{\bear}{\begin{eqnarray}}
\newcommand{\eear}{\end{eqnarray}}
\newcommand{\bearr}{\begin{eqnarray*}}
\newcommand{\eearr}{\end{eqnarray*}}

\newcommand{\beal}{\begin{align}}
\newcommand{\eeal}{\end{align}}
\newcommand{\beall}{\begin{align*}}
\newcommand{\eeall}{\end{align*}}

\newcommand{\CP}{\mathds{C}\mathds{P}}
\newcommand{\CC}{\mathds{C}}
\newcommand{\RR}{\mathds{R}}
\newcommand{\dd}{\partial}

\newcommand{\comment}[1]{}

\newcommand{\dP}{\mathbf{dP}}

\usepackage{xcolor}

\makeatletter
\def\@seccntformat#1{\@ifundefined{#1@cntformat}%
{\csname the#1\endcsname\quad}
{\csname #1@cntformat\endcsname}
}
\def\section@cntformat{{\normalfont\large\thesection.}\quad}
\def\subsection@cntformat{\textsection\, \thesubsection.\quad}
\def\subsubsection@cntformat{\textsection\textsection\, \thesubsubsection.\quad}
\makeatother

\newcommand{\ssection}[1]{%
  \section[#1]{\centering\normalfont\scshape #1}}
\newcommand{\ssubsection}[1]{%
   \subsection[#1]{\raggedright\normalfont  #1}}

\newsavebox\MBox

\begin{document}

\title{Ricci-flat metrics on the cone over $\CP^2 \# \overline{\CP^2}$}
\author{Dmitri Bykov\footnote{Emails:
dmitri.bykov@physik.uni-muenchen.de, dbykov@mi.ras.ru}  \\ \\
{\small $\bullet$ Arnold Sommerfeld Center for Theoretical Physics,}\\{\small Department f\"ur Physik,
Ludwig-Maximilians-Universit\"at M\"unchen,}\\{\small Theresienstra{\ss}e 37, 80333 M\"unchen, Germany,}
\\
{\small $\bullet$ Max-Planck-Institut f\"ur Gravitationsphysik, Albert-Einstein-Institut,} \\ {\small Am M\"uhlenberg 1, D-14476 Potsdam-Golm, Germany} \\ {\small $\bullet$ Steklov
Mathematical Institute of Russ. Acad. Sci.,}\\{\small Gubkina str. 8, 119991 Moscow, Russia \;}}
\date{}
\maketitle
\vspace{-0.8cm}
\begin{center}
\line(1,0){450}
\end{center}
\vspace{-0.3cm}
\textbf{Abstract.} We describe a framework for constructing the Ricci-flat metrics on the 
total space of the canonical bundle over $\CP^2 \# \overline{\CP^2}$ (the del Pezzo surface of rank one). We construct explicitly the first-order deformation of the so-called `orthotoric metric' on this manifold. We also show that the deformation of the corresponding conformal Killing-Yano form does not exist.
\vspace{-0.4cm}
\begin{center}
\line(1,0){450}
\end{center}
\begin{textblock}{4}(9.7,-7.5)
\underline{LMU-ASC 76/17}
\end{textblock}

\noindent Whereas Ricci-flat metrics on \emph{compact} Calabi-Yau manifolds are difficult to construct, there exist many explicitly known Ricci-flat metrics on \emph{noncompact} Calabi-Yau manifolds (the first examples being \cite{EH}, \cite{GH}, \cite{CdO}). The reason is that these latter metrics possess sufficiently many isometries. The role of these metrics is that they describe the geometry of the compact Calabi-Yau manifold in the vicinity of a singularity, after it has been resolved.

\vspace{0.3cm}
\noindent One particular type of singularity that can occur for a complex Calabi-Yau threefold is that of a complex cone over a complex surface. In this article we will be dealing with a particular case, when the surface is the del Pezzo surface of rank one (also known as a Hirzebruch surface $F_1$) --- the blow-up of $\CP^2$ at one point. Topologically, the blow-up of $\CP^2$ at one point is the same as the connected sum $\CP^2 \# \overline{\CP^2}$ \cite{Gompf}, where $\overline{\CP^2}$ means $\CP^2$ with inverted orientation.
In fact, one explicit Ricci-flat metric on
\bea\label{Ydef}
Y:=\;\textrm{Total space of the canonical bundle over}\; \CP^2 \# \overline{\CP^2}
\eea
is known \cite{LuPope1} -- it is a metric that can be obtained by the so-called `orthotoric ansatz'~\cite{Gauduchon} and later will be referred to as the orthotoric metric. This ansatz follows from the requirement that the corresponding metric possesses a conformal Killing-Yano form of type $(1, 1)$ with respect to the Hodge decomposition \cite{Gauduchon, Moroianu}. The main results of the present paper concern the study of the first-order deformation of the orthotoric metric: 

\begin{thm}\label{prop1}
There exists a first-order Ricci-flat deformation $\delta g$ of the orthotoric metric on~$Y$. This deformation corresponds to a change $\delta \omega$ of the K\"ahler class of the metric that lies in the compactly supported cohomology group $\delta \omega\in H^2_c(Y, \mathbb{R})$. The metric, before and after the deformation, is asymptotic to the metric cone over the Sasaki-Einstein manifold~$Y^{2,1}$.
\end{thm}
\begin{thm}
There does not exist a deformation of the conformal Killing-Yano tensor, corresponding to the deformation of the metric.
\end{thm}

\vspace{0.3cm}
\noindent The structure of the paper is as follows.

\vspace{0.3cm}
\noindent In \S\,\ref{torgeom} for completeness of the exposition we recall the salient aspects of toric differential geometry, which are well-known but necessary for the foregoing discussion. Most importantly, we introduce the `master' function that determines the metric on a toric K\"ahler manifold -- the so-called symplectic potential $G$.

\vspace{0.3cm}
\noindent In \S\,\ref{pezzo} we introduce the manifold $Y$ as a toric manifold. 

\vspace{0.3cm}
\noindent In \S\,\ref{diff} we write out a Ricci-flatness equation for the metric on $Y$. In \S\, \ref{biangle} we introduce the moment polytope for a $U(1)^3$ action on $Y$. We explain that most of the information is in fact encoded in a two-dimensional slice of this polytope, which is an unbounded polygon. We describe its topological properties and, in particular, determine the normal bundles of the two $\CP^1$'s embedded in the corners of the polygon.

\vspace{0.3cm}
\noindent In \S\,\ref{3linesec} we review a particular solution of the Ricci-flatness equation -- it has the form of a metric cone, i.e. it defines a metric of the type $ds^2=dr^2+r^2 \widetilde{ds^2}$. The expression for $\widetilde{ds^2}$ can be found explicitly and leads to the Sasakian manifolds $Y^{p, q}$. In \S\,\ref{infasympt} we show how the topology of the underlying del Pezzo cone fixes the Sasakian manifold to be $Y^{2, 1}$. 

\vspace{0.3cm}
\noindent In \S\,\ref{unique} we prove that the solution of the Ricci-flatness equation is unique, once the moment polytope is specified. This is similar in spirit to the proof of \cite{Calabi}, the main difference being in the analysis of the behavior at infinity -- the issue arises due to the non-compactness of the cone. The key technical result is the lower bound for the first non-zero eigenvalue of the Laplacian on $Y^{2,1}$, which is the subject of Lemma\;\ref{eigenvaluelemma}. The result of this section implies that the only potential moduli of the metric are the moduli of the moment polytope.

\vspace{0.3cm}
\noindent In oder to introduce the known metric on $Y$ -- the orthotoric metric -- we come in \S\,\ref{KYforms} to the discussion of conformal Killing-Yano forms (CKYF), with particular emphasis on such forms on Calabi-Yau manifolds. In \S\,\ref{KY20part} we show that the $(2, 0)$ part of such a form is highly constrained -- we show that a vector `dual' to the $(2, 0)$ part of a conformal Killing-Yano form has to be a zero-vector of the Riemann tensor (Proposition\;\ref{Riemannzerovector}). If one insists that the $(2, 0)$ part is zero, i.e. the form is of type $(1, 1)$, one arrives at an object termed twistor form or Hamiltonian 2-form\footnote{These are not exactly the same but related to each other in a simple way \cite{Moroianu}.}, and the existence of such an object severely constrains the metric \cite{Gauduchon}. We review the calculations of \cite{Gauduchon} in \S\,\ref{CKYF11}, the main results summarized in Lemmas\;\ref{gaud1}-\ref{gaud2}. The expression for the orthotoric metric (which is the metric that admits a Hamiltonian 2-form) is given in \S\,\ref{orthometric}.

\vspace{0.3cm}
\noindent In \cite{Coevering}, \cite{Goto} a claim was put forward that the Calabi-Yau theorem holds for asymptotically-conical non-compact Ricci-flat manifolds, of which $Y$ is an example. This is a generalization of the asymptotically-locally-Euclidean (ALE) case previously considered in~\cite{Joyce}. Since we have the explicit orthotoric metric at hand, we may test the proposal directly, by deforming the metric.
 In \S\,\ref{orthodeform} we construct a first-order deformation of the orthotoric metric, compatible with the topological properties of $Y$. We show that the corresponding variation of the K\"ahler form belongs to the compactly-supported cohomology group~$H^2_c(Y, \mathbb{R})$.
 
 \vspace{0.3cm}
 \noindent The next question to be answered is whether the deformed metric as well admits a conformal Killing-Yano form. For that to be the case, the $(2, 0)$-part of the deformed form would have to be non-zero, as the CKYF of type $(1, 1)$ completely fixes the metric to be of orthotoric form. As we proved earlier in \S\,\ref{KY20part}, however, that would imply that the Riemann tensor of the orthotoric metric has a zero-vector. In \S\,\ref{KYdeform} we show that this is not the case. Therefore the deformed metric does not admit a conformal Killing-Yano form.
 
 \vspace{0.3cm}
\noindent The question of whether the first-order deformation of the metric may be extended to a finite one could, at least in principle, also be answered with the help of our methods. An affirmative answer would then constitute (at least locally in K\"ahler moduli space) an alternative proof to the Calabi-Yau theorem for the manifold~$Y$. To this end, one should recall that the first-order deformation can be extended to a finite one by means of an inverse function argument. It turns out, however, that in the language we use -- the one of a symplectic potential $G$ defined on a domain, which is the moment polytope of $Y$, -- the linearized equation is a degenerate elliptic equation (the corresponding quadratic form degenerates at the boundary of the moment polytope), and there does not seem to be a readily available answer to the question of whether this operator may be inverted in the relevant weighted Banach spaces (despite a long history of the subject of boundary-degenerate problems, which started with the seminal work \cite{Keldysh}). 

\vspace{0.3cm}
\noindent
There are several appendices:\\
A. We present an explicit derivation of the metric (\ref{metric}).\\
B. We find a canonical form for the vector fields generating $U(2)\times U(1)$ action on a three-dimensional (complex) manifold.\\
C. Contains some technical results pertaining to \S\,\ref{infasympt}.\\
D. We find a rational parametrization for the space of polynomials of the form $x^3-{3\over 2} x^3+d, d\in\mathbb{R}$, encountered in the Ricci-flat metrics built using the orthotoric ansatz.\\
E. Contains the derivation of a one-parametric generalization of the `unresolved' solution (with a conical singularity), discussed in \S\,\ref{3linesec}.\\
F. We show how the Ricci-flatness equation for a K\"ahler metric with the relevant symmetries may be obtained from a variational problem, akin to the one of optimal transport theory.\\
G. We  review the formal definition of a conformal Killing-Yano tensor (form).\\
H. At the example of Taub-NUT we discuss the possibility of having non-holomorphic Killing vector fields on Calabi-Yau twofolds.

\ssection{Aspects of toric differential geometry}\label{torgeom}

{\normalsize
Most of the statements in this section may be easily generalized to an arbitrary number of dimensions, but for concreteness we will limit ourselves to complex threefolds. On a toric threefold we may choose the complex coordinates $(u_1, u_2, u_3)$ in such a way that the torus $U(1)^3$ acts simply by shifts of these coordinates: $u_k \to u_k+i\,\beta_k$\;($\beta_k\in \mathbb{R}$), i.e. the holomorphic Killing vector fields are $K_j=\mathrm{Re}\left(i\,\frac{\dd}{\dd u_j}\right)$. The K\"ahler potential that is preserved under these shifts has the form
\bea\label{Kahpot1}
K=K(\underbracket[0.6pt][0.6ex]{u_1+\bar{u}_1}_{:=x_1}, \underbracket[0.6pt][0.6ex]{u_2+\bar{u}_2}_{:=x_2}, \underbracket[0.6pt][0.6ex]{u_3+\bar{u}_3}_{:=x_3})\,.
\eea
The moment maps are $\mu_j=\frac{\dd K}{\dd x_j}$. It is convenient to introduce the dual symplectic potential $G$ -- the Legendre dual of $K$:
\bea
G(\mu_1, \mu_2, \mu_3)=\sum\limits_{j=1}^3\,\mu_j\,x_j-K(x_1, x_2, x_3)\,.
\eea
In terms of $G$, the metric corresponding to the K\"ahler potential (\ref{Kahpot1}) has the form (here $\phi_i=\mathrm{Im}\,u_i$)
\bea\label{metr}
ds^2={1\over 4}\,G_{ij}d\mu^i d\mu^j+(G^{-1})^{ij}d\phi_id\phi_j\,.
\eea
The K\"ahler form is $g_K=\sum\limits_{k=1}^3\,d\mu_k\wedge d\phi_k$. The potential $G$ for flat space $\CC^3$ is
\vspace{-0.4cm}
\bea\label{Gflatspace}
G_{\mathrm{flat}}=\sum\limits_{k=1}^3\;\mu_k\,(\log \mu_k-1)\,.
\eea
\noindent
On a K\"ahler manifold the only non-zero Christoffel symbols are $\Gamma^i_{jk}$ and $\Gamma^{\bar{i}}_{\bar{j}\bar{k}}$. The only non-zero components of the Riemann tensor are, accordingly,
\bea\label{Riemcurvkah}
R^i_{\;jk\bar{n}}=-\dd_{\bar{n}}\Gamma^i_{jk}
\eea
and their complex conjugates. The K\"ahler metric, Christoffel symbols and the curvature tensor (\ref{Riemcurvkah}) of a toric manifold have a particularly simple expression in the moment map variables:
\bear\label{GammaRicci}
g_{i\bar{j}}=\frac{\dd^2 K}{\dd x_i \dd x_j}, \quad\quad\quad
\Gamma^i_{jk}=\frac{\dd G^{-1}_{jk}}{\dd \mu_i},\quad\quad\quad
R^i_{\;jk\bar{n}}=-\sum\limits_s\;G^{-1}_{ns}\,\frac{\dd^2 G^{-1}_{jk}}{\dd \mu_s \dd \mu_i}\,.
\eear
Here $G^{-1}_{jk}$ means the $jk$-component of the matrix inverse to the Hessian of $G$. It is also useful to write out the expression for the Riemann tensor with all lower indices:
\bea\label{RiemHerm}
R_{\bar{m}jk\bar{n}}=-\sum\limits_{s, t}\;G^{-1}_{ns}\,\frac{\dd^2 G^{-1}_{jk}}{\dd \mu_s \dd \mu_t}\,G^{-1}_{tm}.
\eea
One can check directly that it has all the correct symmetry properties of the Riemann tensor\footnote{Note also the following additional symmetry property. Since the Riemann tensor is real in real coordinates, one has, in general, $R_{\bar{m}jk\bar{n}}^\ast=R_{\bar{j}mn\bar{k}}$. In the particular toric coordinates that we are using, however, the Hermitian components (\ref{RiemHerm}) of the Riemann tensor are real as well, therefore we have the symmetry property $R_{\bar{m}jk\bar{n}}=R_{\bar{j}mn\bar{k}}$, i.e. a symmetry under the simultaneous exchange $m\leftrightarrow j, k\leftrightarrow n$. It is not immediately obvious from the expression (\ref{RiemHerm}) but can be checked directly.}. A useful immediate check is the verification that the curvature vanishes for the symplectic potential (\ref{Gflatspace}) of flat space $\CC^3$. 

\vspace{0.3cm}
\noindent
The Ricci tensor of the metric (\ref{metr}) is obtained from (\ref{GammaRicci}) by contracting indices:
\bea
R_{i\bar{j}}=\sum\limits_{p, s}\;G_{js}^{-1}\,\frac{\dd}{\dd \mu_s}\,\left(G_{ip}^{-1}\frac{\dd}{\dd \mu_p}\,\log\, \mathrm{Det\; Hess}\,G\right)\,.
\eea
The Ricci-flatness equation $R_{i\bar{j}}=0$ may be integrated to give
\bea\label{Ricciflat1}
\mathrm{Det\,Hess}\,G= a\,e^{\sum\limits_k\, b^k {\dd G\over \dd \mu^k}}
\eea
One of the benefits of using the symplectic potential $G$ in place of the K\"ahler potential $K$ is that the domain in $\mu$-space, on which $G$ is defined, is the moment polytope of the toric manifold. From the perspective of the equation (\ref{Ricciflat1}), it is the singularities of the function $G$ that determine the polytope. It is known 
\cite{Guillemin} that in the simplest case of a (generally non-Ricci-flat) metric induced by a K\"ahler quotient of flat space with respect to an action of a complex torus, the potential $G$ takes the form of a superposition of `hyperplanes':
\bea\label{Guilleminform}
G_{\mathrm{toric}}=\sum\limits_{i=1}^M\,\ell_i \,(\log{\ell_i}-1)\quad\mathrm{with}\quad \ell_i=\sum\limits_k c_{ik}\, \mu_k+d_i\, .
\eea
In general, a potential $G$ satisfying (\ref{Ricciflat}) will not have this form. However, we will assume that it has the corresponding \emph{asymptotic} behavior at the faces of the moment polytope. More exactly, when we approach an arbitrary face $\ell_i$, i.e. when $\ell_i \to 0$, we impose the asymptotic condition
\bea\label{asymptcond}
G=\ell_i \,(\log{\ell_i}-1)+\ldots\quad\mathrm{as}\quad \ell_i \to 0,
\eea
where the ellipsis indicates terms regular at $\ell_i \to 0$. Despite being subleading, they are important for the equation (\ref{Ricciflat}) to be consistent even in the limit $\ell_i \to 0$.

\vspace{0.3cm}
\noindent
Moment polytopes of toric symplectic manifolds are rather constrained -- they must possess Delzant properties \cite{Delzant}:
\begin{itemize}
\item They are simple: at every vertex exactly $N$ faces meet, $2N$ being the dimension of the manifold. In our case $N=3$.
\item The normals to the faces $c_{ik}$ are integer-valued: $c_{ik}\in\mathbb{Z}$. \\Moreover, the normals to the three faces meeting at each vertex form a basis of $\mathbb{Z}^3$.
\end{itemize}
Let us consider a vertex of the polytope and label the three hyperplanes meeting at this vertex as $\ell_1=0$, $\ell_2=0$, $\ell_3=0$. Then, in the notations of (\ref{Guilleminform}) the second Delzant property means that 
$\sum\limits_{i=1}^3 \, f_{ji}\,c_{ik}=\delta_{jk}$, where $f_{ji}\in \mathbb{Z}$. Therefore $c, f\in GL(3, \mathbb{Z})$. In other words, the inverse of the matrix $c$ is integer-valued as well. The meaning of the condition of integrality of the normal vectors may be understood by analyzing the metric~(\ref{metr}) in the vicinity of an angle of the moment polytope, defined by $\ell_1=\ell_2=\ell_3=0$. According to (\ref{asymptcond}), near such an angle the potential is asymptotically approximated by $G=\sum\limits_{k=1}^3\, \ell_k \,(\log{\ell_k}-1)+\ldots$ We may now make a linear change of variables from $(\mu_1, \mu_2, \mu_3)$ to $(\ell_1, \ell_2, \ell_3)$. The metric (\ref{metr}) then reads
\bea
ds^2=\sum\limits_{i=1}^3\,\left(\frac{d\ell_i^2}{4 \,\ell_i}+\ell_i\,d\widetilde{\phi}_i^2\right)+\ldots\,\quad\quad\textrm{where}\quad\quad \widetilde{\phi}_i=\sum\limits_j\,f_{ji}\,\phi_j\,.
\eea
The metric in brackets is the metric of flat space, if $\widetilde{\phi}_i$ have periodicity $2\pi$. Otherwise the metric has a conical singularity. The map $\phi\to \widetilde{\phi}$ is an automorphism of the torus $\mathbb{T}^3$ if and only if $c\in GL(3, \mathbb{Z})$, which is precisely the second Delzant condition.

}

\ssection{The resolved cone over the del Pezzo surface}\label{pezzo}

In this paper we will be constructing Ricci-flat metrics on the manifold $Y$ introduced in~(\ref{Ydef}).
The manifold $\CP^2 \# \overline{\CP^2}$ is diffeomorphic to the del Pezzo surface of rank one (or, equivalently, of degree $8$) \cite{Gompf} -- the blow-up of $\CP^2$ at one point. This surface is further denoted by $\dP_1$, and we will mostly use this abbreviation in what follows.
It is a compact simply-connected K\"ahler manifold of complex dimension 2, such that $H^2(\dP_1, \mathbb{Z})=\mathbb{Z}^2$, and the intersection pairing on $H^2(\dP_1, \mathbb{Z})$ has the form $\left( \begin{array}{ccc}
1 & 0 \\
0 & -1 \end{array} \right)$.
\vspace{0.5cm}
Denoting the corresponding de-Rham generators of $H^2(Y, \mathbb{R})\simeq H^2(\dP_1, \mathbb{R})$ by $\omega_1$, $\omega_{-1}$, we may write the K\"ahler class $\Omega=[g_K]$ of the metric $g$ on $Y$ as follows:
\bea\label{Kahclass}
\Omega=a\,\omega_1+b\,\omega_{-1}\,,\quad a, b\in \mathbb{R}\,.
\eea

\begin{figure}
\floatbox[{\capbeside\thisfloatsetup{capbesideposition={right,top},capbesidewidth=6cm}}]{figure}[\FBwidth]
{\caption{The $(\alpha, \beta)$ section of the moment polytope of $Y$.
    The marked points have coordinates $A=({\eta_2\over 3}, {\eta_2\over 3}), B=({\eta_2\over 3}-\eta_1, {\eta_2\over 3})$. The moment polytope of $Y$ has five faces whenever the projection has three edges. This happens when the two conditions in (\ref{Kahlercone}) are satisfied.} \label{mompol1}} 
{\includegraphics[width=0.58\textwidth]{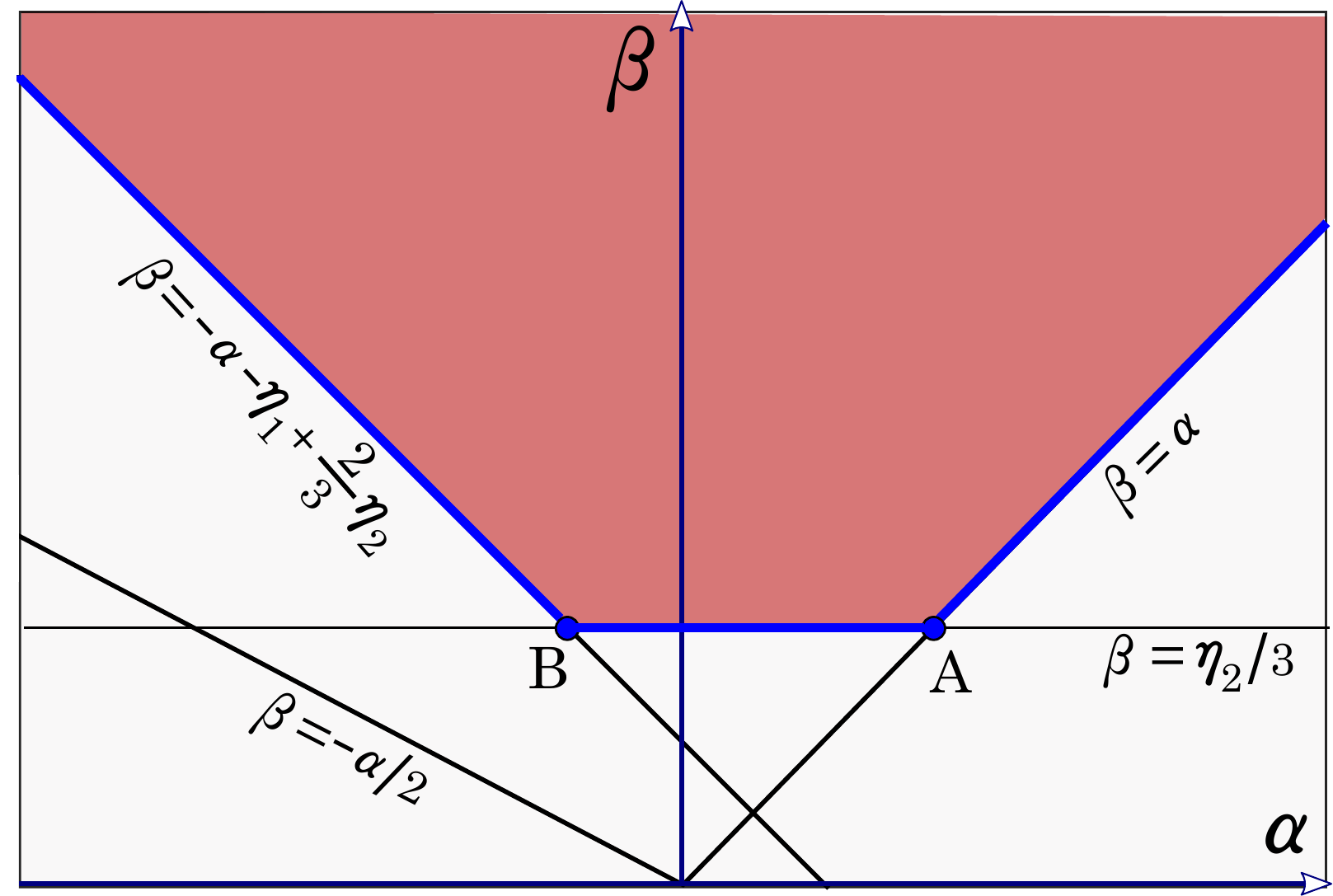}}
\end{figure}

\noindent The space $Y$ has a representation in terms of a GIT quotient (see \cite{Greene}, Table 1):
\bea\label{GIT1}
Y=\CC^5/(\CC^\ast)^2,
\eea
with the charge vectors given by
\bear\label{GIT2}
\vec{v}_1=(0, 0, 1, 1, -2),\quad\quad \vec{v}_2=(1, 1, 1, 0, -3)
\eear

\noindent $Y$ is a toric K\"ahler manifold, and the representation (\ref{GIT1})-(\ref{GIT2}) allows to build the associated moment polytope $\triangle$:
\bear
\triangle=\!\!\!\!\!\!\!\!\!&&\{\mu_3+\mu_4-2\mu_5=\eta_1, \\ \nonumber && \mu_1+\mu_2+\mu_3-3\,\mu_5=\eta_2\}\subset \RR^5_+=\{\mu_i\geq 0\}
\eear
Clearly, the equations in figure brackets define a three-dimensional space, which we will parametrize by means of the coordinates $\alpha, \beta, \gamma$. These are related to $\mu_k$ as follows:
\bear\nonumber
&\mu_1=\alpha+\beta+\gamma,\quad \mu_2=\beta-\gamma,\quad \mu_3=\beta-\alpha,&\\ \label{musols}  &\mu_4=\eta_1-{2\,\eta_2\over 3}+\beta+\alpha,\quad \mu_5=-{\eta_2\over 3}+\beta&
\eear
The inequalities defining the polytope are now $\mu_k\geq0, k=1\ldots 5$. In the $(\alpha, \beta)$ projection we have the following inequalities:
\bear\nonumber
&\mu_1+\mu_2=2\beta+\alpha\geq0,\quad \mu_3=\beta-\alpha\geq 0,&\\ \nonumber &\mu_4=\eta_1-{2\,\eta_2\over 3}+\beta+\alpha\geq 0,\quad \mu_5=-{\eta_2\over 3}+\beta\geq 0&
\eear

\vspace{0.3cm}
\noindent
The relevant chamber in the $(\eta_1, \eta_2)$-space is where the polytope has five faces -- this chamber is defined by (see Fig.~\ref{mompol1})
\bea\label{Kahlercone}
0<\eta_1<\eta_2\,.
\eea
The parameters $\eta_1, \eta_2$ are related to the cohomological parameters $a, b$ of (\ref{Kahclass}). Speaking more invariantly, the K\"ahler moduli are moduli of the moment polytope. To find the relation, we can build a K\"ahler quotient metric on $Y$. The corresponding K\"ahler form is
\bea
\Omega=d\alpha\wedge d\varphi_1+d\beta\wedge d\varphi_2+d\gamma\wedge d\varphi_3,
\eea
where $(\varphi_1, \varphi_2, \varphi_3)$ are global angular variables associated to the moment map variables $(\alpha, \beta, \gamma)$.

\vspace{0.3cm}
\noindent
In section~\ref{biangle} we will find that, as generators of $H_2(Y, \mathbb{R})$ one can take the two spheres, which are the edges of the moment polytope lying at points $A$ and $B$ orthogonal to the section shown in Fig.~1. Under the isomorphism $H_2(Y, \mathbb{R})\simeq H^2(Y, \mathbb{R})$, the corresponding generators are the forms $\omega_1$, $\omega_{-1}$ from (\ref{Kahclass}), obeying the following relations:
\bea\label{omegapm1}
\int\limits_{\CP^1_A}\,\omega_1=1,\quad  \int\limits_{\CP^1_B}\,\omega_{-1}=1,\quad \int\limits_{\CP^1_B}\,\omega_1=0,\quad \int\limits_{\CP^1_A}\,\omega_{-1}=0.
\eea
Therefore we see from (\ref{Kahclass}) that $a$ and $b$ are integrals of the K\"ahler form over the corresponding cycles:
\bea
a=\int\limits_{\CP^1_A}\,\Omega,\quad\quad b=\int\limits_{\CP^1_B}\,\Omega
\eea
Let us calculate, for instance, the integral over $\CP^1_A$. The corresponding edge of the polytope is defined by $\mu_3=\mu_5=0$.
Since on $\CP^1_A$ we have $d\alpha=d\beta=0$, the integral is
\bea
a=2\pi\,\int\limits_{\mu_1=0}^{\mu_2=0}\,d\gamma=2\pi\,\int\limits_{-2\eta_2\over 3}^{\eta_2\over 3}\,d\gamma=2\pi\,\eta_2
\eea
Analogously
\bea
b=2\pi\,(\eta_2-\eta_1)\,,
\eea
hence we have $a>0, \,b>0$ and the ratio $b\over a$ is bounded as follows:
\bea\label{babound}
0<{b\over a}<1\,.
\eea
The bound (\ref{babound}), together with (\ref{Kahclass}), define the K\"ahler cone of $Y$.

\ssubsection{Compactly supported cohomology}\label{compsuppcoh}

An interesting refined description of the cohomology of $Y$ may be found in \cite{Coeveringcompactcoh}. To explain it, we will have to slightly jump ahead in our exposition and accept the fact (explained in subsequent sections) that, at infinity, the metric on $Y$ has the form of a Riemannian cone over a Sasaki-Einstein manifold $S$ (for the particular case that we are considering we will have $S=Y^{2,1}$, see \S\,\ref{3linesec} and \S\,\ref{infasympt} for definitions, as well as~\cite{Sparks} for a comprehensive review of Sasaki-Einstein manifolds), i.e. the metric at infinity is of the form
\bea\label{conemetr1}
(ds^2)_{\infty}=dr^2+r^2\,(\widetilde{ds^2})_S\,.
\eea
Here $r$ is a certain function on $Y$. In particular, this function has the property that, for $r_0$ sufficiently large, the set $Y_{r_0}:=\{r\leq r_0\subset Y\}$ is a compact manifold with boundary $S$, i.e. $\dd Y_{r_0}\simeq S$. One can then consider the relative cohomology $H^2(Y_{r_0}, S, \mathbb{R})$, which, by definition, is the compactly supported cohomology $H^2_c(Y):=H^2(Y_{r_0}, S, \mathbb{R})$. Using the long exact sequence for relative cohomology and certain facts about $Y$ and $S$, one derives~\cite{Coeveringcompactcoh} that the following sequence is exact:
\bea\label{exactseq}
0\rightarrow H^2_c(Y, \mathbb{R})\rightarrow H^2(Y, \mathbb{R})\rightarrow H^2(S, \mathbb{R})\rightarrow 0\,.
\eea
As we will see below, in the case of interest we have $S=Y^{2,1}$, and topologically $Y^{2,1}\simeq S^2\times S^3$, hence $H^2(S, \mathbb{R})\simeq \mathbb{R}$. Since, as we discussed above, $H^2(Y, \mathbb{R})=\mathbb{R}^2$, we deduce that $H^2_c(Y, \mathbb{R})=\mathbb{R}$.

\vspace{0.3cm}
\noindent
One way to distinguish a compactly supported two-form $\varpi$ is by its decay rate at infinity ($r\to \infty$). Indeed, let $g_0$ be the conical metric, i.e. $(ds^2)_{g_0}=dr^2+r^2\,(\widetilde{ds^2})_{S}$. Then we have the following result:

\begin{lem}\label{decaylemma}
\cite{Coeveringcompactcoh}\; Suppose $\|\varpi\|_{g_0}=O(r^{-\alpha})$ for $\alpha>2$. Then $\varpi\in H^{2}_c(Y)$.
\end{lem}

\vspace{0.3cm}
\noindent\underline{Proof.}\\
According to (\ref{exactseq}), a two-form $\varpi$ lies in $H^{2}_c(Y)$ whenever it is in the kernel of the map $H^2(Y, \mathbb{R})\rightarrow H^2(S, \mathbb{R})$. This map, in turn, is the `restriction to the boundary' map. Therefore to check whether $\varpi\in H^{2}_c(Y)$, we need to check whether its restriction $\varpi\big|_S$ is trivial in $H^2(S, \mathbb{R})$. On the other hand, $\varpi\big|_S$ is trivial if for any three-form $\Lambda\in H^3(S, \mathbb{R})$ one has $\int\limits_S \,\varpi\big|_S\wedge \Lambda=0$. Now, here by $\varpi\big|_S$ we actually mean the restriction $\varpi\big|_{r=r_1}$ for some sufficiently large $r_1$. We may now extend the form $\varpi\big|_{r=r_1}\wedge \Lambda$, defined on $S$, to a form $\varpi\wedge \Lambda$ defined on $S\times I_r$, where $I_r$ is a segment with coordinate~$r$: $I_r=[r_1, r_2]$. The form $\Lambda\in H^3(S, \mathbb{R})$ is extended trivially, and the form $\varpi$ is closed on $S\times I_r$, since it was closed from the start. Therefore, by Stokes theorem, $\int \,\varpi\big|_{r=r_1}\wedge \Lambda=\int \,\varpi\big|_{r=r_2}\wedge \Lambda$. We may now use the decay rate of $\varpi$ to calculate the integral in the limit $r_2\to \infty$. Since $\varpi\big|_{r=r_2}\wedge \Lambda=(\varpi, \ast \Lambda)$ is a (point-wise) scalar product between two-forms on $S$, we may use the Cauchy inequality
\bea\label{Cauchyineq}
\big|\int \,\varpi\big|_{r=r_2}\wedge \Lambda\big|\leq \int\limits_{S=\partial Y_{r_2}}\, \|\varpi\|_{\tilde{g}}\cdot \|\ast \Lambda\|_{\tilde{g}}\cdot\mathrm{vol}_S\,,
\eea
where $\tilde{g}$ is the metric on $S$, entering the formula (\ref{conemetr1}) above, and the Hodge star $\ast$ again refers to $\tilde{g}$. Note that the metric $\tilde{g}$ does not depend on $r_2$. On the other hand, we have the bound for $\|\varpi\|_{g_0}$, rather than $\|\varpi\|_{\tilde{g}}$, but we can easily relate the two. Clearly, $\|\varpi\|_{g_0}^2=g_0^{\alpha\beta}g_0^{\mu\nu}\,\varpi_{\alpha\mu}\,\varpi_{\beta\nu}\geq {1\over r^4}\,\|\varpi\|^2_{\tilde{g}}$, hence $\|\varpi\|_{\tilde{g}}\leq r^2\,\|\varpi\|_{g_0}\leq {\mathrm{const.}\over r^{-(\alpha-2)}}$. Substituting this in (\ref{Cauchyineq}) above, we obtain
\bea
\big|\int \,\varpi\big|_{r=r_2}\wedge \Lambda\big|\leq {\mathrm{const.}\over r_2^{-(\alpha-2)}}\,\int\limits_{S}\,  \|\ast \Lambda\|_{\tilde{g}}\cdot\mathrm{vol}_S\,,
\eea
Supposing $\alpha>2$ and sending $r_2\to\infty$, we find that $\int \,\varpi\big|_{S=\dd Y_{r_0}}\wedge \Lambda=0$ for all $\Lambda\in H^3(S, \mathbb{R})$. As a result, we find that $[\varpi\big|_{S=\dd Y_{r_0}}]=0\in H^2(S, \mathbb{R})$, which, as explained earlier, implies $\varpi\in H^2_c(Y, \mathbb{R})$. $\blacksquare$

\vspace{0.3cm}
\noindent
Another view at the compactly supported cohomology group is via Poincar\'e duality. In fact, this can be described more clearly if we slightly generalize the setup. Let $Y$ be the total space of a vector bundle $V$ of rank $m$ over a surface $X$. The surface is embedded in $Y$ as the zero section, $i: X \hookrightarrow Y$. Using Poincar\'e duality, we can construct the dual compactly-supported form $[i(X)]^\vee \in H^m_c(Y, \mathbb{R})$. It is a classic fact that the restriction of this form to the zero section is the Euler class of the bundle: $[i(X)]^\vee\big|_{i(X)}=eu(V)$ (see \cite{BottTu}, Propositions 6.24~(b) and 6.41). In the case that $V$ is a complex vector bundle, $eu(V)=c_m(V)$. Returning back to our case, we have $m=1$, and moreover $V=K_X$ -- the canonical bundle of $X$. Therefore we have the result $[i(X)]^\vee\big|_{i(X)}=c_1(K_X)=-c_1(X)$. Apart from that, one has $H^{2}_c(Y, \mathbb{R})\simeq H_4(Y, \mathbb{R})\simeq \mathbb{R}$, since the homology of $Y$ is the same as that of the base of the bundle, the surface $X$, for which we have of course $H_4(X, \mathbb{R})\simeq \mathbb{R}$. Therefore, as $H^{2}_c(Y, \mathbb{R})$ is one-dimensional,  $[i(X)]^\vee\big|_{i(X)}\in H^2_c(Y, \mathbb{R})$ is its generator over the real numbers. Summarizing, to find out whether a given two-form $\varpi$ belongs to $H^{2}_c(Y, \mathbb{R})$, we may restrict it to $X$ and check whether it is proportional to $c_1(X)$. To facilitate future use, let us express $c_1(X)$ in terms of the generators $\omega_{\pm 1}$ featuring in (\ref{Kahclass}):
\bea
c_1(X)=-3 \,\omega_1-\omega_{-1}\,.
\eea
Here by $\omega_{\pm 1}$ we mean, again, the restrictions of these forms to $X$. This is essentially the formula $K=-3H+E$, where $K$ is the canonical divisor, $H$ is the hyperplane divisor and $E$ the exceptional divisor of the blow-up (the ($-1$)-curve). The relative sign in front of $\omega_{-1}$ is due to our normalizations (\ref{omegapm1}). The result of this discussion may be reformulated as follows: for a form $\varpi$ to belong to $H^{2}_c(Y, \mathbb{R})$, one should have
\bea\label{critcomp}
\frac{\int\limits_{\CP^1_A}\,\varpi|_X}{\int\limits_{\CP^1_B}\,\varpi|_X}=3\,.
\eea
The theory that we have reviewed was used in \cite{Coevering} to formulate a version of the Calabi-Yau theorem relevant for the case of asymptotically-conical manifolds:

\begin{thm} \cite{Coevering}, \cite{Goto}\;Let $Y_0$ be the manifold with a conical singularity, equipped with the metric (\ref{conemetr1}), that we will denote $g_0$. Let $\pi: Y\to Y_0$ be the Ricci-flat resolution of the conical singularity. Then in every K\"ahler class in $H_c^2(Y, \mathbb{R}) \subset H^2(Y, \mathbb{R})$ there is a unique Ricci-flat K\"ahler metric $g$ asymptotic to $g_0$ as follows
\bea\label{estim1}
|\pi_\ast g-g_0|_{g_0}=O\left(\frac{1}{r^6}\right)\quad\quad \textrm{for}\quad\quad r\to \infty\,.
\eea
Furthermore, in every K\"ahler class in $H^2(Y, \mathbb{R}) \setminus H_c^2(Y, \mathbb{R})$ there is a Ricci-flat metric $g$ asymptotic to $g_0$ with the following decay estimate:
 \bea\label{estim2}
|\pi_\ast g-g_0|_{g_0}=O\left(\frac{1}{r^2}\right)\quad\quad \textrm{for}\quad\quad r\to \infty\,.
\eea
In both cases the derivatives of the metric decay appropriately.
\end{thm}

\vspace{0.3cm}
\noindent The decay estimates of the type above were introduced in \cite{Joyce} in a proof of an analogous Calabi-Yau type theorem for asymptotically locally-Euclidean spaces. The Proposition above is a generalization thereof for asymptotically-conical manifolds. 
A significant part of the present paper will be dedicated to certain explicit checks and illustrations for the statements contained in the Proposition.

\ssubsection{Example. The total space of the canonical bundle over $\CP^1\times \CP^1$.}

A simple example where most of the above assertions may be checked directly is that of a cone over the surface $X_0:=\CP^1\times \CP^1$. The explicit Ricci-flat metric on such manifold was constructed in~\cite{PZTmain} by means of the same ansatz that was used earlier in \cite{CdO}. The ansatz for the K\"ahler potential has the form:
\bea\label{KahCP1}
K=a\,\log(1+|z|^2)+K_0\left(\underbracket[0.6pt][0.6ex]{|u|^2(1+|z|^2)(1+|w|^2)}_{:=x}\right)\, ,
\eea
where $a$ is a certain parameter (K\"ahler modulus) whose meaning will be clarified later. The vector fields $k_1=\mathrm{Re}\left(i\,z\,{\dd\over \dd z}\right)$, $k_2=\mathrm{Re}\left(i\,w\,{\dd\over \dd w}\right)$ and $k_3=\mathrm{Re}\left(i\,u\,{\dd\over \dd u}\right)$, generating phase rotations for the local complex variables $z, w, u$, are clearly Killing. If we denote $|z|^2=e^t, |w|^2=e^s, |u|^2=e^v$, the moment maps are simply derivatives of the K\"ahler potential w.r.t. the corresponding real variables:
\bea\label{mux}
\mu_1=\frac{\dd K}{\dd t}=\frac{e^t}{1+e^t}\,\left(a+x\,K_0'\right),\quad
\mu_2=\frac{\dd K}{\dd s}=\frac{e^s}{1+e^s}\,x\,K_0',\quad 
\mu=\frac{\dd K}{\dd v}=x\,K_0'\,.
\eea
The Ricci-flatness equation is most conveniently expressed in terms of the function $\mu(x)$:
\bea
\mu\,(a+\mu)\,\mu'=\beta\,x,\quad\quad \beta=\mathrm{const.},
\eea
which may be integrated to give
\bea\label{mux2}
{\mu^3\over 3}+a\,{\mu^2\over 2}=\beta\,{x^2\over 2}-{\kappa \over 3}\,.
\eea
The $K_0$-part of the K\"ahler potential may be then obtained from the definition (\ref{mux}):
\bea\label{K0}
K_0=\int\,\frac{\mu}{x}\,dx=\int\,\frac{\mu^2\,(a+\mu)}{\beta\, x^2}\,d\mu=\textrm{using}\;(\ref{mux2})={3\over 2}\,\int\,\frac{\mu^2\,(a+\mu)\,d\mu}{\mu^3+{3a\over 2}\,\mu^2+\kappa}
\eea
Upon taking the integral, we obtain the following expression for the symplectic potential $G$ ($\simeq$ means `up to a linear function', which is irrelevant):
\bear\label{GCP1}
&&G:=\mu\,v+\mu_1\,t+\mu_2\,s-K\simeq \\ \nonumber&&\simeq{1\over 2}\,\sum\limits_{i=1}^3\,(\mu-\lambda_i)\,\log(\mu-\lambda_i)-\mu\,\log\,\mu-(\mu+a)\,\log(\mu+a)+\\ \nonumber && +\mu_1\,\log\,\mu_1+\mu_2\,\log\,\mu_2+(\mu-\mu_2)\,\log(\mu-\mu_2)+(\mu-\mu_1+a)\,\log(\mu-\mu_1+a)\, ,
\eear
where $\lambda_i,\,i=1, 2, 3$ are the roots of the polynomial
\bea\label{fpol}
f(\mu)=\mu^3+{3a\over 2}\,\mu^2+\kappa\,,
\eea
which enters the denominator of the integrand in (\ref{K0}). We choose the ordering $\lambda_3 \geq \lambda_2 \geq \lambda_1$ if all roots are real, otherwise $\lambda_3$ denotes the real root.

\vspace{0.3cm}
\noindent
The region in the parameter space, which corresponds to the manifold being the total space of the canonical bundle over $\CP^1\times\CP^1$, is the following:
\bea\label{lambdapos}
\lambda_3>0,\quad\quad a+\lambda_3>0\,.
\eea
Indeed, in this case the metric on the underlying surface $\CP^1\times\CP^1$ may be recovered from (\ref{KahCP1}) by taking the limit $x\to 0$. This corresponds to sending $\mu\to \lambda_3$, see (\ref{mux2}). Since $\mu=x K_0'$, in the limit $x\to 0$ we have $K_0\simeq \lambda_3\,\log\,x+\ldots$, therefore the full K\"ahler potential reduces to\footnote{Omitting the contribution $\lambda_3 \,\log|u|^2$, which does not affect the metric.}
\bea
K\simeq (a+\lambda_3)\,\log(1+|z|^2)+\lambda_3\,\log(1+|w|^2)+\ldots
\eea
We see that $\lambda_3$ and $a+\lambda_3$ are the squared radii of the two spheres and therefore have to be positive, leading to (\ref{lambdapos}). Note in passing, that once this bound is established, the equations defining the moment polytope may be read off from (\ref{GCP1}):
\bea\label{mompolcp1}
\mu\geq \lambda_3,\quad\quad \mu+a\geq \mu_1\geq 0,\quad\quad \mu\geq \mu_2\geq 0\,.
\eea
In particular, going to infinity corresponds to sending $\mu\to \infty$ with the ratios $\mu_1\over \mu$, $\mu_2\over \mu$ bounded. In the formula (\ref{GCP1}) $a$ and $\lambda_3$ are the resolution parameters, and in the limit the are effectively set to zero\footnote{Note that the parameter $\kappa$ may be related to $\lambda_3$ but it is more convenient to treat $\lambda_3$ as the independent parameter.}. The limiting function will be denoted by $G_0$. One can show that this is the symplectic potential defining a metric cone over $T^{1,1}:=\frac{SU(2)\times SU(2)}{U(1)}$ -- the manifold introduced in \cite{Romans}.

\vspace{0.3cm}
\noindent
For the bounds (\ref{estim1})-(\ref{estim2}) to make sense, the difference $\pi_\ast G-G_0$ should be non-singular at infinity in the first place.
The potential singularity is at $\mu=\mu_1$. 
The hyperplane $\mu=\mu_1$ lies outside the moment polytope for $a<0$. Let us first analyze this case. Since $\lambda_3$ is a root of the polynomial $f(\mu)$ from (\ref{fpol}), in the limit $\lambda_3\to 0$, $a\to 0$ the other two roots of the polynomial vanish as well, $\lambda_{1}, \lambda_2\to 0$. Therefore we may expand (\ref{GCP1}) to first order in $\lambda_1, \lambda_2, \lambda_3, a$ as follows
\bear\nonumber
&&G-G_0\simeq-{1\over 2}\,\left(\sum\limits_{i=1}^3\,\mu_i\right)\,\log\,\mu-a\,\log\,\mu+a\log(\mu-\mu_1)+\ldots=\\
&&=-{a\over 4}\,\log\,\mu+a\,\log(\mu-\mu_1)+\ldots
\eear
When passing to the second line, we used the expression $\sum\limits_{i=1}^3\,\mu_i=-{3a\over 2}$ for the sum of roots, which follows from (\ref{fpol}). Let us denote by $g_0$ the metric given by the symplectic potential $G_0$. It is now easy to check that
\bea\label{decay2}
\|g-g_0\|_{g_0}=O\left({a\over \mu}\right)=O\left({a\over r^2}\right)\quad\quad \textrm{for}\quad\quad r\to\infty\quad\textrm{and}\quad a\neq 0\,.
\eea

\vspace{0.3cm}
\noindent
The case $a>0$ may be analyzed similarly, if one makes the change of variables $\mu\to\mu-a$ in the function $G$, before comparing it to $G_0$ (this corresponds to a choice of map $\pi$ in the Proposition). In this case the moment polytope is defined by the inequalities $\mu\geq \lambda_3+a,\; \mu\geq \mu_1\geq 0,\; \mu-a\geq \mu_2\geq 0\,.$ The potential singularity is now at $\mu=\mu_2$ and lies outside of the moment polytope, and the analysis above can be carried through. Note also that, from the point of view of the polynomial $f(\mu)$, the replacement $\mu\to\mu-a$ amounts to a redefinition of $\kappa$ and the required flip of the sign $a\to-a$.

\vspace{0.3cm}
\noindent
The remaining interesting case to be considered is $a=0$. In this case $\lambda_3=(-\kappa)^{1/3}$, and we have from (\ref{GCP1}):
\bea
G-G_0=\frac{\lambda_3^3}{4\,\mu^2}+\ldots\,,
\eea
and one easily obtains
\bea\label{decay3}
\|g-g_0\|_{g_0}=O\left({\lambda_3^3\over \mu^3}\right)=O\left({\lambda_3^3\over r^6}\right)\quad\quad \textrm{for}\quad\quad r\to\infty\quad\textrm{and}\quad a=0\,.
\eea
According to the Proposition, the decay estimates (\ref{decay2}) for $a\neq 0$ and (\ref{decay3}) for $a=0$ correspond to the K\"ahler form being in $H^2(Y, \mathbb{R})\setminus H^2_c(Y, \mathbb{R})$ and in $H^2_c(Y, \mathbb{R})$ respectively. In the case $a=0$ the two spheres at the base of the cone (i.e. the zero section) have equal radii, and therefore the induced metric on the zero section is K\"ahler-Einstein, meaning that its K\"ahler class is indeed proportional to $c_1(\CP^1\times \CP^1)$.

\ssection{The equation of Ricci-flatness}\label{diff}

In the previous section we introduced the variety $Y$ as a K\"ahler quotient of flat space. This definition allows constructing a K\"ahler metric  on~$Y$. However, this metric is by no means Ricci-flat. In the remainder of the paper we will be looking for a Ricci-flat metric on $Y$. To this end, we will start with the most general K\"ahler potential compatible with the symmetries of the problem, and then solve the Ricci-flatness equation that this K\"ahler potential has to satisfy.

\vspace{0.3cm}
\noindent
We mentioned above that the del Pezzo surface of rank one $\dP_1$ may be thought of as the blow-up of one point on $\CP^2$. Without loss of generality let us choose this point to be $(0: 0: 1) \in \CP^2$. The choice of a distinguished point reduces the automorphism group $\mathbb{P}GL(3, \CC)$ of $\CP^2$ to the automorphism group of $\dP_1$:
\bea\label{aut}
Aut (\dP_1)= \mathbb{P}\left\{ G\in GL(3, \CC): G=\left(
\begin{array}{ccc}
\bullet & \bullet & 0  \\
\bullet & \bullet & 0  \\
\bullet & \bullet & \bullet \\
\end{array}
 \right)\right\},
\eea
The linear part of the group of automorphisms of the affine cone over $\dP_1$ (w.r.t. the anti-canonical embedding) is the maximal parabolic subgroup $\mathrm{H}\subset GL(3, \CC)$ defined by matrices of the form (\ref{aut}) (forgetting the projectivization) (see \cite{Prokhorov}, in particular Proposition~2.15 and Theorem~1.5). The resolved cone inherits these automorphisms as well, as the equation of the blow-up is linear in the embedding coordinates.

\vspace{0.3cm}
\noindent
We will be looking for a K\"ahler metric on $Y$ with the isometry group being the maximal compact subgroup of $\mathrm{H}$:
\bea\label{isom}
\mathrm{Isom}(Y)=U(2)\times U(1)
\eea
We will choose local coordinates $z_1, z_2, u$, in which the $\mathfrak{u}(2)\oplus \mathfrak{u}(1)$ action uniformizes, i.e. the holomorphic vector fields generating this action have the form
\bear
&v^{[u(2)]}_{0}=i z_1 {\dd \over \dd z_1}+i z_2 {\dd \over \dd z_2},\quad\quad v^{[u(2)]}_1=z_1 {\dd \over \dd z_2}+z_2 {\dd \over \dd z_1},&\\ \nonumber
&v^{[u(2)]}_2=i z_1 {\dd \over \dd z_2}-i z_2 {\dd \over \dd z_1},\quad\quad v^{[u(2)]}_3=i z_1 {\dd \over \dd z_1}-i z_2 {\dd \over \dd z_2},\quad\quad v^{[u(1)]}=u{\dd \over \dd u}\;.&
\eear
This is always possible: see Appendix \ref{uniformization}. The $U(2)\times U(1)$-invariant K\"ahler potential depends on the two combinations of these variables:
\bea\label{Kahpot}
K=K(|z_1|^2+|z_2|^2, |u|^2)
\eea
The corresponding K\"ahler form is $\Omega= i \dd \bar{\dd} K$ and the metric is $g_{i \bar{j}}=\dd_i\bar{\dd}_j K$.
Since the Ricci tensor is related to the metric of a K\"ahler manifold as $R_{i\bar{j}}=-\dd_i\bar{\dd}_j \log \det g$, the Ricci-flatness (Calabi-Yau) condition $R_{i\bar{j}}=0$ implies that the determinant of the Hermitian metric $g$ has to factorize in a holomorphic and conjugate antiholomorphic pieces: $\det g=|f(z_1, z_2, u)|^2$. As $\det{g}$ is $U(2)\times U(1)$-invariant, it means that $\det g=a\,|u|^{2l}$ for some constants $a, l$. On the other hand, a direct calculation of $\det{g}$ for a metric arising from the K\"ahler potential (\ref{Kahpot}) gives
\bea
\det{g}=8\,e^{-t-s}\;K_t\,\left( K_{tt}K_{ss}-K_{ts}^2\right),
\eea
where
\bea\label{tsvars}
e^{t\over 2}=|z_1|^2+|z_2|^2\quad \mathrm{and}\quad e^s=|u|^2.
\eea
The Ricci-flatness condition is reduced to the following equation:
\bea\label{RicciflatK}
K_t\,\left( K_{tt}K_{ss}-K_{ts}^2\right)={a\over 8}\,e^{t+(l+1)\,s}
\eea
It turns out useful to perform a Legendre transform, passing from the variables $\{t, s\}$ to the new independent variables
\bea\label{munu}
\mu={\dd K \over \dd t},\quad\quad\nu={\dd K \over \dd s}
\eea
and from the K\"ahler potential $K(t, s)$ to the dual potential $G(\mu, \nu)$:
\bea
G=\mu\,t+\nu\,s-K
\eea
The usefulness of the new variables (\ref{munu}) to a large extent relies on the fact that they have a transparent geometric meaning -- these are the moment maps for the following two $\mathfrak{u}(1)$ actions on $Y$:
\bea\label{U1action}
\mathfrak{u}(1)_\mu:\quad \delta z_1=i\,\epsilon_1\,z_1,\quad \delta z_2=i\,\epsilon_1\,z_2,
\quad\quad\quad
\mathfrak{u}(1)_\nu:\quad \delta u= i\,\epsilon_2\,u\;.
\eea

\vspace{0.3cm}
\noindent
In this paper we will leave aside the case $l+1=0$ ($l$ is the parameter entering the exponent in (\ref{RicciflatK})) and assume that $l+1\neq 0$. In this case we can get rid of the $l$ dependence by a rescaling $\nu \to (l+1)\,\nu$, so in what follows we effectively set $l=0$. 
Then we obtain from~(\ref{RicciflatK}) a Monge-Ampere equation for the dual potential $G$ -- a function of two variables $\mu, \nu$ -- of the following form:
\begin{empheq}[box=\fbox]{align}
\hspace{1em}\vspace{1em}
\label{Ricciflat}
e^{\frac{\dd G}{\dd \mu}+\frac{\dd G}{\dd \nu}}\;\left(\frac{\dd^2 G}{\dd \mu^2} \frac{\dd^2 G}{\dd \nu^2}-\left(\frac{\dd^2 G}{\dd \mu \dd\nu}\right)^2\right)=\tilde{a}\,\mu
\hspace{1em}
\end{empheq}

\vspace{0.3cm}
\noindent
Denoting $(\mu, \nu)$ by $(\mu_1, \mu_2)$, we can recover the metric from the dual potential $G$ \cite{Pedersen} using the formula (see also Appendix \ref{metricder})
\bea\label{metric}
ds^2=\mu\, g_{\CP^1}+{1\over 4}\sum\limits_{i, j=1}^2\,\frac{\dd^2 G}{\dd \mu_i \dd \mu_j}\,d\mu_i\,d\mu_j+\sum\limits_{i, j=1}^2\, \left(\frac{\dd^2 G}{\dd \mu^2}\right)^{-1}_{ij}\,\left(d\phi_i - 2 A_i \right)\,\left(d \phi_j - 2 A_j \right),
\eea
where $g_{\CP^1}$ is the standard round metric on $\CP^1$ of volume $2\pi$ (i.e. $g_{\CP^1}={2\,dw\,d\bar{w}\over (1+\|w\|^2)^2}$), $A_2=0$ and $A_1$ is the `K\"ahler current' of $\CP^1$, i.e. a connection, whose curvature is the Fubini-Study form of $\CP^1$: $d A_1={i\,dw\wedge d\bar{w}\over (1+\|w\|^2)^2}$.

\vspace{0.4cm}
\noindent
\emph{Comment 1.} Note that the parameter $\tilde{a}$ in (\ref{Ricciflat}) is irrelevant, since one can effectively set $\tilde{a}=1$ by a \emph{linear} redefinition of the potential $G$, i.e. $G \to G+\nu \log{(\tilde{a})}$. Such a linear redefinition does not affect the metric (\ref{metric}), which depends only on the second derivatives of $G$. The only requirement is that $\tilde{a}>0$, since this is necessary for the positive-definitiveness of the metric (\ref{metric}).

\vspace{0.4cm}
\noindent
\emph{Comment 2.} There is a group of motions in the $(\mu, \nu)$-plane, under which the equation~(\ref{Ricciflat}) is invariant. It is generated by the transformations
\bear\label{gmot1}
&&\nu\to \nu+\delta,\\ \label{gmot2}
&&\mu\to \sigma \mu,\quad \nu \to \sigma\nu,\quad G\to \sigma\, G+\nu\,\log{(\sigma^3)},\\ \nonumber
&& \delta = \mathrm{const.}, \quad 0\neq \sigma=\mathrm{const.} 
\eear
The metrics, which differ by the transformation (\ref{gmot1}), are isometric, whereas the ones, which differ by (\ref{gmot2}), are related by an overall rescaling.

\vspace{0.5cm}
\ssubsection{The moment `biangle'}\label{biangle}

Since $(\mu, \nu)$ are moment maps for the $\mathfrak{u}(1)^2$ action, they define a map to $\mathbb{R}^2$. 
The domain in $\mathbb{R}^2$ on which the potential $G(\mu, \nu)$ is defined is the moment polygon for this $\mathfrak{u}(1)^2$ action. 
In addition, there is yet another $\mathfrak{u}(1)$ action given by
\bea\label{thirdu1}
\delta z_1=i\,\epsilon_3\,z_1,\quad\quad \delta z_2=-i\,\epsilon_3\,z_2\;.
\eea

\begin{figure}[h]
    \centering
    \includegraphics[width=\textwidth]{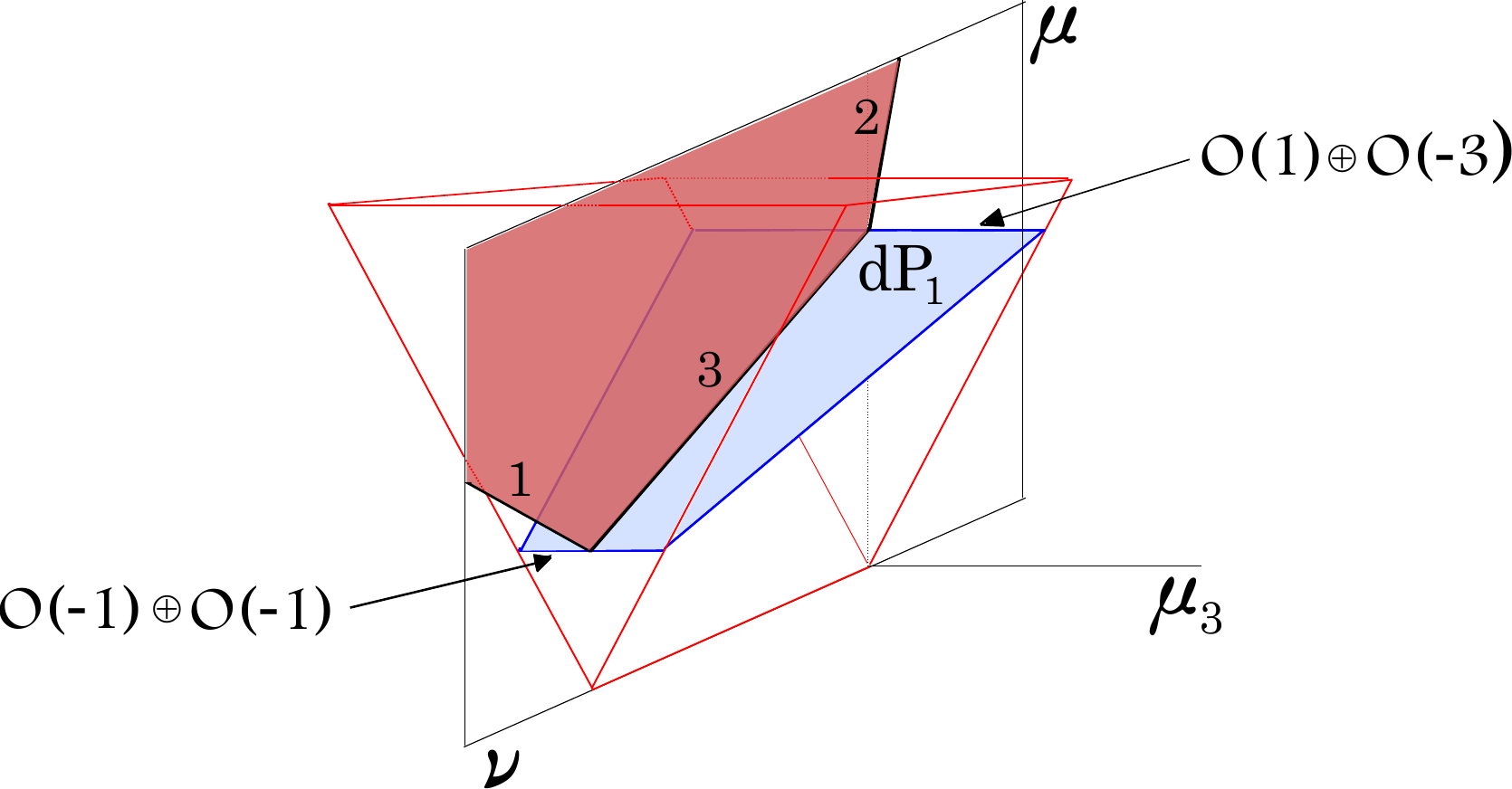}
    \caption{The moment polytope of $Y$. The blue polygon is the moment polygon of the del Pezzo surface -- it may be obtained from the moment polygon of $\CP^2$ (the triangle) by `cutting a corner'.}
    \label{mompol}
\end{figure}
\noindent
Denoting the dual moment map by $\mu_3$, we obtain:
\bea
\mu_3={\dd \over \dd \tau} K(e^\tau |z_1|^2+e^{-\tau} |z_2|^2, |u|^2)\big|_{\tau=0}=\frac{|z_1|^2-|z_2|^2}{|z_1|^2+|z_2|^2}\;\mu\;,
\eea
therefore
$
-\mu\leq \mu_3\leq \mu\, .
$

\vspace{0.3cm}
\noindent
The full three-dimensional moment polytope is shown in Fig. \ref{mompol}. The two-dimensional $(\mu, \nu)$-section, shown in red, is an unbounded domain with two vertices that we will call a `biangle'. The points in $Y$ mapping to a generic point of the biangle -- a point in the interior -- constitute a $\CP^1 \times \mathbb{T}^2$. 
The third $U(1)$ action (\ref{thirdu1}) corresponds to the rotation of the sphere $\CP^1$ around its axis. Now, the points in $Y$ mapping to a point at the edge of the biangle constitute a $\CP^1\times S^1$, and, finally, the points mapping to one of the two corners constitute a $\CP^1$. These two $\CP^1$'s, which map to the corners of the biangle, will be crucial in the foregoing discussion. We will see shortly that their normal bundles fully determine the topology of the moment polytope (i.e. its Delzant properties). For this reason we will not attempt to preserve explicitly the integrality of the normals to the facets of the polytope. The $(\mu, \nu)$ variables, which we are working with, are related to the moment map coordinates, in which the normals are integral, by a linear transformation. This is the reason that the moment biangle shown in Fig.~\ref{mompol} in red color is only congruent to the (integral) one shown in Fig.~\ref{mompol1}. As we just stated, the integrality properties will be automatically accounted for, once we make sure the normal bundles to the spheres at the corners of the polytope are the right ones.

\vspace{0.3cm}
\noindent
Let us analyze what constraint the behavior (\ref{asymptcond}) of $G$ at a facet of the polytope and the Ricci-flatness equation (\ref{Ricciflat}) impose on the facet itself. Suppose the facet is given by ${\ell_i=h_i \mu+ k_i \nu+p_i=0}$. We have the following lemma:

\begin{lem}\label{slopelemma}
The `slope' of the facet is constrained to satisfy  $h_i+k_i=1$.
\end{lem}
\noindent\underline{Proof.}\\
Indeed, let $G=\ell (\log{\ell}-1)+f(\tilde{\ell}, \ell)$ in the vicinity of a facet $\ell=0$, where $\tilde{\ell}$ is another linear combination of $\mu, \nu$ such that $d\ell \wedge d\tilde{\ell} \neq 0$, and $f$ is a smooth function at $\ell\to 0$. Substituting $G$ in (\ref{Ricciflat}), one obtains in the limit $\ell \to 0$: $\ell^{h_i+k_1 -1} \frac{\dd^2 f}{\dd \tilde{\ell}^2}\big|_{\ell=0}\sim \mu(\tilde{\ell})\big|_{\ell=0}$. The non-degeneracy of the induced metric on the facet implies $\frac{\dd^2 f}{\dd \tilde{\ell}^2}\big|_{\ell=0}\neq 0$. We assume $\mu|_{\ell=0} \nequiv 0$, which leads to $h_i+k_i=1$. $\blacksquare$

\vspace{0.3cm}
\noindent
 We will now demonstrate how the angles of the moment polytope are detemined by the normal bundles to the two $\CP^1$'s `located' at the corners.

\vspace{0.3cm}
\noindent
A corner of the moment polytope may be given by the equations
\bea
\ell_1=0,\quad \ell_2=0,
\eea
where
\bea\label{lines}
\ell_i=h_i \mu+ k_i \nu+p_i,\quad i=1, 2
\eea
are two linear forms. Moreover, according to the discussion above we assume that the behavior of the potential $G$ near the corner is as follows:
\bea\label{Gcorner}
G=\ell_1 (\log{\ell_1}-1)+\ell_2 (\log{\ell_2}-1)+\ldots,
\eea
where $\ldots$ denotes less singular terms. Compatibility with the Ricci-flatness condition (\ref{Ricciflat}) implies
\bea
h_i+k_i=1,\quad i=1, 2 
\eea

\vspace{0.3cm}
\noindent
We wish to determine what the behavior (\ref{Gcorner}) implies for the \emph{metric} near a given embedded~$\CP^1$. To this end we will insert the asymptotic form (\ref{Gcorner}) of the symplectic potential (omitting the subleading terms denoted by the ellipsis) into the expression for the metric~(\ref{metric}). To simplify the calculation, it will be useful to pass to the new `moment map' coordinates $\ell_1, \ell_2$ instead of $\mu, \nu$. The Hessian ${\dd^2G\over \dd\mu^2}$ then undergoes the standard transformation ${\dd^2G\over \dd\mu^2}=S^\mathrm{T} {\dd^2G\over \dd\ell^2} S$, where $S={\dd \ell \over \dd \mu}=\left( \begin{array}{cc}
h_1 & k_1  \\
h_2 & k_2  \end{array} \right)$. The virtue of this change of variables, clearly, is that ${\dd^2G\over \dd\ell^2}$ is a diagonal matrix: ${\dd^2G\over \dd\ell^2}=\left( \begin{array}{cc}
1\over \ell_1 & 0  \\
0 & 1\over \ell_2  \end{array} \right)$. The metric (\ref{metric}) acquires the form
\bear
&&ds^2=\mu(\ell_1, \ell_2)\, g_{\CP^1}+\left({d\ell_1^2\over 4 \ell_1}+{d\ell_2^2\over 4 \ell_2}\right)+\ell_1 \mathcal{A}_1^2+\ell_2 \mathcal{A}_2^2\\
&& \left( \begin{array}{c}
\mathcal{A}_1  \\
\mathcal{A}_2   \end{array} \right)=(S^\mathrm{T})^{-1}\circ \left( \begin{array}{c}
d\phi_1-A_1  \\
d\phi_2   \end{array} \right)\\
&&\mu(\ell_1, \ell_2)={\ell_1k_2-\ell_2 k_1\over k_2-k_1}+{p_2k_1-p_1 k_2\over k_2-k_1}
\eear
Introducing the angular variables
\bea
\varphi_1={k_2 \phi_1-h_2\phi_2\over k_2-k_1},\quad\quad \varphi_2={h_1 \phi_2-k_1\phi_1\over k_2-k_1},
\eea
we can write the metric as
\bea\label{metrcorn}
ds^2=\mu(\ell_1, \ell_2)\, g_{\CP^1}+\left({d\ell_1^2\over 4 \ell_1}+{d\ell_2^2\over 4 \ell_2}\right)+\ell_1 (d\varphi_1-n A_1)^2+\ell_2 (d\varphi_2-m A_1)^2,
\eea
where
\begin{empheq}[box=\fbox]{align}
\hspace{1em}\vspace{1em}
\label{normalbundle}
n={2k_2 \over k_2-k_1},\quad m=-{2 k_1 \over k_2-k_1},
\hspace{1em}
\end{empheq}
In appropriate coordinates the K\"ahler potential of the above metric is
\bea\label{Kahlercorner}
K=\kappa\;\log{\left(1+|w|^2\right)}+\left( 1+|w|^2 \right)^n\;|x|^2+\left( 1+|w|^2 \right)^m\;|y|^2,\quad\quad \kappa=\frac{p_2 k_1-p_1 k_2}{k_2-k_1}\,.
\eea
For $\kappa >0$ the formulas (\ref{metrcorn}) or (\ref{Kahlercorner}) imply that the normal bundle $N_{\CP^1}$ to the $\CP^1$ parametrized by the inhomogeneous coordinate $w$ and located in a given corner of the moment polytope is\footnote{See \cite{Bykov} for a detailed discussion of how the K\"ahler potential encodes the normal bundle to a $\CP^1$ in the analogous situation, when the $\CP^1$ is embedded in a complex surface.}
\bea
N_{\CP^1}=\mathcal{O}(-n)\;\oplus\;\mathcal{O}(-m),\quad\quad n+m=2
\eea
Note that $n+m=2$ is essentially a consequence of the Calabi-Yau condition
\bea
\det{N_{\CP^1}}=\;\textrm{the canonical class of}\;\CP^1\;=\mathcal{O}(-2)
\eea

\vspace{0.1cm}
\noindent
In the del Pezzo cone case the two corners of the moment biangle in the $(\mu, \nu)$-plane correspond to the two bases of the trapezium representing the moment polygon of the del~Pezzo surface itself, which serves as the base of the cone. This is emphasized in Fig.~\ref{mompol}, where the moment polygon of the del Pezzo surface is shown in blue.  The two bases of the trapezium correspond to the two $\CP^1$'s embedded in the del Pezzo surface:
\begin{itemize}
\item One $\CP^1$ is inherited from $\CP^2$, i.e. it is the standard embedding $\CP^1 \hookrightarrow \CP^2$, hence the normal bundle inside $\dP_1$ is $N=\mathcal{O}(1)$. This implies that the normal bundle inside the \emph{cone over} $\dP_1$ is $N=\mathcal{O}(1)\,\oplus\,\mathcal{O}(-3)$
\item The second $\CP^1$ is the exceptional divisor of the blow-up and is embedded with normal bundle $N=\mathcal{O}(-1)$. The normal bundle inside the \emph{cone over} $\dP_1$ is therefore $N=\mathcal{O}(-1)\,\oplus\,\mathcal{O}(-1)$. 
\end{itemize}
These two spheres generate the second homology group of the del Pezzo surface, and their intersection matrix is $\left( \begin{array}{ccc}
1 & 0 \\
0 & -1 \end{array} \right)$. The diagonal $\pm 1$ entries encode the normal bundles to the spheres.

\ssection{The `three-line' solution} 

\ssubsection{The metric cone}\label{3linesec}

We start by solving the equation (\ref{Ricciflat}) at infinity. We are looking for solutions, which asymptotically have the form of a metric cone:
\bea\label{coneasympt}
(ds^2)_\infty=dr^2+r^2\,\widetilde{ds^2}\quad\quad \textrm{as}\quad\quad r\to \infty,
\eea
where $\widetilde{ds^2}$ is a Sasakian metric on a 5-manifold. From the point of view of the function~$G$, this behavior translates to the following one:
\bea\label{G0metr}
G_\infty=3 \,\nu \,(\log{\nu}-1)+\nu \,P_0(\xi),\quad\quad\textrm{where}\quad\quad \xi=\frac{\mu}{\nu}
\eea
This leads to a metric with the following `radial' part ($r=2 \,\sqrt{3 \nu}$):
\bea\label{infmetr}
\left[ds^2\right]_\mu:= \frac{\dd^2 G_\infty}{\dd \mu_i \dd \mu_j}\,d\mu_i d\mu_j=3\, {d\nu^2 \over \nu}+\nu \,P_0''(\xi)\, d\xi^2=dr^2+r^2 \,{P_0'' \over 12}\,d\xi^2
\eea
In particular, we see that positivity of the metric requires $P_0''>0$.

\vspace{0.3cm}
\noindent
Substituting (\ref{G0metr}) in (\ref{Ricciflat}), we obtain the ordinary differential equation
\bea
P_0''={a\over 3}\,\xi\,e^{(\xi-1) P_0'-P_0},
\eea
which has the solution
\bea\label{P0func}
P_0(\xi)=\log{\left( {a \over 9}\right)}-\sum\limits_{i=0}^2\;\frac{\xi-\xi_i}{\xi_i-1}\;\log{|\xi-\xi_i|},
\eea
where $\xi_i$ are the roots of the polynomial
\bea\label{Qpolynomial}
Q(\xi)=\xi^3-{3\over 2} \xi^2+d,
\eea
and $d$ is a constant of integration, which plays a crucial geometric role that we will reveal in the next section.

\begin{wrapfigure}{l}{0.45\textwidth}
  \centering
    \includegraphics[width=\textwidth]{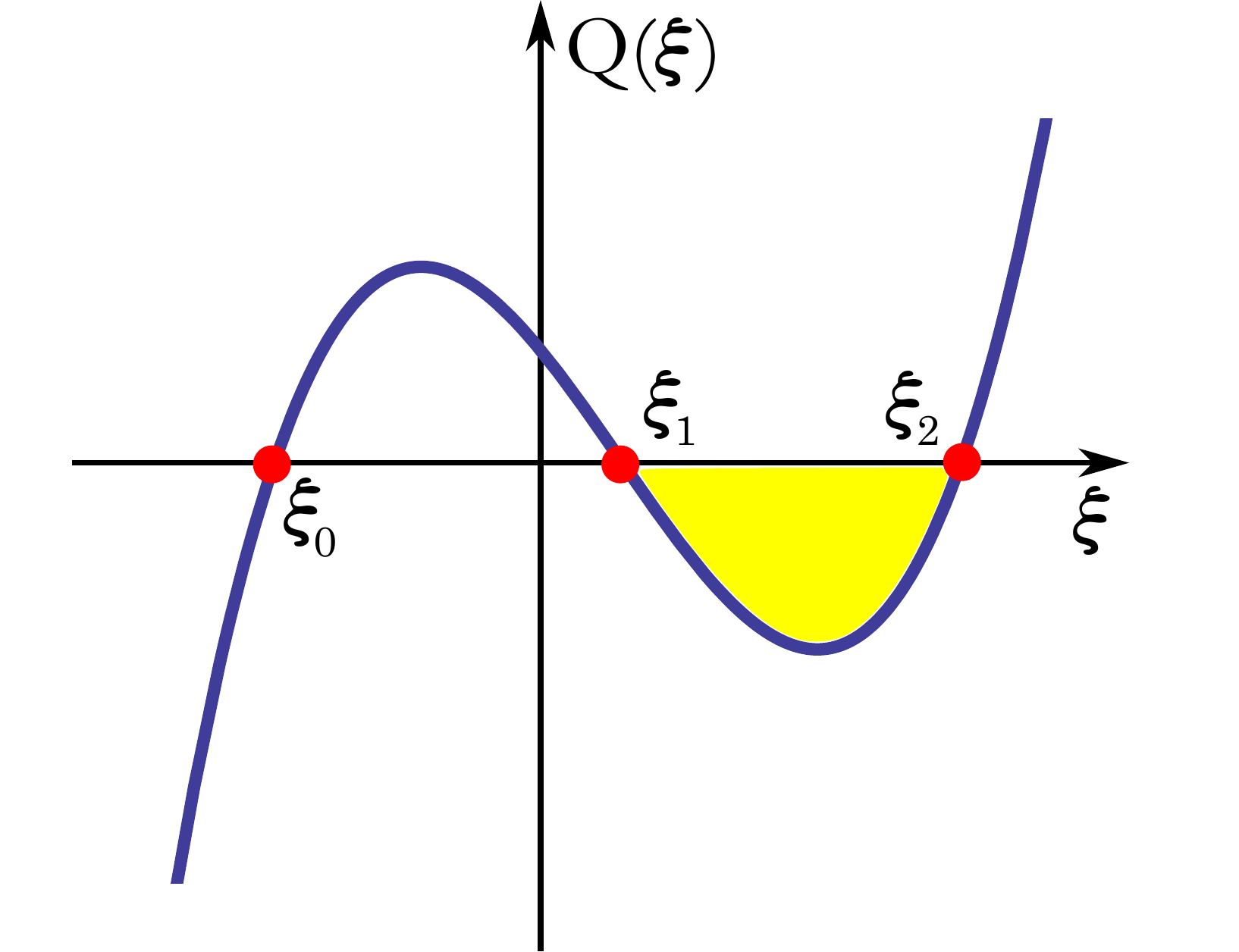}
    \caption{Yellow shading indicates the physical interval $\xi \in (\xi_1, \xi_2)$.\vspace{-1.2cm}}
    \label{Qxfunc}
\end{wrapfigure}

\vspace{0.3cm}
\noindent
The singular case $\xi_1=1$ (and hence $\xi_2=1$) corresponds to the situation, when the physical region shown in Fig. \ref{Qxfunc} shrinks to zero (see next section). We will therefore omit it in our discussion.

\vspace{0.3cm}
\noindent
\emph{Comment}. In Appendix \ref{3lineapp} we construct a one-parametric generalization of the solution (\ref{G0metr}), (\ref{P0func}). The virtue of this generalization is that its isothermal coordinates may be related in a simple way to the `orthotoric variables' that follow from the existence of a conformal Killing-Yano tensor (see \S\,\ref{KYforms} and in particular \S\,\ref{orthometric}).

\ssubsection{Topological considerations}\label{infasympt}

The potential (\ref{G0metr}) may as well be written in the original $(\mu, \nu)$ variables (up to a linear function, which does not affect the metric):
\bea\label{G0}
G_\infty=\sum\limits_{i=0}^2\;\frac{\mu-\xi_i\,\nu}{1-\xi_i}\;\left(\log{|\mu-\xi_i\,\nu|}-1\right)
\eea
One sees that the slopes of the three lines involved are defined by the roots $\xi_i$:
\bea
\mathrm{Slope}_i=\left( {\mu \over \nu}\right)_i=\xi_i
\eea
It is important to mention that the three lines appearing in (\ref{G0}) are \emph{not} the three edges of the $(\mu, \nu)$ moment polytope depicted in Fig. \ref{mompol} in red. (Otherwise we would have already constructed the desired metric.) In fact, two of the lines, associated with the roots $\xi_1, \xi_2$, do correspond to the two semi-infinite edges of the red polygon, however the line associated with the root $\xi_0$ is auxiliary and does not have a direct geometric interpretation.

\vspace{0.3cm}
\noindent
In the notations (\ref{lines}) of the moment polytope, which we used before, one has
\bea
\xi_1=-\frac{k_1}{1-k_1}\quad \mathrm{and}\quad \xi_2=-\frac{k_2}{1-k_2}
\eea
On the other hand, $k_1$ and $k_2$ are both related to $k_3$ (the indices $1, 2, 3$ correspond to the numbering of lines in Fig. \ref{mompol}) through the normal bundle formulas (\ref{normalbundle}), which therefore implies that there is a relation between $\xi_1$ and $\xi_2$. This geometric relation fixes the parameter $d$ of the polynomial $Q(\xi)$.

\vspace{0.3cm}
\noindent
Indeed, from the normal bundle formulas (\ref{normalbundle}) and Fig. \ref{mompol} it follows that
\bea\label{k1k2k3}
1-\frac{k_2}{k_3}=-2,\quad 1-\frac{k_1}{k_3}=2
\eea
Hence $\frac{k_2}{k_1}=-3$. This implies the following relation for $\xi_1, \xi_2$:
\bea\label{relxi}
-\frac{\xi_2}{1-\xi_2}=\frac{3 \xi_1}{1-\xi_1}
\eea

\vspace{0.3cm}
\noindent
One can show (see Appendix \ref{solxi}) that it has two solutions: $(\xi_1^{(1)}, \xi_2^{(1)})$, $(\xi_1^{(2)}, \xi_2^{(2)})$. However, for $\xi \in (\xi_1^{(2)}, \xi_2^{(2)})$ one has $P_0''<0$ and for $\xi \in (\xi_1^{(1)}, \xi_2^{(1)})$ one has $P_0''>0$, so the positivity of the metric requires that we choose the first solution. It corresponds to
\bea\label{dval}
d=\frac{16+\sqrt{13}}{64}\,.
\eea
The third root of $Q(\xi)=0$, which we will denote $\xi_0$, is smaller than the two other roots (see Fig. \ref{Qxfunc}).

\vspace{0.3cm}
\noindent
One can check directly that the Sasakian metric $\widetilde{ds^2}$ in (\ref{coneasympt}), which may be reconstructed from (\ref{G0metr}), (\ref{P0func}) with the value of the parameter $d$ given in (\ref{dval}), defines the Sasaki-Einstein manifold $Y^{2, 1}$, which is one of a family of manifolds found in \cite{GMSW} and termed $Y^{p, q}$. The fact that the $Y^{p,q}$ manifolds are the only compact simply-connected Sasaki-Einstein five-manifolds of cohomogeneity one with respect to the action of the isometry group was proven in~\cite{Conti}. 

\vspace{0.3cm}
\noindent
In what follows we will denote the roots of $Q(\xi)$ by $\xi_0, \xi_1, \xi_2$ so that $Q(\xi)=\prod\limits_{i=0}^2\;(\xi-\xi_i)$ and we will take into account that the `physical' region corresponds to $\xi \in (\xi_1, \xi_2)$.

\ssection{Uniqueness for a fixed polytope}\label{unique}

The goal of this section is to prove the following proposition:

\begin{thm}\label{propunique}
The solution of equation (\ref{Ricciflat}) with the behavior (\ref{asymptcond}) at the edges of the moment polytope and asymptotic to a real cone over $Y^{2,1}$ at infinity cannot be smoothly deformed.
\end{thm}
\noindent\underline{Proof.}\\ First we consider a more general equation $\det{\mathrm{Hess}\, G}=f(\mu, \nu)\,e^{\nabla_a G}$, where $a\in
\mathbb{R}^2$ is a constant vector. Suppose $G_0$ is a solution with the correct asymptotic properties. Construct a first order deformation $G=G_0+\epsilon H$. The linearized equation has the form
\bea
G_0^{ij}\,\dd_i\dd_j H=\nabla_a H
\eea
We can rewrite it as
\bea
\dd_i\left(G_0^{ij}\,\dd_j H \right)=\left(\dd_i(G_0^{ij})+a^j\right) \dd_j H
\eea
We will now use the identity $\dd_i G_0^{ij}=-G_0^{jm}\dd_m \log\det G_0$, which is valid if $(G_0)_{ij}=\dd_i \dd_j G_0$. Since $\log \det G_0=\log(f(\mu, \nu))+a^i \dd_i G_0$ and $a$ is a constant vector, we have $\dd_i(G_0^{ij})=-G_0^{jm}\dd_m \log(f(\mu, \nu))-G_0^{jm} (G_0)_{mi} a^i=-G_0^{jm}\dd_m \log(f(\mu, \nu))-a^j$. Therefore the linearized equation acquires the form
$\dd_i\left(G_0^{ij}\,\dd_j H \right)=-G_0^{jm}\dd_m \log(f(\mu, \nu)) \dd_j H $, which can also be rewritten in divergence form:
\bea\label{selfadj}
{\dd \over \dd \mu_i}\left(f(\mu, \nu)\, G_0^{ij}\,{\dd H\over \dd \mu_j}\right)=0
\eea
It is rather nontrivial that the linearized equation has the self-adjoint form, and this relies on the fact that the r.h.s. of the original equation is an exponential of a linear combination of the derivatives of $G$ with constant coefficients.

\vspace{0.3cm}
\noindent
Multiplying (\ref{selfadj}) by $H$ and integrating over the moment polygon, we obtain upon integration by parts:
\bea
0=-\int\,d\mu\,d\nu\,f(\mu,\nu) \|\nabla H\|_{G_0}^2+\textrm{boundary terms}
\eea
We will now show that, once the asymptotic conditions for $G_0$ are satisfied and $H$ is smooth, the boundary terms vanish. Clearly, the boundary consists of four segments: three edges of the polygon and `a segment at infinity'. First, consider the boundary term for each of the edges. Near such an edge one can make a linear change of coordinates $(\mu, \nu)\to (\ell, \tilde{\ell})$, such that the edge is at $\ell=0$ (same as in the proof of Lemma 3.1). Therefore the relevant boundary term is
\bea
B\big|_{\ell=0}=f(\mu, \nu)\left(\,G_0^{\ell\ell}\,{\dd H\over \dd \ell}+\,G_0^{\ell\tilde{\ell}}\,{\dd H\over \dd \tilde{\ell}}\right)\big|_{\ell=0}
\eea
Since the asymptotic form of $G_0$ is $G_0=\ell (\log \ell-1)+G_0^{reg}(\ell, \tilde{\ell})$, one finds
\bea
G_0^{\ell\ell}=\ell\,(1+\ldots)\quad\quad\textrm{and}\quad\quad G_0^{\ell\tilde{\ell}}=\ell\left(- \frac{(G_0^{reg})_{\ell\tilde{\ell}}}{(G_0^{reg})_{\tilde{\ell}\tilde{\ell}}}+\ldots\right)\,.
\eea
In particular, both vanish at the boundary. Since we have assumed that $H$ is regular at the boundary, it follows that
\bea
B\big|_{\ell=0}=0
\eea
Let us now have a look at the boundary term, corresponding to the boundary segment $\gamma$ at infinity, $B_\infty$:
\bea\label{boundinf}
B_\infty=\int\limits_\gamma\,dl\,\mu\,H\,n_i G_0^{ij}\,{\dd H\over \dd \mu_j},\quad\quad\quad \left(f(\mu, \nu)=\mu\right),
\eea
where $\vec{n}$ is the unit vector, normal to $\gamma$, and $dl$ is the infinitesimal length element along~$\gamma$. In order to estimate the value of $B_\infty$, one needs to know the behavior of $H$ at infinity. Therefore we will start by solving the equation (\ref{selfadj}) at infinity. To this end we need to recall the asymptotic behavior of the solution $G_0$ at infinity, discussed in \ref{infasympt}:
\bea
G_\infty=\sum\limits_{i=0}^2\,\frac{\mu-\xi_i \nu}{1-\xi_i}\,\left(\log|\mu-\xi_i \nu|-1\right)
\eea
One checks that the equation (\ref{selfadj}), with $G_0$ replaced by its asymptotic value $G_\infty$, takes the form
\bea
{\dd \over \dd \mu_i}\left(f(\mu, \nu)\, (G_\infty)^{ij}\,{\dd H\over \dd \mu_j}\right)=
{1\over 3}\,\left[-{\dd \over \dd \xi}\left(Q(\xi)\,{\dd H\over \dd \xi}\right)+{\xi\over \nu}\,{\dd \over \dd \nu}\,\left(\nu^3\,{\dd H\over \dd \nu}\right)\right]=0
\eea
Clearly, the variables separate, so one can use the asymptotic ansatz $H=\nu^{-m}\,h(\xi)$ to~obtain
\bea\label{heuneq}
-{d \over d \xi}\left(Q(\xi)\,{d h\over d \xi}\right)+m(m-2)\,\xi \,h(\xi)=0
\eea
Multiplying by $h(\xi)$ and integrating over the interval $\xi\in (\xi_1, \xi_2)$, one obtains, upon integration by parts
\bea\label{integr1}
\int\limits_{\xi_1}^{\xi_2}\,d\xi\,\left(Q(\xi)\,\left({d h\over d\xi}\right)^2+m(m-2)\,\xi\,h(\xi)^2\right)=0
\eea
In the chosen segment $Q(\xi)<0$, so (\ref{integr1}) leads to the condition
\bea
m(m-2)>0\,.
\eea
Assuming that the metric defined by $G_0$ is subleading to the conical metric defined by~$G_\infty$ at infinity (i.e. $|\mathrm{Hess}\,G_0-\mathrm{Hess}\,G_\infty|_{G_\infty}\to 0$), we wish that $G_0+\epsilon\,H$ is subleading to $G_\infty$ as well. This implies $|\mathrm{Hess}\,H|_{G_{\infty}}\to 0$, and we have two possibilities:
\begin{center}
$\bullet$ I. $m>2$\quad\quad or \quad\quad $\bullet$ II. $-1<m<0$
\end{center}
\emph{Case I.}\newline It is easy to see that the boundary contribution (\ref{boundinf}) vanishes, as
\bea\label{est1}
|B_\infty|<{A\over \nu^{2m-2}}\quad\textrm{for}\quad \nu\to\infty
\eea
Here by $\nu$ we mean the average value of $\nu$ on the boundary $\gamma$. To arrive at (\ref{est1}) one should take into account that $(G_0^\infty)^{ij}\sim \nu$ and the length of $\gamma$ behaves as $\int\limits_\gamma dl \sim \nu$.

\noindent\emph{Case II.}\newline 
We will show that in this case the equation (\ref{heuneq}) cannot have a solution, real analytic at the two singular points in question, $\xi_1$ and $\xi_2$. This is a result of the following lemma:

\begin{lem}\label{eigenvaluelemma}
The smallest non-zero eigenvalue $\lambda$ of the Laplacian $\triangle_\xi=-{d \over d \xi}\left(Q(\xi)\,{d h\over d \xi}\right),$ entering the equation $ \triangle_\xi h+ \lambda \,\xi\, h=0$, is $\lambda=3$.
\end{lem}
\noindent\underline{Proof.}\\ 
The fact that $\lambda=0$ and $\lambda=3$ are eigenvalues of the Laplacian is almost obvious, since one can write out the corresponding eigenfunctions directly: for $\lambda=0$ one has $h=1$, and for $\lambda=3$ one has $h=\xi-1$\,.

\vspace{0.3cm}\noindent
The equation at hand is a Heun equation -- a Fuchsian ODE with four regular singular points: the three roots of the polynomial $Q(\xi)$ and the point at infinity. In order to make a more canonical `centering' of the Heun equation we make a change of variables
\bea
\xi \to \frac{\xi_1+\xi_2}{2}-\frac{\xi_2-\xi_1}{2}\, \xi,
\eea
bringing the equation to the canonical form
\bea\label{Heun}
\frac{d}{d\xi} \left( (1-\xi^2) (\xi-t) \frac{d h}{d \xi}\right)-\lambda\, (s-\xi)\, h=0
\eea
with
\bea\label{thpar}
t=\frac{\xi_1+\xi_2-2\,\xi_0}{\xi_2-\xi_1}\quad \textrm{and}\quad s=\frac{\xi_2+\xi_1}{\xi_2-\xi_1}
\eea
We will use the method of solving the eigenvalue problem for the Heun equation using an expansion in hypergeometric (Jacobi) polynomials, which goes back to Svartholm \cite{Svartholm} (see also \cite{Slavyanov} as a general reference on Heun's equations). In our case, since the exponents of the corresponding singular points are zero, the Jacobi polynomials reduce to Legendre polynomials. We expand $h$ in the Legendre polynomials
\bea\label{Legendre}
h=\sum\limits_{k=0}^{\infty}\;a_k \,L_k(\xi).
\eea
For a function $h(\xi)$, \emph{analytic on the closed segment} $\xi\in[-1, 1]$, the expansion (\ref{Legendre}) is convergent in an ellipse having $\pm 1$ as its foci (\cite{Szego}, p. 245; \cite{Whittaker}, p. 322). Note that the shape of the ellipse depends on the nearest singularities of $h(\xi)$.

\vspace{0.3cm}
\noindent
Substituting the expansion (\ref{Legendre}) in the equation (\ref{Heun}), obtain the recurrence relation
\bea\label{recursion}
g_k\;a_{k+1}-f_k\; a_k+j_k\;a_{k-1}=0
\eea
with
\bear\label{hk}
&&g_k=\frac{(k+1)\left( (k+1)^2-\lambda-1\right)}{2k+3}\\ \label{fk}
&& f_k = t\,k(k+1)-s\,\lambda\\ \label{jk}
&& j_k= \frac{k\left(k^2-\lambda-1\right)}{2k-1}
\eear
Introducing the new variable $\tau_k=\frac{a_{k-1}}{a_k}$, we can rewrite the recurrence relation (\ref{recursion}) as follows:
\bea\label{rec}
\frac{g_k}{\tau_{k+1}}+j_k \tau_k-f_k=0
\eea
and take
\bea
\tau_0=0
\eea
as the initial condition for our recursion.

\vspace{0.3cm}
\noindent
It is easy to solve the recurrence relation in the limit $k \to \infty$. Indeed, in this case we obtain a quadratic equation for $\tau_{\infty}$:
\bea
\tau_\infty^2-2t \tau_\infty+1=0,
\eea
which has the solutions
\bea
(\tau_\infty)_\pm=t\pm \sqrt{t^2-1}
\eea
The solution of the recurrence relation (\ref{recursion}) therefore behaves at large $k$ as
\bea
a_k\sim s_+ \left(\frac{1}{(\tau_\infty)_+}\right)^k+s_- \left( \frac{1}{(\tau_\infty)_-}\right)^k
\eea
It is easy to check, using (\ref{thpar}), that $t>1$, therefore $(\tau_\infty)_-<1$ and $(\tau_\infty)_+>1$. Looking back at the expansion (\ref{Legendre}), and taking into account that $L_k(1)=1, L_k(-1)=(-1)^k$, we see that the requirement of regularity of the function $h$ at the points $\xi=0, 1$ is equivalent to the condition $s_-=0$. We will prove below that this is not so, i.e. that the solution in fact grows as $a_k \sim \left( \frac{1}{(\tau_\infty)_-}\right)^k$, where $\frac{1}{(\tau_\infty)_-}>1$. The proof is by induction: assuming that $0<\tau_k<a$ for a suitable constant $a$, we will show that $0<\tau_{k+1}<a$. If one can take $a<1$, this is sufficient to prove that the sequence $\{a_k\}$ is exponentially growing.

\vspace{0.3cm}
\noindent
The key technical inequality that we will need to prove is the following:
\bea\label{masterineq}
f_k-a j_k-{1\over a} g_k >0\quad \textrm{for all}\quad k\geq 2\quad{\textrm{and some}\;\; a:}\quad 0<a<1\, ,
\eea
where $f_k, j_k, g_k$ have been defined in (\ref{hk})-(\ref{jk}). The relevant values of $s$ and $t$ are
\bea\label{st}
s={4+\sqrt{13}\over 3}\quad\quad \textrm{and}\quad\quad t=\sqrt{13}\,.
\eea
Once we have proven (\ref{masterineq}), suppose $0<\tau_k<a$. Then
\bea\label{reck1}
\tau_{k+1}=\frac{g_k}{f_k-j_k \tau_k}>0,
\eea
since $j_k>0$ and $g_k>0$ for $k>2\geq\sqrt{\lambda+1}$, and it follows from (\ref{masterineq}) that ${f_k\over j_k}>a>\tau_k$. Besides, since, according to (\ref{masterineq}), $f_k-\tau_k j_k> f_k - a j_k> {1\over a} g_k$, (\ref{reck1}) implies
\bea
\tau_{k+1}<a .
\eea
\begin{center}
\line(1,0){50}
\end{center}

In order to prove (\ref{masterineq}), first of all we make some elementary estimates:
\bear
&&g_k<\frac{1}{2} \left((k+1)^2-\lambda-1 \right)\\
&&j_k< \left({1\over 2}+\epsilon\right)\,(k^2-\lambda-1)<\left( (k+1)^2-\lambda-1\right),\quad\quad0<\epsilon\ll 1,
\eear
\vspace{-0.2cm}
hence
\bea\label{phik}
f_k-a j_k-{1\over a} g_k >  t \,k (k+1) - s\, \lambda-b\, ((k+1)^2-\lambda-1):= \phi_k
\eea
with
\bea
b=a+{1\over 2 a},\quad (b\geq\sqrt{2})\,.
\eea
$\phi_k$, defined in (\ref{phik}), is a quadratic function of $k$, so in order to prove that $\phi_k>0$ for $k\geq 2$ we will show that $\phi_{2}>0$ and $\phi'_k>0$ for $k>0$. First of all,
\bea
\phi_2=6\,t-s\,\lambda-b\,(8-\lambda)
\eea
Since we are interested in the segment $\lambda\in(0, 3)$, and $\phi_2$ is a linear function of $\lambda$, it suffices to require that the values of $\phi_2$ at the ends of the segment are positive: $\phi_2\big|_{\lambda=0}=6\,t-8\,b>0$ and $\phi_2\big|_{\lambda=3}=6\,t-3\,s-5\,b>0$. Therefore we need to take $b<\mathrm{min}\left({3\,t\over 4}, {6\,t-3\,s \over 5}\right)={3\,t\over 4}$ for the values of $t$ and $s$ given in (\ref{st}).
To ensure that $\phi_k$ is a growing parabola we require $b<t$ and, since $\phi_k'=(2k+1)\,t-2b\,(k+1)$, for $b<{1\over2} t$ the bottom of the parabola lies at $k<0$. Therefore for $b<{1\over2} t$ we have $\phi_k>0$ for $k\geq 2$, implying
\bea
f_k-a j_k-{1\over a} h_k >0.
\eea
Now, the requirement $b<{1\over2} t$ means that
\bea
a^2-{t\over 2} a+{1\over 2}<0
\eea
This is easily satisfied for $a={1\over 2}$, since $t=\sqrt{13}>3$. What remains to be checked is that $\tau_2<a={1\over 2}$. This is true, since  $\tau_1={1\over 3s}$ and $\tau_2={2\over 5}\,{1\over s-{1\over 3s}}<{1\over 2}$. Therefore we have proven that $\tau_k < {1\over 2}$, so that ${a_{k}\over a_{k-1}}>2$, which implies in particular $\underset{k\to \infty}{\lim} {a_{k}\over a_{k-1}} >1$. The expansion~(\ref{Legendre}) is thus divergent at the two singular points of interest: $\xi=0$, $\xi=1$.

\vspace{0.3cm}
\noindent
This completes the proof of the \textbf{Lemma} $\blacksquare$

\vspace{0.3cm}
\noindent
In equation (\ref{heuneq}) one has $\lambda=m(m-2)$. Case II corresponds to the range $-1<m<0$, which is equivalent to $0<\lambda<3$. It follows from the lemma that there are no eigenvalues of  $\triangle_\xi$ lying in this range. This completes the proof of \textbf{Proposition}\;\ref{propunique} $\blacksquare$

\vspace{0.3cm}
\noindent\emph{Comment.} Proposition 3 implies that the equation (\ref{Ricciflat}) does not require the specification of any boundary values, apart from (\ref{asymptcond}) and the condition that the metric is asymptotic to a real cone over $Y^{2,1}$. This is related to the singular nature of the boundary condition~(\ref{asymptcond}). A Monge-Ampere equation with a similar boundary behavior of the solution was considered in \cite{Urbas}.

\ssection{The Killing-Yano forms}\label{KYforms}

Now that we have proven that the solution is unique, once the moment polytope has been specified, we may ask if a solution exists. It turns out that there is a closed expression for $G$, and hence for the metric, in the case of a particularly chosen moment polytope --- this is the metric obtained in \cite{LuPope1}, as well as in \cite{MS}, and it has the so-called `orthotoric' form~\cite{Gauduchon}. This form of metric arises naturally from the requirement of existence of a conformal Killing-Yano form of type $(1, 1)$ on the manifold. We therefore start by reviewing the concept of conformal Killing-Yano forms on K\"ahler manifolds. For a general review of Killing and Killing-Yano tensors the reader is referred to \cite{Chervonyi}, \cite{Santillan}.

\subsection{Conformal Killing-Yano forms on a Calabi-Yau threefold}

First we consider a manifold $\mathcal{M}$ of arbitrary dimension $D$. By definition, a conformal Killing-Yano form (CKYF) is a 2-form $\omega_{jk}$ on $\mathcal{M}$ satisfying an equation of the form (see the derivation in Appendix \ref{KYapp})
\bear
&&\widetilde{\mathscr{D}}\omega=0,\quad\quad\textrm{where}\\ \label{ckyt3}
&& (\widetilde{\mathscr{D}}\omega)_{ijk}:=\nabla_i\omega_{jk}-{1\over 3} T_{ijk}+{1\over D-1}\,\left(g_{ik}\,g^{mn}\nabla_m \omega_{nj}-g_{ij} g^{mn} \nabla_m \omega_{nk}\right)\\ \label{Tdef}&&
\textrm{and}\quad\quad T_{ijk}=\nabla_i \omega_{jk}+\nabla_k \omega_{ij}-\nabla_j \omega_{ik}\,.
\eear
The tensor $T$ here, which is anti-symmetric in all pairs of indices, is proportional to the exterior derivative of $\omega$, i.e. $T\;\propto\; d\omega$.

\vspace{0.3cm}\noindent
Let us now specialize to the case of a Calabi-Yau manifold $\mathcal{M}$ of complex dimension~$3$, i.e. $D=6$. Since $\mathcal{M}$ is Calabi-Yau, its volume form may be decomposed as
\bea\label{volume}
\mathrm{vol}_{\mathcal{M}}=i\,\Omega\wedge \widebar{\Omega}\,,
\eea
where $\Omega$ is a holomorphic non-vanishing 3-form
\bea
\Omega:=\Omega_{abc}(z)\, dz^a\wedge dz^b\wedge dz^c\,.
\eea
It also follows from the above two equalities that $\Omega$ is covariantly constant:
\bea
\nabla_m \Omega=0,\quad\quad \nabla_{\bar{m}} \Omega=0\,.
\eea
Let us introduce a poly-vector $\widetilde{\Omega}^{abc}$ by raising the indices of the form $\Omega$. This poly-vector is `inverse' to $\Omega$ in the following sense:
\bear\label{inverseform1}
&&\widetilde{\Omega}^{ijk} \Omega_{i'j'k}= \delta^i_{i'}\,\delta^j_{j'}-\delta^i_{j'}\,\delta^j_{i'},\quad\quad \widetilde{\Omega}^{\bar{i}\bar{j}\bar{k}} \Omega_{\bar{i}'\bar{j}'\bar{k}}= \delta^{\bar{i}}_{\bar{i}'}\,\delta^{\bar{j}}_{\bar{j}'}-\delta^{\bar{i}}_{\bar{j}'}\,\delta^{\bar{j}}_{\bar{i}'},\quad\quad\\
\label{inverseform}
&&\widetilde{\Omega}^{ijk} \Omega_{i'jk}=2\, \delta^i_{i'},\quad\quad\quad \widetilde{\Omega}^{\bar{i}\bar{j}\bar{k}} \Omega_{\bar{i}'\bar{j}\bar{k}}=2\, \delta^{\bar{i}}_{{\bar{i}'}}\,.
\eear
To see that this is the case, we write $\Omega$ in components as $\Omega_{ijk}=\epsilon_{ijk}\, q(z)$, where $|q|^2=\mathrm{Det}(g)$. Analogously, $\Omega_{\bar{i}\bar{j}\bar{k}}=\epsilon_{\bar{i}\bar{j}\bar{k}}\, \bar{q}(\bar{z})$. Therefore the dual poly-vector $\widetilde{\Omega}^{ijk}=\epsilon^{ijk}\,(q(z))^{-1}$ is holomorphic, and $\widetilde{\Omega}^{\bar{i}\bar{j}\bar{k}}=\epsilon^{\bar{i}\bar{j}\bar{k}}\,(\bar{q}(\bar{z}))^{-1}$ is anti-holomorphic.

\vspace{0.3cm}
\noindent
Using the form $\Omega$ and its inverse $\widetilde{\Omega}$, we can dualize vectors to forms and vice versa, for example we can view $\widetilde{\Omega}$ as a map
\bea
\widetilde{\Omega}:\quad (T^\ast)^{(1, 0)}\wedge (T^\ast)^{(1, 0)} \to T^{(1, 0)}
\eea
On the other hand, on a K\"ahler manifold, the Killing-Yano form may be disassembled into its Hodge components:
\bea
\omega=\omega^{(2, 0)}\oplus \omega^{(1, 1)}\oplus \omega^{(0, 2)}\,.
\eea
We will be denoting the vector of type $(1,0)$, dual to $\omega^{(2, 0)}$, by the same letter $\omega$:
\bea\label{omegadual}
\omega^s:=\widetilde{\Omega}^{sjk}\omega_{jk}\,.
\eea
The goal of the following paragraphs \S\S \,\ref{KY20part} and \ref{KY11part} will be in proving the following proposition:

\begin{thm}\label{Riemannzerovector}
Let $\mathcal{M}$ be a Ricci-flat complex manifold, $\mathrm{dim}_{\CC}\,\mathcal{M}=3$, without parallel vector fields. 
Then the vector $\omega^m\,\frac{\dd}{\dd z_m}$ of type $(1, 0)$, dual to the $(2, 0)$-part $\omega^{(2, 0)}$ of a conformal Killing-Yano two-form on $\mathcal{M}$, satisfies the following equation:
\bea\label{propcurv}
R^n_{\;mj\bar{k}}\,\omega^m=0\,.
\eea
\end{thm}

\subsubsection{The $(2, 0)$-part of the Killing-Yano form}\label{KY20part}

We first concentrate on the $(2, 0)$ component of the Killing-Yano form. The equation~(\ref{ckyt3}) with all indices holomorphic gives
\bea\label{omegaT}
\nabla_i\omega_{jk}={1\over 3} T_{ijk}
\eea
Here $T$ is a totally skew-symmetric tensor of type $(3, 0)$.

\vspace{0.3cm}
\noindent \textbf{I.} $T$ has to be proportional to the Calabi-Yau 3-form:
\bea\label{TfO}
T=f\cdot\Omega,
\eea
where $f$ is a scalar function on $\mathcal{M}$. Dualizing the eq. (\ref{omegaT}) in the $(j, k)$ indices and using the notation (\ref{omegadual}), we obtain
\bea\label{omegaT2}
\nabla_i\omega^s={2\over 3}\,f\cdot \delta_i^s\,.
\eea
Since on a K\"ahler manifold the holomorphic covariant derivatives commute, $[\nabla_i, \nabla_j]=0$, we obtain a consistency condition
\bea
\dd_j f\,\delta_i^s-\dd_i f\,\delta_j^s=0\quad\quad \Rightarrow \quad \dd_j f=0\,,
\eea
i.e. $f=f(\bar{z})$ is anti-holomorphic.

\vspace{0.3cm}
\noindent \textbf{II.} To proceed further let us introduce the one-form
\bea\label{lamform}
\tilde{\lambda}:=g^{s\bar{t}}\,\nabla_{\bar{t}}\omega_{js}\,dz^j+g^{t\bar{s}}\,\nabla_{t}\omega_{\bar{j}\bar{s}}\,d\bar{z}^j
\eea
and the dual vector field
\bea\label{lamvector}
\lambda=\nabla_{\bar{t}}\tilde{\omega}^{\bar{j}\bar{t}}\,\frac{\dd}{\dd \bar{z}^j}+\nabla_{t}\tilde{\omega}^{jt}\,\frac{\dd}{\dd z^j}\,.
\eea
Let us now act on (\ref{omegaT}) by $\nabla_{\bar{j}}$, contract the equation with $g^{j\bar{j}}$ and commute the covariant derivatives to obtain an equation for the divergence of $\omega$, i.e. for $\lambda$:
\bea
g^{j\bar{j}}\nabla_{\bar{j}}\nabla_i\omega_{jk}=g^{j\bar{j}}\,R^{p}_{\;j\bar{j}i}\,\omega_{pk}+g^{j\bar{j}}\,R^{p}_{\;k\bar{j}i}\,\omega_{jp}-\nabla_i \tilde{\lambda}_k={1\over 3}g^{j\bar{j}} \dd_{\bar{j}}f(\bar{z})\cdot  \Omega_{ijk}
\eea
Both terms involving the Riemann tensor are zero. The first one is zero since it is proportional to the Ricci tensor, and the second one is equal ${R^{p}_{\;k}}^j_{\;i}\,\omega_{jp}=0$ as a contraction of the symmetric (in the $(p, j)$ indices) Riemann tensor with the skew-symmetric tensor $\omega$. Therefore we get
\bea
\nabla_i \tilde{\lambda}_k=-{1\over 3}g^{j\bar{j}} \dd_{\bar{j}}f(\bar{z})\cdot  \Omega_{ijk}
\eea
Since $\Omega$ is skew-symmetric, we see that
\bea\label{Killing2}
\nabla_i \tilde{\lambda}_k+\nabla_k \tilde{\lambda}_i=0\,.
\eea
Moreover, the corresponding dual vector field is holomorphic ($\dd_{\bar{i}}\lambda^j=0 \Leftrightarrow \nabla_i \tilde{\lambda}_k=0$) if and only if $f(\bar{z})=\mathrm{const.}=f_0$.

\subsubsection{The $(1, 1)$-part of the Killing-Yano form}\label{KY11part}

Equation (\ref{omegaT2}) (and its complex-conjugate) is therefore the only constraint on the $(2, 0)$ and $(0, 2)$ parts of the Killing-Yano form. We now turn to the analysis of the remaining equations (\ref{ckyt3}), which constrain the $(1, 1)$ part of the form.
\bear \nonumber
&&\!\!\!\!\!\!\!\!\!\!\!\!\!\!\!i, j\,\textrm{hol.}, \quad k \,\textrm{anti-hol.}:\quad\quad {2\over 3}\,\nabla_i\,\omega_{j\bar{k}}+{1\over 3}(\nabla_j \omega_{i\bar{k}}-\nabla_{\bar{k}}\omega_{ij})+{1\over 5} g_{i\bar{k}}\,g^{\mu\nu}\nabla_\mu\omega_{\nu j}=0\,\\ \nonumber
&&\!\!\!\!\!\!\!\!\!\!\!\!\!\!\!i\,\textrm{anti-hol.},\quad \quad j, k \,\textrm{hol.},\quad i\leftrightarrow k:\\ \nonumber &&\quad -{2\over 3}\,\nabla_{\bar{k}}\omega_{ij}+{1\over 3}(\nabla_i\omega_{j\bar{k}}-\nabla_j\omega_{i\bar{k}})+{1\over 5} (g_{i\bar{k}}\,g^{\mu\nu}\nabla_\mu \omega_{\nu j}-g_{j\bar{k}}\,g^{\mu\nu}\nabla_{\mu}\omega_{\nu i})=0\,.
\eear
Expressing $\nabla_j\omega_{i\bar{k}}$ from the second equation and substituting in the first one, we obtain an equation for $\omega_{j\bar{k}}$:
\bea\label{11compeqmain0}
\nabla_i\omega_{j\bar{k}}=\nabla_{\bar{k}}\omega_{ij}-{2\over 5} g_{i\bar{k}}\,g^{\mu\nu}\nabla_{\mu}\omega_{\nu j}+{1\over 5} g_{j\bar{k}}\,g^{\mu\nu}\nabla_{\mu}\omega_{\nu i}\,.
\eea
Contracting it with $g^{j\bar{k}}$, we obtain the following:
\bea\nonumber
{1\over 5} \,g^{\mu\nu}\nabla_{\mu}\omega_{\nu i}=\dd_i h-\tilde{\lambda}_i\,,\quad\quad h:=g^{j\bar{k}}\omega_{j\bar{k}}=\mathrm{Tr}(\omega)\,.
\eea
Substituting this in (\ref{11compeqmain0}), we get
\bea\label{11compeqmain}
\nabla_i\omega_{j\bar{k}}=g_{j\bar{k}}\,\dd_i h-2 g_{i\bar{k}}\,\dd_j h+\left(\nabla_{\bar{k}}\omega_{ij}+2 g_{i\bar{k}}\,\tilde{\lambda}_j-g_{j\bar{k}}\,\tilde{\lambda}_i\right)\,.
\eea
Note that the terms in brackets depend only on the $(2, 0)$ part of the KY-form. Another important equation is the complex-conjugate one. To write it, note that, since $\omega=\omega_{ij}\,dz_i\wedge dz_j+\omega_{\bar{i}\bar{j}}\,d\bar{z}_i\wedge d\bar{z}_j+2\,\omega_{i\bar{j}}\,dz_i\wedge d\bar{z}_j$ is a real form, $\omega_{i\bar{j}}^\ast=-\omega_{j\bar{i}}$ and $\omega_{ij}^\ast=\omega_{\bar{i}\bar{j}}$. By analogous arguments, $g_{i\bar{j}}^\ast=g_{j\bar{i}}$ and therefore $h^\ast=-h$. Hence  the complex conjugation of (\ref{11compeqmain}) gives, upon the interchange $j\leftrightarrow k$,
\bea\label{11compeqmain2}
\nabla_{\bar{i}}\omega_{j\bar{k}}=g_{j\bar{k}}\,\dd_{\bar{i}} h-2 g_{j\bar{i}}\,\dd_{\bar{k}} h-\left(\nabla_{j}\omega_{\bar{i}\bar{k}}+2 g_{j\bar{i}}\,\tilde{\lambda}_{\bar{k}}-g_{j\bar{k}}\,\tilde{\lambda}_{\bar{i}}\right)\,.
\eea
A potential obstruction to the solvability of equations (\ref{11compeqmain})-(\ref{11compeqmain2}) lies in the commutators $[\nabla_i, \nabla_{\bar{i}}]$ and $[\nabla_i, \nabla_j]=0$. We will first analyze the commutator $[\nabla_i, \nabla_{\bar{i}}]$. In particular, its trace is the Ricci tensor $\mathrm{Ric}=g^{i\bar{i}}[\nabla_i, \nabla_{\bar{i}}]$, acting on the two-form $\omega$. Since the manifold $\mathcal{M}$ is Calabi-Yau, we have $\mathrm{Ric}=0$, therefore we get the necessary condition for the solvability of (\ref{11compeqmain})-(\ref{11compeqmain2}):
\bear\label{constr1}
&&g^{i\bar{i}}\,\nabla_{\bar{i}}\left(g_{j\bar{k}}\,\dd_i h-2 g_{i\bar{k}}\,\dd_j h+\left(\nabla_{\bar{k}}\omega_{ij}+2 g_{i\bar{k}}\,\tilde{\lambda}_j-g_{j\bar{k}}\,\tilde{\lambda}_i\right)\right)=\\ \nonumber
&&=g^{i\bar{i}}\,\nabla_{i}\left(g_{j\bar{k}}\,\dd_{\bar{i}} h-2 g_{j\bar{i}}\,\dd_{\bar{k}} h-\left(\nabla_{j}\omega_{\bar{i}\bar{k}}+2 g_{j\bar{i}}\,\tilde{\lambda}_{\bar{k}}-g_{j\bar{k}}\,\tilde{\lambda}_{\bar{i}}\right)\right)\,.
\eear
The terms involving $h$ cancel out. We may rewrite the equation (\ref{constr1}) term by term as
\bea \nonumber
-\nabla_{\bar{k}}\tilde{\lambda}_j+2\,\nabla_{\bar{k}}\tilde{\lambda}_j-g_{j\bar{k}}\,g^{i\bar{i}}\,\nabla_{\bar{i}}\tilde{\lambda}_i=\nabla_j\tilde{\lambda}_{\bar{k}}-2\,\nabla_j\tilde{\lambda}_{\bar{k}}+g_{j\bar{k}}\,g^{i\bar{i}}\,\nabla_{i}\tilde{\lambda}_{\bar{i}}\,.
\eea
It is easily seen from (\ref{lamform}) that $g^{i\bar{i}}\,\nabla_{\bar{i}}\tilde{\lambda}_i=g^{i\bar{i}}\,\nabla_{i}\tilde{\lambda}_{\bar{i}}=0$ (it is the `double divergence' of the two-form $\omega^{(2,0)}$ or $\omega^{(0,2)}$). Therefore what we get is
\bea\label{Killing}
\nabla_{\bar{k}}\tilde{\lambda}_j+\nabla_j\tilde{\lambda}_{\bar{k}}=0\,.
\eea
The equations (\ref{Killing2}), (\ref{Killing}) imply the following lemma:

\begin{lem}
The vector field $\lambda$ defined in (\ref{lamvector}), i.e. the divergence of the two-form $\omega^{(0, 2)}+\omega^{(2, 0)}$, is Killing.
\end{lem}

\vspace{0.3cm}\noindent
As discussed earlier, $\lambda$ is holomorphic if and only if $f=\mathrm{const.}=f_0$. We will now prove the following statement:

\begin{thm}\label{CYholvec}
On a Calabi-Yau threefold without parallel vectors any Killing vector field is holomorphic\footnote{On a Calabi-Yau twofold the situation is different, see Appendix \ref{TNUT} for more details.}.
\end{thm}
\noindent \underline{Proof.}\\
A vector $v$ is Killing if $
\nabla_\mu\,v_\nu+\nabla_\nu\,v_\mu=0
$, 
where $v_\mu$ is the dual one-form. On a K\"ahler manifold this equation may be split into two:
\bear\label{ijkilleq}
&&\nabla_i\,v_j+\nabla_j\,v_i=0\\ \label{killeq2}
&&\nabla_i\,v_{\bar{j}}+\nabla_{\bar{j}}\,v_i=0\,.
\eear
A holomorphic Killing field is the one that satisfies $\dd_i \,v^{\bar{j}}=0$ or, in terms of the dual one-form, as $\nabla_i\,v_{j}=0$. Therefore for a holomorphic vector field the two terms in (\ref{ijkilleq}) are separately zero.

\vspace{0.3cm}
\noindent
Imagine, however, that the Killing vector field $v$ is not necessarily holomorphic. In this case the quantity characterizing its non-holomorphicity is $F_{ij}=\nabla_i\,v_j-\nabla_j\,v_i$. In fact, it arises naturally in the Lie derivative of the K\"ahler form $\varpi$ w.r.t. the Killing field $v$. Indeed, one calculates $i_v\,\varpi=\varpi_{a\bar{a}}\,(v^{\bar{a}}\,dz^a-v^a\,dz^{\bar{a}})=i\,(v_a\,dz^a-v_{\bar{a}}\,dz^{\bar{a}})$, where we used the fact that the Hermitian components of the K\"ahler form and of the metric are related simply as $\varpi_{a\bar{a}}=i\,g_{a\bar{a}}$. therefore
\bea
\mathfrak{L}_v\,\varpi=i\,\dd_b\,v_a\,dz^b\wedge dz^a-i\,\dd_{\bar{b}}\,v_{\bar{a}}\,dz^{\bar{b}}\wedge dz^{\bar{a}}-i\,(\dd_a\, v_{\bar{b}}+\dd_{\bar{b}}\,v_a)\,dz^a\wedge dz^{\bar{b}}\,.
\eea
The term in brackets vanishes due to one of the Killing conditions\footnote{The covariant derivative in that expression may in fact be replaced by an ordinary derivative as the mixed Christoffel symbols are zero.} (\ref{killeq2}). As a result,
\bea\label{killF}
\mathfrak{L}_v\,\varpi={i\over 2}\left(F-F^\ast\right)\,.
\eea
We see that the Lie derivative $\mathfrak{L}_v\,\varpi\in \Omega^{(2, 0)}(\mathcal{M})\oplus \Omega^{(0, 2)}(\mathcal{M})$ is uniquely characterized by the two-form $F$. Let us derive constraints on this form, starting from the defining equations (\ref{ijkilleq}), (\ref{killeq2}). Since on a K\"ahler manifold $[\nabla_i, \nabla_k]=0$ we have from (\ref{ijkilleq})
\bea\label{eq1}
0=\nabla_k\nabla_j\,v_i-\nabla_i\nabla_j\,v_k=\nabla_j\,F_{ki}.
\eea
By the same token from (\ref{killeq2}) we get
\bea
0=\nabla_k\nabla_{\bar{j}}\,v_i-\nabla_i\nabla_{\bar{j}}\,v_k=\nabla_{\bar{j}}\,F_{ki}+R^n_{\;ik\bar{j}}\,v_n-R^n_{\;ki\bar{j}}\,v_n
\eea
On a K\"ahler manifold, since the $(2, 0)$-components of the Riemann tensor are zero, ${R^n_{\;\bar{j}ik}=0}$, the cyclic Bianchi identity implies the symmetry property $R^n_{\;ik\bar{j}}=R^n_{\;ki\bar{j}}$, therefore the above equation is simplified to
\bea\label{eq2}
\nabla_{\bar{j}}\,F_{ki}=0\,.
\eea
The two equations (\ref{eq1}), (\ref{eq2}) together imply that the $(2, 0)$ two-form $F$ is parallel:
\bea\label{parallelform}
\nabla_\mu\,F_{ij}=0\,.
\eea
Clearly, its complex conjugate, which is a form of type $(0, 2)$, is parallel as well: ${\nabla_\mu\,F_{\bar{i}\bar{j}}=0\,.}$

\vspace{0.3cm}
\noindent
On a Calabi-Yau 3-fold there is a nowhere-vanishing holomorphic 3-form $\Omega_{ijk}$. If it is normalized so that the volume form is $\mathrm{vol}=i\,\Omega \wedge \widebar{\Omega}$, then $\Omega$ is also parallel: $\nabla_\mu\, \Omega=0$. Raising the indices, we also obtain the dual poly-vector $\widetilde{\Omega}^{abc}$, which is parallel as well. Using this poly-vector, we can dualize the $(2, 0)$ two-form $F$ to a $(1, 0)$ vector field $f^a:={1\over 2}\,\widetilde{\Omega}^{abc}\,F_{bc}$. It follows from (\ref{parallelform}) that this vector field is parallel:
\bea
\nabla_\mu\,f^a=0\,.
\eea
On a K\"ahler manifold the parallel vector fields come in pairs, since $I\circ f$ is parallel as well due to the fact that $\nabla\,I=0$. For $f\neq 0$ this implies the reduction of the holonomy group $SU(3)\to SU(2)$, and the manifold is $\mathcal{M}_3\simeq\mathbb{R}^2\times \mathcal{M}_2$. $\blacksquare$

\vspace{0.3cm}
\noindent
We have thus proven that the manifold $\mathcal{M}$ has no non-holomorphic isometries, so that we may set $f=f_0$ and assume that $\lambda$ is a \emph{holomorphic} vector field.

\vspace{-0.5cm}
\begin{center}
\line(10,0){460}
\end{center}
\vspace{-0.3cm}
{\small
To derive a further constraint on $\lambda$, let us calculate $\Omega(\bullet, \bullet, \lambda)$:
\bear\nonumber
\Omega_{abc}\,\lambda^c=\textrm{by}\;(\ref{lamvector})=\Omega_{abc}\,\nabla_t\omega^{ct}=\textrm{by}\;(\Omega-\textrm{dualization})=\Omega_{abc}\widetilde{\Omega}^{cts}\,\nabla_t\omega_s=\\ \nonumber=\textrm{by}\;(\ref{inverseform1})=\nabla_a\omega_b-\nabla_b\omega_a=(\dd\omega)_{ab}\,.
\eear
Since $\Omega(\bullet, \bullet, \lambda):=\Omega_{abc}\,\lambda^c\,dz_a\wedge dz_b$ is a holomorphic 2-form (both $\Omega$ and $\lambda$ are holomorphic), $\bar{\dd}\Omega(\bullet, \bullet, \lambda)=0$. According to the above, one also has $\dd \Omega(\bullet, \bullet, \lambda)=0$. To summarize,
\bea\label{presholform}
\mathcal{L}_\lambda \Omega=0\,,
\eea
i.e. $\lambda$ is a holomorphic Killing vector field that preserves the Calabi-Yau 3-form $\Omega$.

\vspace{0.3cm}
\noindent
The requirement (\ref{presholform}) is an additional condition on $\lambda$, i.e. it is not satisfied for an arbitrary holomorphic Killing field: consider the case of $\mathcal{M}=\CC$ with K\"ahler form (=volume form) $\mathrm{vol}=i\,dz\wedge d\bar{z}$, holomorphic one-form $\Omega:=dz$ and holomorphic Killing field $\lambda:=\mathrm{Re}\,(i\,z\,{\dd \over \dd z})$. One has $\mathcal{L}_\lambda\Omega=i\,\Omega\neq 0$.
}
\vspace{-0.7cm}
\begin{center}
\line(10,0){460}
\end{center}
\vspace{-0.3cm}

\vspace{0.3cm}
\noindent
Since $\lambda$ is a holomorphic Killing vector field, it preserves the K\"ahler form and one can introduce the corresponding moment map $\tau$ by means of the following equation
\bea\nonumber
d\tau=\mathcal{J}\circ \tilde{\lambda},
\eea
where $\mathcal{J}$ is the complex structure. In components, $\tilde{\lambda}_k=i\,\dd_k\tau$ and $\tilde{\lambda}_{\bar{k}}=-i\,\dd_{\bar{k}}\tau$. We can now rewrite the equation (\ref{11compeqmain}) as
\bear\label{11parteq5}
\nabla_i\omega_{j\bar{k}}=g_{j\bar{k}}\,\dd_i \widehat{h}-2 g_{i\bar{k}}\,\dd_j \widehat{h}+\nabla_{\bar{k}}\omega_{ij}\,,\\
\textrm{where}\quad\quad \widehat{h}=h-i\,\tau\,.
\eear
Let us now discuss the integrability conditions for (\ref{11parteq5}). If $i$ and $m$ are both holomorphic indices, one has $[\nabla_i, \nabla_m]=0$, therefore one has the following condition:
\bea\label{cons1}
\nabla_m(g_{j\bar{k}}\,\dd_i \widehat{h}-2 g_{i\bar{k}}\,\dd_j \widehat{h}+\nabla_{\bar{k}}\omega_{ij})-\nabla_i(g_{j\bar{k}}\,\dd_m \widehat{h}-2 g_{m\bar{k}}\,\dd_j \widehat{h}+\nabla_{\bar{k}}\omega_{mj})=0
\eea

\noindent
Contracting this with $g^{i\bar{k}}$, we get
\bea\label{cons2}
4\nabla_m\dd_j\widehat{h}+\nabla_m\tilde{\lambda}_j+g^{i\bar{k}}\,\nabla_i\nabla_{\bar{k}}\omega_{mj}=0
\eea
Since $\lambda$ is a holomorphic vector field, $\nabla_m\tilde{\lambda}_j=0$. Noting the following equality:
\bea\label{antiholholomega}
\nabla_{\bar{k}}\nabla_i\omega_{mj}=0,
\eea
which follows from (\ref{omegaT}), (\ref{TfO}), the assumption $f=f_0=\mathrm{const.}$ and the fact that $\Omega$ is a holomorphic form, we deduce that $g^{i\bar{k}}\,\nabla_i\nabla_{\bar{k}}\omega_{mj}=g^{i\bar{k}}\,\nabla_{\bar{k}}\nabla_i\omega_{mj}+(\mathrm{Ric}\circ \omega)_{mj}={f_0\over 3}\,g^{i\bar{k}}\,\nabla_{\bar{k}}\Omega_{imj}=0$. Hence (\ref{cons2}) leads to
\bea\label{mommaphol}
\nabla_m\dd_j\widehat{h}=0\,.
\eea
Substituting this in (\ref{cons1}), we obtain
\bea \nonumber
\nabla_m\nabla_{\bar{k}}\omega_{ij}-\nabla_i\nabla_{\bar{k}}\omega_{mj}=0
\eea
Commuting the derivatives, using (\ref{antiholholomega}) and the symmetry property of the Riemann tensor on a K\"ahler manifold: $R^n_{\;jm\bar{k}}=R^n_{\;mj\bar{k}}$ (which follows from the cyclic Bianchi identity), we get
\bea
\label{RiemOmega}
R^n_{\;mj\bar{k}}\,\omega_{in}=R^n_{\;ij\bar{k}}\,\omega_{mn}\,.
\eea
Dualizing the two-form $\omega$ to a vector using the three-form $\Omega$, i.e. $\omega_{ij}={1\over 2}\Omega_{ijk}\,\omega^k$, subsequently dualizing the equation (\ref{RiemOmega}) (which is skew-symmetric in the $(i, m)$ indices) by means of multiplication by $\widetilde{\Omega}^{ims}$ and using (\ref{inverseform1}), we get
\begin{empheq}[box=\fbox]{align}
\hspace{1em}\vspace{2em}
\label{RiemOmega2}
R^n_{\;mj\bar{k}}\,\omega^m=0\,
\hspace{1em}
\end{empheq}
i.e. the Riemann tensor has a `null-vector'. This completes the proof of the \textbf{Proposition}.~$\blacksquare$

\subsection{CKYF of type $(1, 1)$}\label{CKYF11}

For the time being we will make an additional assumption that the CKYF 2-form $\omega$ is of type $(1, 1)$, i.e. we set $\omega^{(2,0)}=\omega^{(0,2)}=0$ (which trivially satisfies (\ref{RiemOmega2})). We therefore find ourselves in the situation studied in \cite{Gauduchon, Moroianu} -- in this section we mainly review the results of these papers. The equation (\ref{11compeqmain}) simplifies to
\bea
\nabla_a \omega_{b\bar{c}}=\left(g_{b\bar{c}} \dd_ah-2\, g_{a\bar{c}} \dd_b h\right)\;,\quad \mathrm{where}\quad h=g^{a\bar{b}} \omega_{a\bar{b}}
\eea
It is convenient to introduce the `shifted' 2-form of type $(1,1)$
$\Omega_{b\bar{c}}=\omega_{b\bar{c}}-\,h\,g_{b\bar{c}}$,
which brings the equation to the form
\bea\label{ckyteq1}
\nabla_a \Omega_{b\bar{c}}=-2\,g_{a\bar{c}}\,\dd_b h\,.
\eea
The complex conjugate equation is\footnote{Using the notation $X=X^a\,\dd_a+X^{\bar{a}}\,\dd_{\bar{a}}$ for a vector field $X$ as well as for the corresponding dual 1-form $X=X_a dz^a+X_{\bar{a}} d z^{\bar{a}}$, one has
\bea
\nabla_X \Omega_{b\bar{c}}=2\,\left(X_{\bar{c}}\,\dd_b h+X_{b}\,\dd_{\bar{c}} h\right)\,.
\eea
This equation is the defining equation of a so-called Hamiltonian 2-form \cite{Gauduchon}, and may be rewritten invariantly as follows:
\bea\label{ham2form}
\nabla_X \Omega=\mathcal{J}\circ dh \wedge X-dh\wedge \mathcal{J}\circ X\,.
\eea
}
\bea\label{ckyteq2}
\nabla_{\bar{a}} \Omega_{b\bar{c}}=-2\,g_{b\bar{a}}\,\dd_{\bar{c}} h\,.
\eea
The equations defining a Killing-Yano tensor of type $(1, 1)$ have the form (\ref{ckyteq1})-(\ref{ckyteq2}) for a manifold $\mathcal{M}$ of arbitrary complex dimension $M>2$, up to a rescaling of the function $h$. Therefore for the moment we will relax the assumption that $\mathrm{dim}_{\CC}\;\mathcal{M}=3$ and consider this more general situation. Let us assume that the Hamiltonian 2-form $\Omega$ has a maximum number of \emph{distinct} (real) eigenvalues $\lambda_1\, \ldots\, \lambda_M$ with corresponding eigenvectors $v^{(1)}\, \ldots\, v^{(M)}$, i.e.
\bea\label{eigenveceq}
\Omega_{m\bar{n}}\,(v^{(i)})^{\bar{n}}=\lambda_i\,g_{m\bar{n}}\,(v^{(i)})^{\bar{n}},\quad\quad\quad (v^{(i)})^{m}\,\Omega_{m\bar{n}}\,=\lambda_i\,(v^{(i)})^{m}\,g_{m\bar{n}}\,.
\eea
We will also assume that the eigenvalues $\lambda_1\, \ldots\, \lambda_M$, which are functions on $Y$, are functionally independent, i.e. $d\lambda_1\wedge \ldots \wedge d\lambda_M\notequiv 0$.

\begin{lem}\label{gaud1} \cite{Gauduchon}\;
The gradients of the eigenvalues $\lambda_i$ are mutually orthogonal. The eigenvalues $\lambda_i$ are in involution w.r.t. the Poisson bracket.
\end{lem}

\vspace{0.3cm}
\noindent \underline{Proof.}\\ Multiplying the first equality in (\ref{eigenveceq}) by $(v^{(j)})^m$ and using the second equality, we get
\bea
(\lambda_i-\lambda_j)\,(v^{(j)})^m\,g_{m\bar{n}}\,(v^{(i)})^{\bar{n}}=0\,,
\eea
hence the eigenvectors corresponding to different eigenvalues are orthogonal (in the Hermitian sense):
\bea\label{eigenortho}
(v^{(j)})^m\,g_{m\bar{n}}\,(v^{(i)})^{\bar{n}}=0\quad\quad \textrm{for}\quad i\neq j\,.
\eea
Let us now multiply the equation (\ref{ckyteq1}) defining the Hamiltonian 2-form $\Omega$ by $v^b\,v^{\bar{c}}$, where $v$ is a unit-normalized eigenvector of $\Omega$ ($\|v\|^2:=v^b\,v^{\bar{c}}\,g_{b\bar{c}}=1$) corresponding to eigenvalue~$\lambda$. Using
\bear \nonumber
&&v^b\,v^{\bar{c}}\,\nabla_a \Omega_{b\bar{c}}=\dd_a(\lambda\,\|v\|^2)-\Omega_{b\bar{c}}\,(\nabla_a v^b\,v^{\bar{c}}+ v^b\,\nabla_a v^{\bar{c}})=\\ \nonumber &&=\dd_a\,(\lambda\,\|v\|^2)-\lambda\,\dd_a\|v\|^2=(\textrm{since}\,\|v\|^2=1)=\dd_a\lambda\,,
\eear
we then get
\bear
 \label{gradlam}
\dd_a\lambda=-2\,g_{a\bar{c}}\,v^{\bar{c}}\cdot(v^b\dd_b h)\,.
\eear
This formula, together with (\ref{eigenortho}), implies that the gradients of the eigenvalues $\lambda_i$ are mutually orthogonal.
Recall also that the Poisson bracket defined by the K\"ahler form is $\{f_1, f_2\}=g^{a\bar{b}}\,\dd_af_1\,\dd_{\bar{b}}f_2-g^{a\bar{b}}\,\dd_af_2\,\dd_{\bar{b}}f_1$. It follows easily from (\ref{eigenortho}) and (\ref{gradlam}) that the $\lambda_i$ are in involution with respect to this Poisson bracket. $\blacksquare$

\vspace{0.3cm}\noindent
We will now show that it is possible to associate to the $\lambda_i$ a set of commuting holomorphic Killing vector fields (i.e. vector fields preserving both the metric and the complex structure). To this end, we construct the elementary symmetric polynomials of $\lambda_i$ (up to $\pm$ signs), which we call $\mu_k$:
\bea\label{mulambdadef}
\prod\limits_{k=1}^n\;(\vartheta-\lambda_k)=\sum\limits_{k=0}^n\,\vartheta^k\,\mu_{k+1},\quad\quad \mu_{n+1}=1.
\eea

\begin{lem}\label{commholkill}\cite{Gauduchon}\;
The vector fields $\xi_i:=\mathcal{J}\circ \nabla \mu_i$ are commuting holomorphic Killing vector fields.
\end{lem}

\vspace{0.3cm}
\noindent \underline{Proof.}\\ First, let us write out these vector fields and the dual one-forms in components:
\bea\label{xiveccomp}
(\xi_i)^a=i\,g^{a\bar{b}}\dd_{\bar{b}}\mu_i,\quad (\xi_i)^{\bar{b}}=-i\,g^{a\bar{b}}\dd_{a}\mu_i\quad \Rightarrow \quad (\xi_i)_a=-i\,\dd_a \mu_i,\quad (\xi_i)_{\bar{b}}=i \,\dd_{\bar{b}}\mu_i\;.
\eea
 Let us start with $\mu_n=-\sum\limits_{k=1}^n\,\lambda_k=-g^{m\bar{n}}\Omega_{m\bar{n}}:=-\mathrm{Tr}(\Omega)$ (Here we have used the partition of unity
 \bea\label{partunity}
g^{m\bar{n}}=\sum\limits_{j=1}^M\,(v^{(j)})^m\,(v^{(j)})^{\bar{n}}
 \eea
 formed out of the eigenvectors $v^{(i)}$ of $\Omega$). From the definition of the Hamiltonian 2-form we obtain:
\bea\label{2formdef2}
\nabla_a\Omega_{b\bar{c}}=-g_{a\bar{c}}\dd_b \mu_n
\eea
We have already seen in (\ref{mommaphol}) that $\nabla_a\dd_b \mu_n=0$, which may be also written as the holomorphicity of the vector field $\xi_n$ (defined in (\ref{xiveccomp})):
\bea
\dd_a(\xi_n)^{\bar{c}}=0.
\eea
Note that the vector field $\xi_i := \mathcal{J}\circ \nabla \mu_i$ is Hamiltonian by definition, i.e. it preserves the K\"ahler form. The Killing condition $\mathfrak{L}_\xi g_{\mu\nu}=\nabla_\mu\xi_\nu+\nabla_\nu\xi_\mu=0$ is then automatically satisfied:
\bear
\nabla_a (\xi_n)_{\bar{b}}+\nabla_{\bar{b}} (\xi_n)_a=0 \Rightarrow \nabla_a \dd_{\bar{b}} \mu_n-\nabla_{\bar{b}} \dd_a \mu_n\equiv 0\quad \textrm{since}\; \Gamma_{a\bar{b}}^{\bar{c}}=0\\
\nabla_a (\xi_n)_b+\nabla_b (\xi_n)_a=0\quad \textrm{(follows from}\; \dd_a(\xi_n)^{\bar{c}}=0\,).
\eear
So far we have shown that $\xi_n:=\mathcal{J}\circ\nabla \mu_n$ is a holomorphic Killing vector field. Now we will prove inductively that $\nabla \mu_i$ are holomorphic Killing for all $i$. Suppose that $\dd_b (\xi_{j})^{\bar{a}}=0$ for $j=n, \ldots, k$.

\begin{lem}
One has the following recurrence relation:
\bea\label{killrecurr}
(\xi_{k-1})^{\bar{a}}=\mu_k (\xi_n)^{\bar{a}}+\Omega^{\bar{a}}_{\;\bar{c}} (\xi_{k})^{\bar{c}},
\eea
where $\Omega^{\bar{a}}_{\;\bar{c}}:=g^{a\bar{a}}\,\Omega_{a\bar{c}}$. (This gives the matrix elements of the operator $\Omega$ in the basis of vectors $\{\xi_k\}$.)
\end{lem}
\noindent\underline{Proof.}\\ According to (\ref{gradlam}), one has for the gradients of $\lambda_i$ the following formula:
\bea\label{grad1}
\nabla^{\bar{a}}\lambda_i:=g^{a\bar{a}}\,\dd_a\lambda_i=-2\,v_i^{\bar{a}}\cdot(v_i^b\,\dd_b h)\,.
\eea
\noindent Using it, we calculate the gradient of the logarithm of (\ref{mulambdadef}):
\bea
\sum\,\frac{-\nabla^{\bar{a}}\lambda_k}{\vartheta-\lambda_k}=\frac{\sum\,\vartheta^k\,\nabla^{\bar{a}}\mu_{k+1}}{\sum\,\vartheta^k\,\mu_{k+1}}\,.
\eea
Acting on it by $-i\,(\vartheta\,\delta^{\bar{c}}_{\bar{a}}-\Omega^{\bar{c}}_{\;\bar{a}})$, using (\ref{grad1}), the definition (\ref{eigenveceq}) in the form $\Omega^{\bar{c}}_{\;\bar{a}}\,v_i^{\bar{a}}=\lambda_i\,v_i^{\bar{c}}$ and the definition (\ref{xiveccomp}) ($-i\,\nabla^{\bar{a}}\mu_{k+1}\equiv (\xi_{k+1})^{\bar{a}}$), we have
\bea\label{recrel1}
-2i\,\nabla^{\bar{c}}h=\frac{\vartheta\,\sum\,\vartheta^k\,(\xi_{k+1})^{\bar{c}}-\sum\,\vartheta^k\,\Omega^{\bar{c}}_{\;\bar{a}}\,(\xi_{k+1})^{\bar{a}}}{\sum\,\vartheta^k\,\mu_{k+1}}
\eea
To simplify the l.h.s., we have used the partition of unity (\ref{partunity}). Recalling the definition $(\xi_n)^{\bar{a}}=-i\,\nabla^{\bar{a}}\mu_n=-2i\,\nabla^{\bar{a}} h$, we get from (\ref{recrel1}) the recurrence relation (\ref{killrecurr}). $\blacksquare$

\vspace{0.5cm}
\noindent\underline{Continuation of proof of Lemma \ref{commholkill}.}\\
Let us now calculate the holomorphic derivative of (\ref{killrecurr}):
\bea
\dd_b(\xi_{k-1})^{\bar{a}}=\dd_b \mu_k (\xi_n)^{\bar{a}}+(\dd_b \Omega^{\bar{a}}_{\;\bar{c}}) (\xi_{k})^{\bar{c}}=ig_{b\bar{c}}\,(\xi_k)^{\bar{c}}\,(\xi_n)^{\bar{a}}+(\dd_b \Omega^{\bar{a}}_{\;\bar{c}}) (\xi_{k})^{\bar{c}}
\eea
In order to evaluate the last term we multiply the definition (\ref{2formdef2}) of the Hamiltonian 2-form by $ (\xi_{k})^{\bar{c}}$ to obtain $\dd_b(\Omega^{\bar{a}}_{\bar{c}})\xi_k^{\bar{c}}=-g_{b\bar{c}}(\xi_k)^{\bar{c}}\nabla^{\bar{a}} \mu_n=-ig_{b\bar{c}}(\xi_k)^{\bar{c}}\,(\xi_n)^{\bar{a}}$. Plugging this in the above expression, we get
\bea
\dd_b(\xi_{k-1})^{\bar{a}}=0\,.
\eea
We have proven that the vector fields $\xi_i$ are Killing. Moreover, they commute, as these are Hamiltonian vector fields, whose corresponding Hamiltonians are $\mu_i=\mu_i(\{\lambda_k\})$, and we have shown above that $\lambda_k$'s are in involution. (We use the property of the Hamiltonian vector fields $[X_f, X_g]=X_{\{f, g\}}$.) This completes the proof of \textbf{Lemma 6.2}. $\blacksquare$

\vspace{0.5cm}
\noindent
Let us analyze the consequences of lemmas\;\ref{gaud1} and \ref{commholkill}, first at the example of the toric metric~(\ref{metr}) and assuming the maximal number (three) of linearly independent vector fields~$\xi_k$.

\begin{lem}\label{gaud2}\cite{Gauduchon}\;
Suppose the metric (\ref{metr}) admits a Killing-Yano form of type~$(1, 1)$, and the Killing vector fields $\xi_i, i=1, 2, 3,$ generated by this form, as defined in Lemma\;\ref{commholkill}, coincide with $\dd \over \dd \phi_i$. Then in the metric (\ref{metr}) one has
{\small
\bea\label{metrorthoij}
G_{ij}=\sum\limits_{k=1}^n\,Q_k(\lambda)\,\frac{\dd \lambda_k}{\dd \mu_i}\,\frac{\dd \lambda_k}{\dd \mu_j}\,, \quad\quad
\textrm{where}\quad\quad
Q_p=f_p(\lambda_p)\;\prod\limits_{t\neq p}\,(\lambda_p-\lambda_t)\,.
\eea
}
The potential $G$ can be expressed as (up to a function linear in $\{\mu_k\}$)
{\small
\bea\label{orthopot2}
G=-\sum\limits_m\, \int\limits^{\lambda_m}\,dy_m\,f_m(y_m)\,\prod\limits_{k}\,(y_m-\lambda_k)\;. 
\eea
}
\end{lem}
\noindent\underline{Proof.}\\
We proved in Lemma 5.1 that the gradients of $\lambda_k$ are orthogonal, therefore in the new variables $\{\lambda_k\}$ the metric should be of orthogonal form:
\bea\nonumber
G_{ij}d\mu^i d\mu^j=\sum\limits_{k=1}^n\,Q_k(\lambda)\,d\lambda_k^2\,.
\eea
This is clearly the same as (\ref{metrorthoij}). Let us now use the condition $\dd_k G_{ij}=\dd_i G_{kj}$. Multiplying the resulting equation by $\mathfrak{J}^j_p\cdot \mathfrak{J}^i_s\cdot \mathfrak{J}^k_t$, where $\mathfrak{J}$ is the Jacobian $\mathfrak{J}^i_s:={\dd \mu_i\over \dd \lambda_s}$, we obtain:
{\small
\bea\nonumber
\delta_{ps}{\dd Q_p\over \dd \lambda_t}-\delta_{pt}{\dd Q_p\over \dd \lambda_s}+Q_s\,T_{pt}^s-Q_t\,T^t_{ps}=0,\quad\quad T_{pt}^s:=\sum\limits_{i, j}\,{\dd \mu_j\over \dd \lambda_p} {\dd \mu_i\over \dd \lambda_t}\,{\dd^2 \lambda_s\over \dd \mu_i \dd \mu_j}
\eea
}
Setting in the above equation $p=s\neq t$, we get
\bea\label{maineqs}
t\neq p\quad\Rightarrow\quad{\dd Q_p\over \dd \lambda_t}+Q_p\,T^p_{pt}-Q_t\,T^t_{pp}=0
\eea
In order to calculate $T^p_{pt}$ and $T^t_{pp}$ we use the defining equation $\prod\limits_{k=1}^n\;(\vartheta-\lambda_k)=\sum\limits_{k=0}^n\,\vartheta^k\,\mu_{k+1}$. Differentiating it w.r.t. $\mu_j$ and sending $\vartheta\to \lambda_i$, we get
\bea\nonumber
{\dd \lambda_i\over \dd \mu_j}=-\frac{\lambda_i^{j-1}}{\prod\limits_{k\neq i}\,(\lambda_i-\lambda_k)}
\eea
It is also easy to calculate the second derivative ${\dd \over \dd \lambda_k}\left({\dd \lambda_i\over \dd \mu_j}\right)$ for $k\neq i$:
\bea\nonumber
k\neq i\quad \Rightarrow \quad {\dd \over \dd \lambda_k}\left({\dd \lambda_i\over \dd \mu_j}\right)=\frac{1}{\lambda_i-\lambda_k}\,\frac{\dd \lambda_i}{\dd \mu_j}\,.
\eea
Using this, we get
\bear\nonumber
t\neq p \quad\Rightarrow\quad T^p_{pt}=\frac{\dd \mu_i}{\dd \lambda_p}\,\frac{\dd}{\dd \lambda_t}\left(\frac{\dd \lambda_p}{\dd \mu_i}\right)=\frac{1}{\lambda_p-\lambda_t},\quad\quad T_{pp}^t=0
\eear
The equations (\ref{maineqs}) therefore are $\quad{\dd Q_p\over \dd \lambda_t}+\frac{1}{\lambda_p-\lambda_t}\,Q_p=0\, (p\neq t)$ and have the solution
\bea\label{Qp}
Q_p=f_p(\lambda_p)\;\prod\limits_{t\neq p}\,(\lambda_p-\lambda_t)
\eea
Using the above results, one can integrate (\ref{metrorthoij}) to obtain an expression for the symplectic potential $G$. Indeed, since $G_{ij}=\frac{\dd \lambda_m}{\dd \mu_j}\,\frac{\dd}{\dd \lambda_m}\left(\frac{\dd G}{\dd \mu_i}\right)$, we have from (\ref{metrorthoij}):
\bea\nonumber
\frac{\dd}{\dd \lambda_m}\left(\frac{\dd G}{\dd \mu_i}\right)=\frac{\dd \lambda_m}{\dd \mu_i}\,Q_p(\lambda)=-\lambda_m^{i-1}\,f_m(\lambda_m)\,.
\eea
Integrating, we get $\frac{\dd G}{\dd \mu_i}=-\sum\limits_m\, \int\limits^{\lambda_m}\,dy_m\,y_m^{i-1}\,f_m(y_m)\,.$ Once again passing to the $\lambda$-variables in the l.h.s., we get
\bear\nonumber
&&\frac{\dd G}{\dd \lambda_n}=-\sum\limits_i\,\frac{\dd \mu_i}{\dd \lambda_n}\,\sum\limits_m\, \int\limits^{\lambda_m}\,dy_m\,y_m^{i-1}\,f_m(y_m)=
 -\sum\limits_m\, \int\limits^{\lambda_m}\,dy_m\,f_m(y_m)\,\left(\sum\limits_i\,\frac{\dd \mu_i}{\dd \lambda_n} y_m^{i-1}\right)=\\ \nonumber &&=\textrm{using the definition}\,(\ref{mulambdadef})=\sum\limits_m\, \int\limits^{\lambda_m}\,dy_m\,f_m(y_m)\,\prod\limits_{k\neq n}\,(y_m-\lambda_k)\,.
\eear
This is easily integrated to give (\ref{orthopot2}). $\blacksquare$

\vspace{0.3cm}
\noindent
Let us now see what form the metric (\ref{metr}) takes. To this end, recall that we have already seen that the matrix $\mathfrak{J}$ diagonalizes the metric $G_{ij}$, i.e. $\mathfrak{J}^T\circ G\circ \mathfrak{J}=\mathrm{Diag}\{Q_1, \ldots, Q_n\}$. Therefore
\bea\label{orthotoricmetric}
ds^2={1\over 4}\,\sum\limits_{k=1}^3\,Q_k\,d\lambda_k^2\,+\,\sum\limits_{k=1}^3\,\frac{1}{Q_k}\,(\mathfrak{J}^i_k\,d\phi_i)\,(\mathfrak{J}^j_k\,d\phi_j ),
\eea
where the functions $Q_k$ are of the form (\ref{Qp}).
The Jacobian $\mathfrak{J}={\dd \mu\over \dd \lambda}$ is easily found by differentiating the definition (\ref{mulambdadef}) 
w.r.t. $\lambda_m$:
\bea\nonumber
-\prod\limits_{k=1, k\neq m}^n\;(\vartheta-\lambda_k)=\sum\limits_{k=0}^{n-1}\,\vartheta^k\,{\dd\mu_{k+1}\over \dd \lambda_m}
\eea
It follows that $\mathfrak{J}^{k+1}_m={\dd\mu_{k+1}\over \dd \lambda_m}$ is minus the elementary symmetric polynomial of degree $n-1-k$ in the variables $\lambda_1, \ldots, \widehat{\lambda_m},\ldots, \lambda_n$ with $\lambda_m$ omitted.

\vspace{0.3cm}
\noindent 
The expression for the metric (\ref{metrorthoij}) and for the potential (\ref{orthopot2}) essentially reproduce the formulae of Proposition 11 in \cite{Gauduchon}. We also note that an expression identical to (\ref{Qp}) was obtained in the investigation of metrics possessing Killing tensors in \cite{Chervonyi}. 

\vspace{0.3cm}\noindent
Now, it is not difficult to see that the fully orthotoric metric (\ref{orthotoricmetric}) cannot describe the geometry (\ref{metric}) of~$Y$. Indeed, consider the sphere $\CP^1\subset Y$ lying at the intersection of two hyperplanes, say $\ell_1=0$ and $\ell_2=0$. According to (\ref{asymptcond}), at each of these hyperplanes one should have $\mathrm{Det} (\mathrm{Hess}\,G)^{-1}=0$. From the point of view of the orthotoric metric, this means that one of the functions $\{{1\over f_p(\lambda_p)}\}$ has to vanish at this hyperplane. Therefore at the intersection of two hyperplanes two functions vanish, and we may assume ${1\over f_1(\lambda_1^{(0)})}={1\over f_2(\lambda_2^{(0)})}=0$. The induced metric on the sphere is therefore determined by the remaining function $f_3(\lambda)$. For this metric to be the standard round metric one has to require $Q_3(\lambda_1^{(0)}, \lambda_2^{(0)}, \lambda_3)\sim \frac{1}{1-\lambda_3^2}$. This choice is however incompatible with the Ricci-flatness of the metric (\ref{orthotoricmetric}) (the conditions that Ricci-flatness imposes on the functions $f_p(\lambda_p)$ are given in~\cite{Gauduchon}, see the Theorem in the introduction). The resolution of this problem lies in relaxing the requirement of having three linearly independent vector fields $\{\xi_k\}$ and considering instead the situation when one of the eigenvalues of the Killing-Yano tensor is constant, and the other two give rise to two linearly independent Killing vector fields. In this case, as it was shown in~\cite{Gauduchon}, the metric takes the form
\bear\label{orthotoricmetric2}
\!\!\!\!\!\!\!\!\!\!\!\!\!\!\!ds^2=&&-x\,y\,\cdot \left(g_V\,d\zeta\,d\widebar{\zeta}\right)+{y-x\over 4}\,\left(f_2(y)\,dy^2-f_1(x)\,dx^2\right)+\\ \nonumber &&+\frac{1}{y-x}\left(\frac{1}{f_2(y)}\,(d\phi_1+x\,\omega)^2-\frac{1}{f_1(x)}\,(d\phi_1+y\,\omega)^2\right)\\ \nonumber
&& 
\omega=d\phi_2-A,\quad dA=i\,g_V\,d\zeta\wedge d\widebar{\zeta}\,.
\eear
Here $g_V(\zeta, \widebar{\zeta})\,d\zeta\,d\widebar{\zeta}$ is a K\"ahler metric on a Riemann surface, which may be seen as the metric on the K\"ahler quotient w.r.t. the vector fields ${\dd \over \dd \phi_i}$. 
If one assumes that $g_V$ is the standard round metric on $\CP^1$, $g_V\,d\zeta\,d\widebar{\zeta}=\frac{2 \,d\zeta\,d\widebar{\zeta}}{(1+|\zeta|)^2}$,the metric (\ref{orthotoricmetric2}) becomes compatible with (\ref{metric}), and in fact may be obtained from it by carrying out the above derivations word-by-word for the part of (\ref{metric}), `transverse' to the~$\CP^1$. In particular, the formula~(\ref{orthopot2}) now gives the `transverse', or `reduced', symplectic potential $G$ referred to in (\ref{metric}).

\ssubsection{The orthotoric metric}\label{orthometric}

The dual potential for the Ricci-flat metric may be obtained from (\ref{orthopot2}) by setting
\bear\nonumber
&&\!\!\!\!\!\!\!\!\!\!\!\!\!\!\!\!\!\!\!\!\!\!\!\!f_1(x):=-\frac{3\,x}{P(x)},\quad\quad f_2(y):=-\frac{3\,y}{Q(y)},\quad\quad\textrm{where}\\ \label{cubicpols2}
&&\!\!\!\!\!\!\!\!\!\!\!\!\!\!\!\!\!\!\!\!\!\!\!\!P(x)=x^3-{3\over 2} x^2+c=\prod\limits_{i=1}^3\,(x-x_i),\quad\quad Q(y)=y^3-{3\over 2} y^2+d=\prod\limits_{i=1}^3\,(y-y_i)\,.
\eear
This form of the functions follows from the Ricci-flatness equation. Upon the identification $\lambda_1=x$ and $\lambda_2=y$ we get
\begin{empheq}[box=\fbox]{align}
\hspace{0.5em}
\label{Gortho}
G_{\mathrm{o}}\!\!=\!-\sum\limits_{i=1}^3\;\frac{(x-x_i)(y-x_i)}{1-x_i}\,\log{|x-x_i|}\!-\!\sum\limits_{i=1}^3\;\frac{(x-y_i)(y-y_i)}{1-y_i}\,\log{|y-y_i|}\!+\!3\,(x+y)\,.
\end{empheq}
The roots are labeled in increasing order, i.e. $x_-:=x_1<x_2<x_3$ and $y_1<y_2<y_3$. Note that $y_i=\xi_i$ are the roots of $Q(y)$ that we encountered before, in \S\S\,\ref{3linesec} and \ref{infasympt}. The domain in the $(x, y)$ space is as follows (for details see \cite{MS}):
\bea
y\in [y_1, y_2],\quad\quad x\leq x_-<0\,.
\eea
Note that in this domain $f_1(x)\leq 0$ and $f_2(y)\geq 0$, therefore the metric (\ref{orthotoricmetric2}) is positive-definite. The moment maps $\mu, \nu$ are related to the auxiliary `orthotoric' variables $x, y$ by means of the following formulas, which follow essentially from (\ref{mulambdadef}):
\bea\label{change}
\mu=-x\,y,\quad \nu= -(x+y)\,.
\eea
The meaning of these variables was explained in detail in the previous paragraphs. The minus signs are needed, since in the $(\mu, \nu)$ variables `infinity' corresponds to $\mu, \nu\to+\infty$, whereas in the $(x, y)$ variables it corresponds to $x\to -\infty$, $y$ bounded and positive. The potential (\ref{Gortho}), expressed in terms of $\mu, \nu$, satisfies the Ricci-flatness equation (\ref{Ricciflat}) with~$a=9$. 

\vspace{0.3cm}
\noindent
Let us first of all expand the potential  $G_{\mathrm{ortho}}$ at `infinity', i.e. at $\nu \to \infty$ with $\xi={\mu \over \nu}$ fixed. It is easy to see from (\ref{change}) that this corresponds, in terms of the $x, y$ variables, to the limit $x\to-\infty$, $y$ fixed. We obtain:
\bea
G_{\mathrm{o}} \to -3\,x\, \log{|x|}-x\,\sum\limits_{i=1}^3\;\frac{(y-y_i)}{1-y_i}\,\log{|y-y_i|}+\ldots
\eea
This should be compared with formula (\ref{G0metr}). In particular, this means that the parameter $d$ of the orthotoric metric coincides with the corresponding parameter $d$ from (\ref{Qpolynomial}). Its value is therefore given by (\ref{dval}):
\bea\label{dval2}
d=\frac{16+\sqrt{13}}{64}\,.
\eea
It might seem that the orthotoric potential $G_{\mathrm{o}}$ still possesses one nontrivial parameter~$c$, the free term of the polynomial $P(x)$. However, it turns out that this parameter has to be fixed to a particular value by the requirement that the 3-rd line of the biangle in Fig.~\ref{mompol} passes at a correct angle with respect to the other two lines (meaning that the topology of the manifold is indeed the one of a cone over $\dP_1$). In fact, the value of $c$ may be deduced from the above formulas (\ref{k1k2k3}). Indeed, we calculate $k_1=-{\xi_1^{(1)}\over 1-\xi_1^{(1)}}, k_2=-{\xi_2^{(1)}\over 1-\xi_2^{(1)}}$, where the values of $\xi_{1,2}^{(1)}$ are given in (\ref{xisols}). Assuming that the lower line of the moment polygon is given by $x=x_-$ and remembering that $k_3=-{x_- \over 1- x_-}$, we can then calculate $x_-$ from either of the relations (\ref{k1k2k3}): $x_-={1\over 2}(4+\sqrt{13})$. Since $x_-$ has to be a root of the polynomial $P(x)$, we obtain $c={3\over 2}\,x_-^2-x_-^3$, which numerically turns out to be
\bea\label{cval2}
c=-{1\over 8}(133+37\sqrt{13})\,.
\eea

\ssection{Deformation of the moment polytope}\label{orthodeform}
\vspace{0.2cm}

In this section we will show directly that there is a first-order deformation of the orthotoric metric that reflects an infinitesimal deformation of the moment polytope. The deformation of the polytope in question is the $\epsilon$-shift of its lower side, as shown in Fig.~\ref{linemovpic}. The canonical way to deal with this problem is to keep the domain unchanged and to introduce the explicit dependence on $\epsilon$ in the equation itself. This can be done by explicitly mapping the new domain to the old one, which can be achieved as follows:
\begin{figure}[h]
    \centering 
    \includegraphics[width=0.65\textwidth]{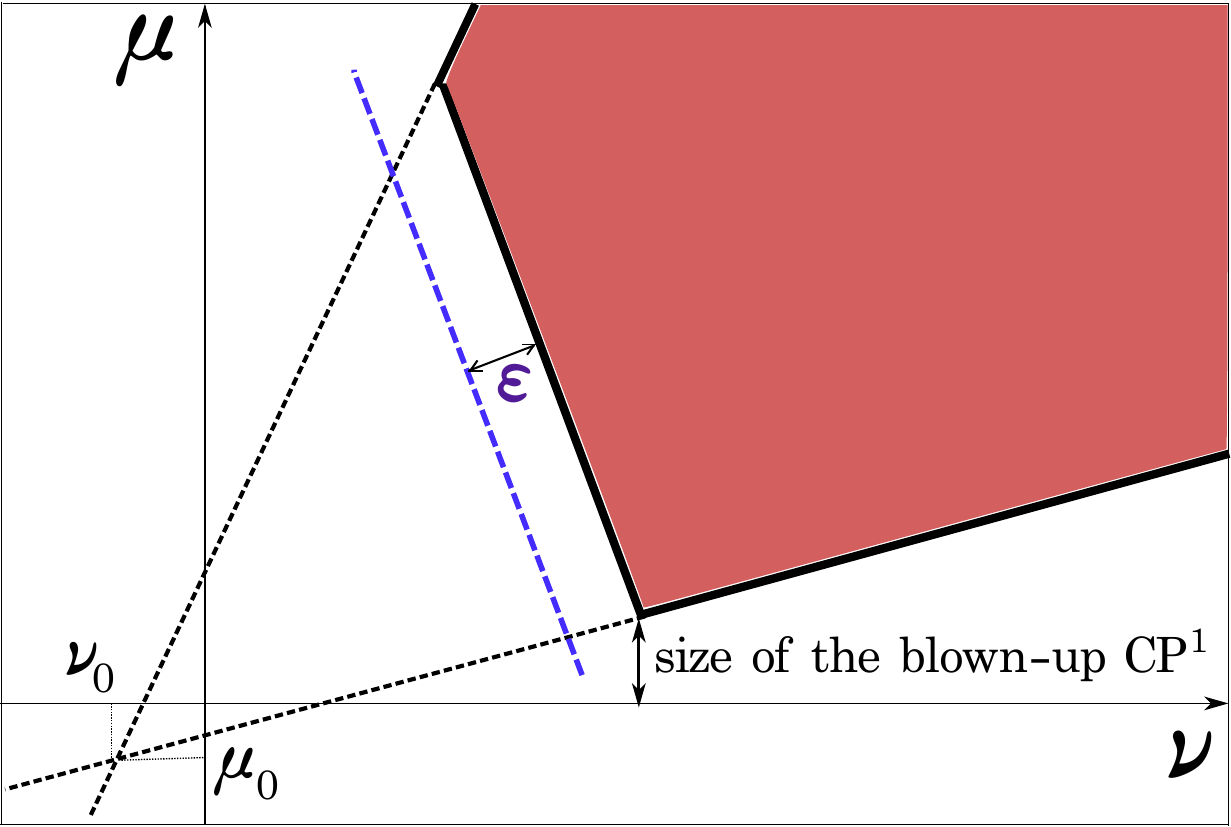}
    \caption{Deformation of the moment polytope.}
    \label{linemovpic}
\end{figure}
\begin{itemize}
\item Shift the variables $(\mu, \nu)$ so that the new origin is located at the intersection point of the dashed lines shown in Fig. \ref{linemovpic}
\item Rescale the variables infinitesimally i.e. $\mu\to(1+\epsilon)\,\mu, \;\;\nu\to (1+\epsilon)\,\nu$. Clearly, this maps the dashed lines to themselves (since they pass through the origin) and moves the lower line of the polytope parallel to itself by a distance of order $\epsilon$.
\item Shift the variables $(\mu, \nu)$ back\,.
\end{itemize}
The net effect is in the following change of variables:
\bea\label{mompolshift}
\bar{\mu}=\frac{\mu+\epsilon\,\mu_0}{1+\epsilon},\quad\quad\quad \bar{\nu}=\frac{\nu+\epsilon\,\nu_0}{1+\epsilon}\,,
\eea
where $(\mu_0, \nu_0)$ are the coordinates of the intersection point of the dashed lines in the original coordinates:
\bea
\mu_0=-\xi_1\xi_2,\quad\quad\quad \nu_0=-(\xi_1+\xi_2)\,.
\eea
It is also convenient to pass to a new unknown function $\widebar{G}$:
\bea\label{GGbar}
G=(1+\epsilon)\,(3\,\bar{\nu}\,\log(1+\epsilon)+\widebar{G})\,.
\eea
One then has the following equation for $\widebar{G}$:
\bea\label{Gbareq}
e^{\frac{\dd \widebar{G} }{\dd \bar{\mu}}+\frac{\dd \widebar{G} }{\dd \bar{\nu}}}\times \mathrm{Det}\,\widebar{\mathrm{Hess}}\,\widebar{G}=a\,\left(\bar{\mu}-\frac{\epsilon \mu_0}{1+\epsilon}\right)\,.
\eea
What we have achieved is that the domain of definition here is the same as the original moment polytope. In particular, for $\epsilon=0$ we know the (unique) solution to (\ref{Gbareq}) -- it is given by the orthotoric potential (\ref{Gortho}) above (for the case $a=9$). Since the latter is most conveniently expressed in the orthotoric $(x, y)$ coordinates, let us pass to these coordinates in the eq.~(\ref{Gbareq}), using the formulas (\ref{change}). We then obtain:
\bea\label{Ricortho1}
\!\!\!\!\!\!\!\!\!\mathcal{R}[\widebar{G}, \epsilon]=0,\quad\quad\textrm{where}
\eea
{\small
\bea
\nonumber 
 \!\!\!\!\!\!\!\!\!\mathcal{R}[\widebar{G}, \epsilon]:=e^{\left(\frac{1-y}{y-x}\widebar{G}_y-\frac{1-x}{y-x}\widebar{G}_x\right)}\;
\underbracket[0.6pt][0.6ex]{\frac{1}{(x-y)^2}\;\left(\widebar{G}_{xx}\widebar{G}_{yy}-\left(\widebar{G}_{xy}+\frac{\widebar{G}_x-\widebar{G}_y}{x-y}\right)^2\right)}_{=\mathrm{det}\widebar{\mathrm{Hess}}_{(\bar{\mu}, \bar{\nu})} \widebar{G}}
+9 \,\left(xy+\frac{\epsilon\, \mu_0}{1+\epsilon}\right)\,. 
\eea
}
\!\!\!The first order of perturbation theory in $\epsilon$ for the solution $\widebar{G}$ may be constructed as follows:
\bear\label{GHdef}
&&\widebar{G}=G_{\mathrm{o}}+\epsilon\,H(x,y),\\ \nonumber
&& \frac{\delta \mathcal{R}[\widebar{G}, 0]}{\delta \widebar{G}}\big|_{\widebar{G}=G_{\mathrm{o}}}\circ (\epsilon\,H)+\mathcal{R}[G_{\mathrm{o}}, \epsilon]=0\,.
\eear
Taking into account that {\small
\bea
\frac{\delta \mathcal{R}[\widebar{G}, 0]}{\delta \widebar{G}}\big|_{\widebar{G}=G_{\mathrm{o}}}\circ H=\frac{3}{x-y}\,\left(y\frac{\dd}{\dd x}\left(P(x)\frac{\dd H}{\dd x}\right)-x\frac{\dd}{\dd y}\left(Q(y)\frac{\dd H}{\dd y}\right)\right),
\eea}
we can write the linearized equation as 
{\small\bea\label{linH1}
\frac{1}{x-y}\,\left(y\frac{\dd}{\dd x}\left(P(x)\frac{\dd H}{\dd x}\right)-x\frac{\dd}{\dd y}\left(Q(y)\frac{\dd H}{\dd y}\right)\right)+3\,\mu_0=0\,.
\eea}
The variables separate, and one can look for the solution in the form
\bea\label{Hh1h2}
H(x, y)=h_1(x)+h_2(y)\,.
\eea
The functions $h_1(x), h_2(y)$ then satisfy the following equations:
\bea
h_1'(x)=\frac{\mathrm{B}+3\,\mu_0\, x+\mathrm{A}\, x^2}{P(x)},\quad\quad\quad h_2'(y)=\frac{\widetilde{\mathrm{B}}+3\,\mu_0\, y+\mathrm{A}\, y^2}{Q(y)},
\eea
where $\mathrm{A}, \mathrm{B}, \widetilde{\mathrm{B}}$ are constants. Since we wish the function $H$ to be regular at the sides of the moment polytope, i.e. at $y=\xi_1, \xi_2$ and $x=x_-$, we have to require that the numerators of the fractions in the right hand sides of the equations vanish at the prescribed points, which can be formulated as
\bea
\widetilde{\mathrm{B}}+3\,\mu_0\, y+\mathrm{A}\, y^2=\mathrm{A}\,(y-\xi_1)\,(y-\xi_2),\quad\quad\quad \mathrm{B}+3\,\mu_0\, x_-+\mathrm{A}\, x_-^2=0\,.
\eea
Using the relations between the roots $\xi_0, \xi_1, \xi_2$ summarized in Appendix \ref{solxi}, we find
\bea
\mathrm{A}=-3\,\xi_0,\quad \mathrm{B}=3\,x_-\, \xi_0\,(x_--\xi_1-\xi_2),\quad \widetilde{\mathrm{B}}=3\,d,
\eea
so that
\bea\label{h1h2sols}
h_1'(x)=-\frac{3\,\xi_0\,(x-\tilde{x})}{(x-x_1)(x-x_2)},\quad h_2(y)=-3 \,\xi_0\,\log{|y-\xi_0|}\,,\quad \tilde{x}=\xi_1+\xi_2-x_-\,.
\eea
The formulas (\ref{GGbar}), (\ref{GHdef}), (\ref{Hh1h2}) and (\ref{h1h2sols}) together give the first-order Ricci-flat deformation of the metric, corresponding to the change of the polytope depicted in Fig.~\ref{linemovpic}.

\vspace{0.2cm}
\ssubsection{Asymptotic behavior at infinity}

We wish to quantify the deviation of the $\epsilon$-corrected metric from the conical metric (\ref{coneasympt}) at infinity. 
To do so, first we write the symplectic potential as (see (\ref{GGbar}))
\bea\label{orthopotcorr}
G=(1+\epsilon)\,(G_{o}(\bar{x}, \bar{y})-3\,\log(1+\epsilon)\,(\bar{x}+\bar{y})+\epsilon\,H(\bar{x}, \bar{y})),
\eea
where $\bar{x}, \bar{y}$ are the orthotoric variables which themselves depend on $\epsilon$ via the formulas~(\ref{mompolshift}):
\bea\label{tildex}
-\bar{x}\,\bar{y}=\frac{\mu+\epsilon\,\mu_0}{1+\epsilon},\quad\quad\quad -(\bar{x}+\bar{y})=\frac{\nu+\epsilon\,\nu_0}{1+\epsilon}
\eea
Since we are interested in the first order in $\epsilon$, we may construct a perturbation theory for the variables $\bar{x}, \bar{y}$. If we set $\bar{x}=x+\epsilon\,\delta x, \quad\bar{y}=y+\epsilon\,\delta y$\, , from (\ref{tildex}) we easily find
\bea
\delta x=\frac{(x-\xi_1)(x-\xi_2)}{y-x},\quad\quad \delta y=-\frac{(y-\xi_1)(y-\xi_2)}{y-x}\,.
\eea
From (\ref{orthopotcorr}) we find that the first order (in $\epsilon$) correction to the orthotoric potential $G_{o}(x, y)$ can be expressed as
\bea
G=G_{o}(x, y)+\epsilon\,\left(\underbracket[0.6pt][0.6ex]{G_{o}(x, y)+ \frac{\dd G_{o}(x, y)}{\dd x}\,\delta x+\frac{\dd G_{o}(x, y)}{\dd y}\,\delta y+H(x, y)-3\,(x+y)}_{:=\delta G(x, y)}\right)+\ldots
\eea
Now, $x$ and $y$ are related to $\mu, \nu$ by means of the standard formulas (\ref{tildex}) with $\epsilon=0$, i.e. $\mu=-x\,y, \nu=-(x+y)$. We are interested in the behavior of $\delta G(x, y)$ at infinity, i.e. $x\to-\infty$, $y$ bounded. A direct calculation shows that
\bea
\delta G(x, y)=\frac{\alpha}{x^2}+o\left(\frac{1}{x^2}\right),\quad\quad \alpha=\mathrm{const.}
\eea
Since for large negative $x$ we have $|x|=O(\mu)=O(\nu)=O(r^2)$, we find
\bea\label{decay1}
|g-g_{\mathrm{ortho}}|_{g_\mathrm{ortho}}=O\left({1\over r^6}\right)\,.
\eea
According to the general theory introduced in \S\,\ref{compsuppcoh} (see Lemma\;\ref{decaylemma}), this implies that the variation of the K\"ahler form
\bea
[\delta \omega]\in H^2_c(Y, \mathbb{R})
\eea
lies in the compactly supported cohomology group.

\ssubsection{An alternative derivation of the deformation}

There is also a simpler way to evaluate the decay rate of the first-order deformation. Indeed, instead of first mapping the new moment polytope to the old one, one can try to construct directly a deformation of the potential $G$ as follows:
\bea\label{Gdeform}
G=G_{\mathrm{o}}+\epsilon\,H
\eea
Expanding the Ricci-flatness equation (\ref{Ricciflat}) around the orthotoric solution to the first order in the deformation $H$, or simply using (\ref{linH1}) that follows from the linearization of~(\ref{Ricortho1}), we obtain the following remarkably simple linear equation:
\bea\label{Hlin}
\frac{1}{x}\,\frac{\dd}{\dd x} \left( P(x)\,\frac{\dd H}{\dd x}\right)-\frac{1}{y}\,\frac{\dd}{\dd y} \left( Q(y)\,\frac{\dd H}{\dd y}\right)=0
\eea

\vspace{0.3cm}
\noindent
The price that we will have to pay for not working in a fixed domain (as we did in the previous paragraph, by mapping the new domain to the old one) is that the deformation $H$ will be affected by the domain shift in a singular way. Indeed, this has to be the case, since we proved in \S\,\ref{unique} that there cannot be a deformation that is smooth at all sides of the polytope (this would imply that there is a deformation of the Ricci-flat metric with the same moment polytope, i.e. within the same K\"ahler class).

\vspace{0.3cm}
\noindent
We recall that near any one of its edges, let us say the one defined by $\ell=0$, the potential $G$ behaves as follows:
\bea
G=\ell \,(\log \ell-1)+\ldots
\eea
For our application we think of $\ell=0$ as being the lower line of the moment polytope depicted in Fig.~\ref{linemovpic}. Transporting the line defined by $\ell=0$ parallel to itself means changing $\ell$ by $\ell+\epsilon$, where $\epsilon$ is a constant. Therefore after the shift
\bea
G=(\ell+\epsilon)\,(\log(\ell+\epsilon)-1)+\ldots=\ell \,(\log \ell-1)+\epsilon \, \log{\ell}+\ldots
\eea
One sees that the deformation is formally proportional to $\log{\ell}$, which is singular, however the important point is that the coefficient of proportionality is a constant ($\epsilon$), whereas in general it could be a function of $\mu, \nu$. We come to the conclusion that the admissible deformation of the potential $G$, i.e. the one that can be resummed into a smooth potential $G$ defined on a deformed moment polygon, is the one which has the form
\bea
H\sim \beta\,\log{\ell}+\ldots,\quad \beta=\textrm{const.}
\eea
In the $x, y$ variables this means that we are looking for a deformation of the form
\bea\label{Hasympt}
H \sim \beta\,\log{|x-x_-|}+\ldots \quad\mbox{as}\quad x\to x_-
\eea
Since the left and right sides of the moment polytope in Fig.~\ref{linemovpic} are not shifted, we are looking for solutions of (\ref{Hlin}), non-singular at $y=y_1, y=y_2$. The general solution of~(\ref{Hlin}) has the form:
\bea
H(x,y)=\sum\limits_\lambda\;h_\lambda(x)\,g_\lambda(y),
\eea
where $h_\lambda$ and $g_\lambda$ are eigenfunctions of the Heun operators,
\bear\label{SL1}
&&\frac{\dd}{\dd x}\left( P(x)\,\frac{\dd h_\lambda(x)}{\dd x}\right)-\lambda\, x\, h_\lambda(x)=0,\\ \label{SL2}
&&\frac{\dd}{\dd y}\left( Q(y)\,\frac{\dd g_\lambda(y)}{\dd y}\right)-\lambda\, y\, g_\lambda(y)=0
\eear
Moreover, $g_\lambda(y)$ is an eigenfunction of the Sturm-Liouville problem, namely it has to be real-analytic at $y=y_1, y=y_2$. It  then follows from Lemma\;\ref{eigenvaluelemma} that $\lambda=0$ or $\lambda \geq 3$. Then the standard Frobenius analysis of the equation (\ref{SL1}) shows that $h_\lambda(x)$ decays at least as $1\over x^2$ at infinity\footnote{If we exclude the growing case, which would then not be subleading to the conical metric at infinity.}. Now, if we assume $h_\lambda(x)$ regular at $x=x_-$, multiplying (\ref{SL1}) by $h_\lambda(x)$ and integrating by parts, we find that $h_\lambda(x)=\mathrm{const.}$ for $\lambda=0$ and $h_\lambda(x)=0$ for $\lambda>0$. Therefore a regular deformation (different from a constant) does not exist. To have a nontrivial deformation, we have to assume that $h_\lambda(x)$ is singular at $x=x_-$, moreover the Frobenius analysis shows that it behaves as
\bea
h_\lambda(x)=a_\lambda\,\log{|x-x_-|}+\ldots,\quad\quad a_\lambda \neq 0
\eea
Thus, we see that the solution $H(x, y)$ behaves at $x=x_-$ as
\bea
H(x, y)\sim q(y)\,\log{|x-x_-|}+\ldots\quad\mathrm{with}\quad q(y)=\sum\limits_\lambda\,a_\lambda g_\lambda(y)
\eea
According to the condition (\ref{Hasympt}), $q(y)=\beta=\mathrm{const.}$, so that
\bea\label{eps}
\beta=\sum\limits_\lambda\,a_\lambda g_\lambda(y)
\eea
It is clear from (\ref{SL2}) that one can take $g_0(y)=1$. Moreover, by standard Sturm-Liouville theory arguments, the eigenfunctions $g_{\lambda_1}(y), g_{\lambda_2}(y)$ for $\lambda_1\neq \lambda_2$ are orthogonal with respect to the weight function $y>0$:
\bea
\int\limits_{y_1}^{y_2}\,dy\,y\,g_{\lambda_1}(y)\, g_{\lambda_2}(y)=0\quad\mathrm{for}\quad \lambda_1\neq \lambda_2
\eea
It follows that in (\ref{eps}) $a_0=\beta$, $a_{\lambda\neq 0}=0$, hence
\bea
H(x, y)=-\beta\,\int\limits_x^\infty\,\frac{d\hat{x}}{P(\hat{x})}=-\frac{\beta}{2\,x^2}+\ldots\,\quad\quad\textrm{for}\quad x\to -\infty\,.
\eea
This confirms the decay estimate (\ref{decay1}).

\vspace{0.3cm}
\noindent
We may also use an alternative criterion, given by formula (\ref{critcomp}), to confirm that the variation of the K\"ahler form lies in $H^2_c(Y, \mathbb{R})$. It involves the calculation of the integrals of $\delta \omega$ over the homologically non-trivial cycles, which we called $\CP^1_A$ and $\CP^1_B$. These spheres are embedded in the del Pezzo surface with normal bundles $\mathcal{O}(1)$ for $\CP^1_A$ and $\mathcal{O}(-1)$ for $\CP^1_B$. From the point of view of the plot shown in Fig.~\ref{linemovpic}, the sphere $\CP^1_A$ is located transversely to the plot at the upper angle of the polygon, and the sphere $\CP^1_B$ -- at the lower angle. It follows from the expression for the metric (\ref{metric}) that the integrals of the K\"ahler form over these cycles are proportional to their $\mu$-coordinates in the plot. The integrals of $\delta \omega$ -- the deformation of the K\"ahler form -- are then the differences of the $\mu$-coordinates of the corners in the original and shifted polytopes. Therefore we arrive at the following equations:
\bear\nonumber
\CP^1_A:\quad\quad\delta\mu_A-\xi_2 \,\delta\nu_A=0,\quad\quad \delta \mu_A-x_-\,\delta\nu_A-\epsilon=0\quad\Rightarrow\quad \delta\mu_A=\frac{\xi_2\,\epsilon}{\xi_2-x_-}\\ \nonumber
\CP^1_B:\quad\quad\delta\mu_B-\xi_1 \,\delta\nu_B=0,\quad\quad \delta \mu_B-x_-\,\delta\nu_B-\epsilon=0\quad\Rightarrow\quad \delta\mu_B=\frac{\xi_1\,\epsilon}{\xi_1-x_-}
\eear
Using the actual values for the roots $\xi_{1,2}$ and $x_-$\footnote{These values are: $x_-={1\over 2}(4+\sqrt{13}),\quad \xi_1={1\over 8} (1+\sqrt{13}),\quad \xi_2={1\over 8} (7+\sqrt{13})$.}, which may be obtained from (\ref{cubicpols2}), (\ref{dval2}), (\ref{cval2}), we find
\bea
\frac{\delta\mu_A}{\delta\mu_B}=3\,,
\eea 
which, according to the criterion (\ref{critcomp}), implies $\delta\omega \in H^2_c(Y, \mathbb{R})$.

\vspace{0.3cm}
\noindent \emph{Comment.} Interestingly, the same calculation shows that the moment polygon, corresponding to the compactly supported K\"ahler class, is the one where the two semi-infinite sides in Fig.~\ref{linemovpic} intersect precisely at $\mu=0$. The polygon may be freely translated in the $\nu$-direction, so we may assume that in this case the two semi-infinite sides intersect at the origin. The three lines are then given by the equations $\mu=\xi_1 \,\nu, \mu=\xi_2\,\nu, \mu=x_-\,\nu+a$, and the calculation again gives the answer $\frac{\delta\mu_A}{\delta\mu_B}=3$ for the ratio of the volumes of the two $\CP^1$'s located at the angles of the polygon.

\vspace{0.2cm}
\ssubsection{Deformation of the Killing-Yano form}\label{KYdeform}

\begin{thm}
The curvature tensor of the orthotoric metric does not possess a null vector, i.e. equation (\ref{RiemOmega2}) is only satisfied for $\omega=0$.
\end{thm}
\noindent \underline{Proof.}\\
The statement that the Riemann tensor has a null vector can be formulated in two equivalent ways:
\bea
R^i_{\;jk\bar{n}}\,\omega^k=0,\quad\quad R^{i}_{\;jk\bar{n}}\,\omega^{\bar{n}}=0\,.
\eea
The two are effectively related by complex conjugation and invoking the symmetry properties of the Riemann tensor. We will use the second form and the expression (\ref{RiemHerm}) for the Riemann tensor, which can be seen to imply
\bea\label{nullvec1}
\sum\limits_{t}\;\frac{\dd^2 G^{-1}_{jk}}{\dd \mu_s \dd \mu_t}\,\widehat{\omega}^t=0 \quad \textrm{for all}\quad j, k, s,\quad\quad \widehat{\omega}^t=G^{-1}_{tn}\,\omega^{\bar{n}}\,.
\eea
In particular, the necessary condition is
\bea
\mathrm{Det}\,\left\{\frac{\dd^2 G^{-1}_{jk}}{\dd \mu_s \dd \mu_t} \right\}_{s, t}=\mathrm{Det}\,\mathrm{Hess}(G^{-1}_{jk})=0\quad\quad \textrm{for all}\quad j, k\,.
\eea
We will now prove that this does not hold for the orthotoric metric. First, we specialize to the case that the metric has $U(2)\times U(1)$ symmetry, rather than $U(1)^3$, i.e. we assume the form (\ref{Kahpot}) of the K\"ahler potential. One can check that, for the dual potential, this implies the following form:
\bear\nonumber
&&G=\left({\mu\over 2}+\tau\right)\log{\left({\mu\over 2}+\tau\right)}+\left({\mu\over 2}-\tau\right)\log{\left({\mu\over 2}-\tau\right)}-\mu\log{\mu}+\widetilde{G}(\mu, \nu)\\ \nonumber
&& \mu=\mu_1+\mu_2,\quad\quad \tau=\frac{\mu_1-\mu_2}{2},\quad\quad \nu=\mu_3\,.
\eear
Here $\widetilde{G}(\mu, \nu)$ is the `reduced' potential used everywhere above -- it does not depend on $\tau$. In our application we have in mind, of course, that
\bea
\widetilde{G}=G_{\mathrm{o}}.
\eea
The Hessian $G_{ij}$ for the potential $G$ of the above form is (the ordering of rows/columns is $\tau, \mu, \nu$):
\bear
&&\mathrm{Hess}\,G=\{G_{ij}\}=\left( \begin{array}{ccc}
\frac{\mu}{{\mu^2\over 4}-\tau^2} & \frac{-\tau}{{\mu^2\over 4}-\tau^2} & 0 \\
\frac{-\tau}{{\mu^2\over 4}-\tau^2} & \frac{\tau^2}{\mu({\mu^2\over 4}-\tau^2)}+\widetilde{G}_{\mu\mu} & \widetilde{G}_{\mu\nu} \\
0 & \widetilde{G}_{\mu\nu} & \widetilde{G}_{\nu\nu} \end{array} \right)\\
&&\mathrm{Det}\,\mathrm{Hess}\,G=\frac{\mu}{{\mu^2\over 4}-\tau^2}\;\mathrm{Det}\,\mathrm{Hess}\,\widetilde{G}\,.
\eear
One easily calculates
\bea\label{probefunc1}
G^{-1}_{33}\equiv ((\mathrm{Hess}\,G)^{-1})_{33}=\frac{\widetilde{G}_{\mu\mu}}{\mathrm{Det}\,\mathrm{Hess}\,\widetilde{G}}\,.
\eea
Since $G^{-1}_{33}$ is independent of $\tau$, $\mathrm{Hess}\,G^{-1}_{33}$ is degenerate and has a null-vector $\left( \begin{array}{ccc}
1  \\
0  \\
0  \end{array} \right)$. Let us check that this is the only null-vector of $\mathrm{Hess}\,G^{-1}_{33}$. To this end, we need to show that
\bea
\mathrm{Det}\,\underset{\mu, \nu}{\mathrm{Hess}}\;(G^{-1}_{33})\nequiv 0\,.
\eea 
Since in the case of the orthotoric metric everything is expressed in terms of the $(x, y)$ variables, we will be using the following formulas describing the change of variables, valid for an arbitrary function $F(\mu, \nu)$:
{\footnotesize
\bear\nonumber
&&F_{\mu\mu}=\frac{F_{xx}+F_{yy}}{(x-y)^2}-\frac{2}{(x-y)^2}\left(F_{xy}+\frac{F_x-F_y}{x-y}\right),\\  \nonumber&&
F_{\nu\nu}=\frac{x^2 \,F_{xx}+y^2\,F_{yy}}{(x-y)^2}-\frac{2\,x\,y}{(x-y)^2}\left(F_{xy}+\frac{F_x-F_y}{x-y}\right) \\ 
\nonumber
&&F_{\mu\nu}=-\frac{x\,F_{xx}+y\,F_{yy}}{(x-y)^2}+\frac{x+y}{(x-y)^2}\left(F_{xy}+\frac{F_x-F_y}{x-y}\right)\,,\\ \label{dethessF}
&&\mathrm{Det}\,\mathrm{Hess}\,F=\frac{1}{(x-y)^2}\,\left(F_{xx}F_{yy}-\left(F_{xy}+\frac{F_x-F_y}{x-y}\right)^2\right)\,.
\eear}We write out the derivatives of the orthotoric symplectic potential $G_{\mathrm{o}}$:
\bea\label{orthoders}
(G_{\mathrm{o}})_{xx}=\frac{3x(x-y)}{P(x)},\quad (G_{\mathrm{o}})_{yy}=\frac{3y(y-x)}{Q(y)},\quad (G_{\mathrm{o}})_{xy}+\frac{(G_{\mathrm{o}})_{x}-(G_{\mathrm{o}})_{y}}{x-y}=0\,,
\eea
where $P(x)=\prod\limits_{i=1}^3\,(x-x_i)=x^3-{3\over 2} x^2+c$, $Q(y)=\prod\limits_{i=1}^3\,(y-y_i)=y^3-{3\over 2} y^2+d$.

\vspace{0.3cm}
\noindent
Setting in (\ref{probefunc1}) $\widetilde{G}=G_{\mathrm{o}}$ and using the above formulas,  we obtain
\bea
G^{-1}_{33}=\frac{1}{3\,(x-y)}\,\left(\frac{P(x)}{x}-\frac{Q(y)}{y}\right)\,.
\eea
Substituting in (\ref{dethessF}) $F=G^{-1}_{33}$, one finds explicitly
{\small
\bear
&&\!\!\!\!\!\!\!\!\!\!\mathrm{Det}\,\underset{\mu, \nu}{\mathrm{Hess}}\;G^{-1}_{33}=\\ \nonumber &&\!\!\!\!\!\!\!\!\!\!=\frac{4 (c-d) \left(y^3 \left(c\, (x+y) \left(10 x^2-5 x y+y^2\right)+3 x^3 (x-y)^5\right)-d\, x^3 (x+y) \left(x^2-5 x y+10 y^2\right)\right)}{3 x^3
   y^3 (x-y)^{10}}\nequiv 0
\eear
}\noindent
To summarize, we have proven that the only null-vector of $\mathrm{Hess}\;G^{-1}_{33}$ is $\left( \begin{array}{ccc}
1  \\
0  \\
0  \end{array} \right)$. For equation (\ref{nullvec1}) to have a non-zero solution $\widehat{\omega}$, this vector would have to be a null-vector of \emph{all} matrices $\mathrm{Hess}\;G^{-1}_{ij}$, which would imply $\dd^2_{\tau} G^{-1}_{ij}=0$ for all $i, j$. We compute, however,
\bea
G^{-1}_{11}=\frac{1}{\mu}\left({\mu^2\over 4}-\tau^2\right)+\frac{\tau^2}{\mu^2}\,\frac{\widetilde{G}_{\nu\nu}}{\mathrm{Det} \,\mathrm{Hess}\, \widetilde{G}}\,.
\eea
Therefore
\bea
\dd^2_{\tau} G^{-1}_{11}=\frac{2}{\mu^2}\left(\frac{\widetilde{G}_{\nu\nu}}{\mathrm{Det} \,\mathrm{Hess}\, \widetilde{G}}-\mu\right)\nequiv 0
\eea
meaning that the only solution of (\ref{nullvec1}) is $\widehat{\omega}=0=\omega$. $\blacksquare$

\section{Summary}
\noindent In the present paper we analyzed the space of Ricci-flat metrics on the non-compact manifold $Y$ -- the total space of the canonical bundle over the del Pezzo surface of rank one. This surface is the blow-up of the projective plane $\CP^2$ at one point. As we explained, a version of the Calabi-Yau for the space $Y$ requires that the metric contain two real parametes, which are the K\"ahler moduli representing the sizes of the original $\CP^2$ and the blown-up sphere $\CP^1$. The only explicitly known (so-called `orthotoric') metric, however, has just one parameter (the overall scale). In the paper we have related this to the fact that the orthotoric metric admits a conformal Killing-Yano form. We have shown that, although the metric allows a first-order deformation, which preserves the Ricci-flatness (as it should, by the Calabi-Yau theorem), the deformed metric will no longer admit a conformal Killing-Yano tensor.  

\vspace{0.6cm}
\textbf{Acknowledgements.}
{\footnotesize
I would like to thank Dmitri Ageev for a collaboration at an initial stage of this project and for many useful conversations. I am grateful to Sergey Frolov, Ulrich Menne, Osvaldo Santillan, Armen Sergeev, Stefan Theisen, Konstantin Zarembo for discussions. I would like to thank the Institut des Hautes \'Etudes Scientifiques and in particular Vasily Pestun for hospitality during my stay, during which a part of this work was done. I am indebted to Prof.~A.A.Slavnov and to my parents for support and encouragement. My work was supported by the ERC Advanced Grant No. 320045 ``Strings and Gravity'' (Principal Investigator Prof.~D.~L\"ust).}
\vspace{-0.3cm}
\begin{center}
\line(1,0){450}
\end{center}
\vspace{0.2cm}
\appendix
\begin{center}
{\normalfont\scshape \large \underline{Appendices}}
\end{center}

\section{Derivation of the metric (\ref{metric})}\label{metricder}

Here we will derive the formula (\ref{metric}) for the metric, starting from the $U(2)\times U(1)$-invariant K\"ahler potential (\ref{Kahpot}): $K=K(|z_1|^2+|z_2|^2, |u|^2)$. Denoting by $K_0$ the K\"ahler potential of the sphere $\CP^1$, $K_0=\log(|z_1|^2+|z_2|^2)$, and introducing the real variables $t, s$ via $e^{t\over 2}=|z_1|^2+|z_2|^2$ and $e^s=|u|^2$, we obtain the following formulas:
{\footnotesize
\bear
&&{\dd^2 K\over \dd z_i \dd \widebar{z}_j}=4 {\dd^2 K\over \dd t^2} \dd_i K_0 \widebar{\dd}_j K_0+2 {\dd K\over \dd t} \dd_i \widebar{\dd}_j K_0,\\
&& {\dd^2 K\over \dd u \dd \widebar{u}}={1\over |u|^2} {\dd^2 K\over \dd s^2},\quad\quad {\dd^2 K\over \dd z_i \dd \widebar{u}}= 2 {\dd^2 K \over \dd t \dd s} {1\over \widebar{u}} \dd_i K_0\,.
\eear
}Taking into account that ${\dd K\over \dd t}=\mu$ and $2\, \dd_i \widebar{\dd}_j K_0\, dz_i d\widebar{z}_j=g_{\CP^1}$, we obtain the following expression for the line element:
{\footnotesize
\bear\label{metrsimpl}
&
 \!\!\!\!\!\!\!\!\!\!\!\!\!\!\!ds^2=\mu\,g_{\CP^1}+4 {\dd^2 K\over \dd t^2} \dd_i K_0 \widebar{\dd}_j K_0\, dz_i d\widebar{z}_j+{1\over |u|^2} {\dd^2 K\over \dd s^2} du d\widebar{u}+2 {\dd^2 K \over \dd t \dd s} {1\over \widebar{u}} \dd_i K_0 \,dz_i d\widebar{u}+2 {\dd^2 K \over \dd t \dd s} {1\over u} \widebar{\dd}_i K_0 \,du d\widebar{z}_j  &
\eear
}
We now introduce the following combinations:
\bear
\mathcal{A}={i\over 2}\,(\widebar{\dd}_i K_0\, d\widebar{z}_i-\dd_i K_0\, dz_i)\\
dt=2\, (\dd_i K_0\, dz_i+\widebar{\dd}_i K_0\, d\widebar{z}_i)
\eear
Therefore
\bea
\widebar{\dd}_i K_0\, d\widebar{z}_i={dt\over 4}-i\, \mathcal{A},\quad\quad \dd_i K_0 \,dz_i={dt\over 4}+i\, \mathcal{A}
\eea
We also parametrize the variable $u$ as follows:
\bea
u=e^{{s\over 2}-i \phi}
\eea
Substituting these expressions into (\ref{metrsimpl}), we obtain
{\small
\bear\nonumber
&&ds^2=\mu\,g_{\CP^1}+{\dd^2 K\over \dd t^2} \left({dt^2\over 4}+4\mathcal{A}^2\right)+ {\dd^2 K\over \dd s^2} \left({ds^2\over 4}+d\phi^2\right)+{\dd^2 K \over \dd t \dd s} \left({ds \,dt\over 2}-4\, d\phi\, \mathcal{A}\right)=\\ \label{metr2}
&& = \mu\,g_{\CP^1}+ {1\over 4} {\dd^2 K\over \dd t_i \dd t_j}dt_i dt_j+{\dd^2 K\over \dd t_i \dd t_j} \mathcal{A}_i \mathcal{A}_j,
\eear
}where $(t_1, t_2)=(t, s)$ and $\mathcal{A}_1=-2\mathcal{A}, \mathcal{A}_2=d\phi$. Let us choose the following parametrization for $(z_1, z_2)$: $(z_1, z_2)=\rho\, e^{-i\tilde{\phi}\over 2}\,(1, w)$. Then the current $\mathcal{A}$ can be rewritten as
{\small
\bea
\mathcal{A}={i\over 2} \frac{z_i d\widebar{z}_i-\widebar{z}_i dz_i }{|z_1|^2+|z_2|^2}=-{d\tilde{\phi}\over 2}+{i\over 2} \frac{w d\widebar{w}-\widebar{w} dw }{1+|w|^2}
\eea}Taking into account that ${\dd^2 K\over \dd t_i \dd t_j}dt_i dt_j={\dd^2 G\over \dd \mu_i \dd \mu_j}d\mu_i d\mu_j$, we may rewrite (\ref{metr2}) as
\bea
ds^2=\mu\,g_{\CP^1}+ {1\over 4}{\dd^2 G\over \dd \mu_i \dd \mu_j}d\mu_i d\mu_j+\left({\dd^2 G\over \dd\mu^2}\right)^{-1}_{ij} (d\phi_i-2 A_i)(d\phi_j-2 A_j),
\eea
where $(\phi_1, \phi_2)=(\tilde{\phi}, \phi)$, $A_1={i\over 2} \frac{w d\widebar{w}-\widebar{w} dw }{1+|w|^2}, A_2=0$. We have thus arrived at the desired result, formula (\ref{metric}). Note that, in these notations, $g_{\CP^1}=\frac{2\,dw d\widebar{w}}{(1+|w|)^2}$.

\section{Vector fields generating the $\mathfrak{u}(2)\oplus \mathfrak{u}(1)$ action}\label{uniformization}

To see what restrictions the enhanced $U(2)\times U(1)$ symmetry imposes on the toric metric~(\ref{metr}), let us start from the holomorphic Killing vector fields generating the toric subgroup\footnote{Representations of various Lie algebras through vector fields in 3 variables were thoroughly studied in~\cite{Morozov}. The discussion presented here is sufficient for us, due to the fact that we have two additional $\mathfrak{u}(1)$ actions on top of the $\mathfrak{su}(2)$.}, i.e. $v_1:=w_1{\dd\over \dd w_1}, v_2:=w_2{\dd\over \dd w_2}, v_3:=w_3{\dd\over \dd w_3}$. Moreover, let us assume that $v_1$ is the generator of the Cartan subgroup $\mathfrak{u}(1)\subset \mathfrak{su}(2)$ (one can always make a change of variables to make sure this is fulfilled). We wish to construct the remaining generators $L_{\pm}$ of $\mathfrak{su}(2)$, defined by the following properties:
\bea\label{su2comm}
[v_1, L_{\pm}]=\pm L_{\pm},\quad [L_+, L_-]=2 \,v_1,\quad [L_{\pm}, v_2]=[L_{\pm}, v_3]=0\;.
\eea
The first of these commutation relations leads to the following form of $L_\pm$:
\bear
L_+=w_1^2{\dd \over \dd w_1}+w_1\, \left(a \,w_2\,{\dd\over \dd w_2}+b \,w_3\,{\dd \over \dd w_3}\right),\\
L_-=-{\dd \over \dd w_1}-{1\over w_1}\, \left(c \,w_2\,{\dd\over \dd w_2}+d \,w_3\,{\dd \over \dd w_3}\right)\,.
\eear
A priori $a, b, c, d$ are functions of $w_2, w_3$, however the last two commutation relations in (\ref{su2comm}) imply that they are constants. It follows from the remaining commutation relation that $c=-a, d=-b$. There are two distinct possibilities:

\vspace{0.3cm}
\noindent A) $a=b=0$, in which case $L_+=w_1^2{\dd \over \dd w_1}$, $L_-=-{\dd \over \dd w_1}$. The orbit of the $SU(2)$ action is given by familiar fractional-linear transformations, $w_1\to \frac{a w_1+b}{c w_1+d}$.

\vspace{0.3cm}
\noindent
B) $a\neq 0$ or $b\neq 0$. A linear change of $(\log w_2, \log w_3)$-variables brings the vector fields to the form
$L_+=w_1^2{\dd \over \dd w_1}+w_1 w_2\,{\dd\over \dd w_2}$, $L_-=-{\dd \over \dd w_1}+ {w_2\over w_1} \,{\dd \over \dd w_2}$. Changing variables again according to $w_2'={1\over (w_1 w_2)^{1/2}}, w_1'=w_1$, we arrive at the canonical form of the generators:
\bea
L_-=-{\dd \over \dd w_1'},\quad L_+=w_1'^2\,{\dd\over \dd w_1'}- w_1'\,w_2'\,{\dd \over \dd w_2'},\quad v_1=w_1'\,{\dd\over \dd w_1'}-{1\over 2} w_2'\,{\dd\over \dd w_2'}
\eea
The orbit is $\big\{ w_1'\to \frac{a w_1'+b}{c w_1'+d},\; w_2' \to (c w_1'+d)\,w_2\big\}$.

\vspace{0.3cm}
\noindent
In order to build a K\"ahler metric with the corresponding isometries, one can construct a K\"ahler potential $K(w)$, which, under the transformations, is shifted as $K\to K+f(w)+\widebar{f(w)}$. The only such possibility in case A is to have a potential of the form $K=\log{(1+|w_1|^2)}+\tilde{K}(|w_2|^2, |w_3|^2)$, but this implies that the manifold is a product of a sphere and some complex surface. In case B, however, the most general choice is $K=K(|w_2|^2(1+|w_1|^2), |w_3|^2)$, which coincides with (\ref{Kahpot}) after an obvious change of variables.

\section{Determining the physical roots $\xi_1, \xi_2$ of $Q(\xi)=0$} \label{solxi}

We showed in \S\, \ref{infasympt} that the normal bundles of the spheres embedded in the cone require that
\bea
-\frac{\xi_2}{1-\xi_2}=\frac{3 \xi_1}{1-\xi_1},
\eea
where $\xi_1$ and $\xi_2$ are \emph{both} roots of the polynomial $Q(\xi)$. This means that
\bear
&\xi_1+\xi_2+\xi_0={3\over 2},&\\
&\xi_1\xi_2+\xi_1\xi_0+\xi_2\xi_0=0,&\\
&\xi_2=\frac{3 \xi_1}{4 \xi_1-1}&
\eear
Eliminating the variables $\xi_0$ and $\xi_2$ we arrive at a cubic equation for $\xi_1$, which, however, factorizes:
\bea
(\xi_1-1)(16 \xi_1^2-4 \xi_1-3)=0
\eea
As we mentioned in \S \, \ref{infasympt}, the case $\xi_1 = 1$ corresponds to the case when the physical region shrinks to zero (i.e. the planes $1, 2$ in Fig. \ref{mompol} merge), so we assume that $\xi_1 \neq 1$. Then we have the two solutions:
\bear\label{xisols}
\xi_1^{(1)}={1\over 8} (1+\sqrt{13}),\quad \xi_2^{(1)}={1\over 8} (7+\sqrt{13})\\
\xi_1^{(2)}={1\over 8} (1-\sqrt{13}),\quad \xi_2^{(2)}={1\over 8} (7-\sqrt{13})
\eear
Since $P_0''=-\frac{3 \xi}{Q(\xi)}$, in order for the metric at infinity (\ref{infmetr}) to be positive-definite, we ought to determine in which of these segments $(\xi_1^{(i)}, \xi_2^{(i)})$ the function $\frac{\xi}{Q(\xi)}$ is negative (in the whole segment). An elementary check shows that this is so only for the first segment, $(\xi_1^{(1)}, \xi_2^{(1)})$. This leads to the following value of $d$:
\bea
d=\frac{16+\sqrt{13}}{64}\,.
\eea

\section{The space of polynomials $y^3-{3\over 2} y^2+d$}

In most calculations one encounters the roots $\xi_i$ of the polynomials of the form
\bea\label{poltyp}
Q(y)=y^3-{3\over 2} y^2+d
\eea
These can be written out explicitly in terms of Cardano's formula, however this expression is rather complicated. A better approach is to use a \emph{rational} parametrization for the space of polynomials of the form (\ref{poltyp}). Indeed, denoting the roots of such a polynomial by $\xi_0, \xi_1, \xi_2$ (as we did in the body of the paper), polynomials of the type (\ref{poltyp}) are defined by the following relations:
\bea
\xi_0+\xi_1+\xi_2={3\over 2},\quad\quad \xi_0\xi_1+\xi_0 \xi_2+\xi_1\xi_2=0
\eea
Reparametrizing the roots as $\xi_1 = \lambda_1 \xi_0,\;\xi_2=\lambda_2 \xi_0$, we arrive at a simple equation $(\lambda_1+1)(\lambda_2+1)=1$, which can be `solved' as follows: $\lambda_1+1=u,\; \lambda_2+1={1\over u}$, where $u$ is a new variable. In terms of this variable the roots are parametrized as
\bea
\xi_0={3\over 2}\,\frac{1}{u+{1\over u}-1},\quad \xi_1={3\over 2}\,\frac{u-1}{u+{1\over u}-1},\quad \xi_2={3\over 2}\,\frac{{1\over u}-1}{u+{1\over u}-1},
\eea
whereas the parameter $d$ of the polynomial $Q(y)$ is expressed as
\bea
d={27 \over 8}\,\frac{(u-1)^2}{u}\,\frac{1}{\left( u+{1\over u}-1\right)^3}
\eea

\section{The general three-line solution}\label{3lineapp}

In sections \ref{3linesec}, \ref{infasympt} we studied the simplest solution of the Monge-Ampere equation, which is the metric cone over a Sasakian manifold.  The structure of the solution (\ref{G0}) hints at the possibility of using the following more general ansatz:
\bea\label{3lines}
G=\sum\limits_{i=0}^2\;\ell_i\,(\log{|\ell_i|}-1)-(\log{\kappa})\,\nu,\quad\quad\quad \kappa>0\,,
\eea
where $\ell_i=0, i=0, 1, 2$ are three a priori arbitrary lines in the $(\mu, \nu)$-plane:
\bea
\ell_i=a_i\,\mu+b_i\,\nu+c_i
\eea
and $\kappa$ is a constant. Asymptotically, when $\mu, \nu\to\infty$ with $\xi={\mu\over \nu}$ fixed,
\bea
G=\nu \log(\nu)\,\left(\sum\limits_{i=0}^2\,b_i+\xi \sum\limits_{i=0}^2\,a_i\right)+\nu \,\left(\sum\limits_{i=0}^2\,(b_i+a_i\xi)\,\left(\log|b_i+a_i\xi|-1\right)-\log{\kappa}\right)+\ldots
\eea
Compatibility with the conical asymptotics (\ref{coneasympt}), (\ref{G0metr}) requires
\bea\label{abeqs}
\sum\limits_{i=0}^2\,a_i=0,\quad\quad\sum\limits_{i=0}^2\,b_i=3\,.
\eea
Substituting the ansatz (\ref{3lines}) in the Ricci-flatness condition (\ref{Ricciflat}) (with $\tilde{a}=1$), one arrives at the following system of equations for the parameters of the ansatz:
\bear\label{CYcond}
&&a_i+b_i=1\quad \textrm{(already encountered in Lemma\;\ref{slopelemma})}\\ \label{aeq}
&&(a_1-a_2)^2\,a_3+(a_1-a_3)^2\,a_2+(a_2-a_3)^2\,a_1=\pm\kappa\\ \label{beq}
&&(a_1-a_2)^2\,b_3+(a_1-a_3)^2\,b_2+(a_2-a_3)^2\,b_1=0\\
&&(a_1-a_2)^2\,c_3+(a_1-a_3)^2\,c_2+(a_2-a_3)^2\,c_1=0\,.
\eear
The sign $\pm$ in the r.h.s. of (\ref{aeq}) is defined by $\pm=\mathrm{sgn}(\ell_0\,\ell_1\,\ell_2)$. Using the equation
$
\sum\limits_{i=0}^2\,a_i=0
$
from (\ref{abeqs}) (the second equation in (\ref{abeqs}) now being a consequence of the first one and (\ref{CYcond})), eq.~(\ref{aeq}) can be rewritten as follows:
\bea\label{eq2}
(a_1+a_2+a_3)(a_1a_2+a_1a_3+a_2a_3)-9\,a_1a_2a_3=\pm\kappa \quad \Rightarrow \quad a_1a_2a_3=\mp{\kappa\over 9}
\eea
Eq. (\ref{beq}) can be brought to the following form, using $b_i=1-a_i$:
\bear\nonumber
&&\pm\kappa=(a_1-a_2)^2+(a_1-a_3)^2+(a_2-a_3)^2=2(a_1+a_2+a_3)^2-6(a_1a_2+a_1a_3+a_2a_3)\\ &&\Rightarrow \quad a_1a_2+a_1a_3+a_2a_3=\mp{\kappa\over 6} \label{eq3}
\eear
Since $\kappa>0$, we see that the first equation has real solutions, only if one chooses the sign~$+$ in the l.h.s. This implies
\bea\label{detpos}
\ell_0\,\ell_1\,\ell_2>0\,.
\eea

\vspace{0.3cm}\noindent
It follows from (\ref{abeqs}), (\ref{eq2}), (\ref{eq3}) that $a_1, a_2, a_3$ are roots of the equation
\bea
a^3-{\kappa\over 6}a+{\kappa\over 9}=0
\eea
Upon a change of variables $a={1\over 1-\xi}$, we obtain
\bea
Q(\xi):=\xi^3-{3\over 2}\xi^2+d=0,\quad\textrm{where}\quad d={1\over 2}-{9\over \kappa}\,.
\eea
We therefore obtain the following parametrization for the constants $a_i, b_i, c_i$:
\bear
a_i=\frac{1}{1-\xi_i},\quad\quad b_i=-\frac{\xi_i}{1-\xi_i},\quad\quad c_i=\frac{\sigma_1 \xi_i^2+ \sigma_2 \xi_i}{1-\xi_i},
\eear
where $\sigma_{1,2}$ are arbitrary constants, and $\xi_i$ are the solutions of the equation $Q(\xi)=0$.

\vspace{0.3cm}
\noindent
If $\sigma_1=0$, the solution differs from (\ref{G0}) by a trivial shift of $\nu\to\nu+\sigma_2$. The interesting case is $\sigma_1\neq0$ -- in this situation we can as well shift $\nu$ to set $\sigma_2=0$, arriving at the solution
\bea
G_{3L}=\sum\limits_{i=0}^2\;\frac{\mu-\xi_i\,\nu+\sigma_1\xi_i^2}{1-\xi_i}\;\left(\log{|\mu-\xi_i\,\nu+\sigma_1\xi_i^2|}-1\right)
\eea
This is a one-parametric generalization of (\ref{G0}). For the case of the manifold $Y^{2,1}$, taking into account that $a_0>0, a_1>0, a_2<0$, the analogue of our former requirement $\xi\in(\xi_1, \xi_2)$ is $\ell_0>0, \ell_1>0, \ell_2>0$, which is compatible with (\ref{detpos}).

\vspace{0.3cm}
\noindent
Since the `radial' part of the metric $\left[ds^2\right]_\mu:= \frac{\dd^2 G_{3L}}{\dd \mu_i \dd \mu_j}\,d\mu_i d\mu_j$ is two-dimensional, one may introduce isothermal coordinates to simplify it. One can check that the orthotoric coordinates $(x, y)$ (see \S\,\ref{orthometric}) serve this purpose. Indeed, if one makes the change of variables
\bea\label{orthocoord}
\mu=\sigma_1 x y, \nu=\sigma_1 (x+y)\, ,
\eea
the radial part of the metric acquires the form
\bea\label{orthoform}
\frac{\dd^2 G_{3L}}{\dd \mu_i \dd \mu_j}d\mu_i d\mu_j= \frac{3x(x-y)\sigma_1}{Q(x)}\,dx^2+\frac{3y(x-y)\sigma_1}{Q(y)}\,dy^2\,.
\eea
This is a special case of the orthotoric metric, which arises if one makes the polynomials $P(x), Q(y)$ in (\ref{cubicpols2}) identical, i.e. if one equates the parameters $c=d$. 

\section{The variational problem}

Interestingly, the Monge-Ampere equation (\ref{Ricciflat}) may be obtained from a variational principle. In fact, although the Ricci-flatness equation $R_{ij}=0$ can be obtained through the extremization of the Einstein-Hilbert functional $\mathcal{S}_{EH}=\int\,d^n x\,\sqrt{g}\, R=\int\,d^n x\,\mathcal{L}$, this is no longer true if one restricts to the class of K\"ahler manifolds. It turns out that in this case the Lagrangian $\mathcal{L}$ is a total derivative:
\bea
\mathcal{L}\left(g\;\;\textrm{K\"ahler}\right)=\sqrt{g}\, R=-\det g_H\cdot g_H^{i \bar{j}}\,\dd_i \bar{\dd}_j \log\det g_H=-\bar{\dd}_j (g_H^{i\bar{j}} \dd_i \det g_H)
\eea
Here $g_H$ is the Hermitian metric associated with the real metric $g$. One concludes that the action only depends on the values of the (derivatives of the) K\"ahler potential at the boundary and does not give rise to any equation in the bulk.

\vspace{0.3cm}
\noindent
In order to obtain an equation of the type (\ref{Ricciflat}) one should consider the following action:
\bea\label{action1}
\mathcal{S}=\int\,d\mu\,d\nu\,f(\mu, \nu)\,G(\mu, \nu)+\int \,ds\,dt\,g(s,t)\,K(s, t),
\eea
where the variables $(\mu, \nu)$ and $(s, t)$, as well as the functions $G(\mu, \nu), K(s, t)$, are Legendre dual to each other, just as in Section~\ref{diff}. Indeed, passing to a single set of variables, say $(\mu, \nu)$, we obtain:
\bea
\mathcal{S}=\int\,d\mu\,d\nu\,\left(f(\mu, \nu)\,G(\mu, \nu)+\left(G_{\mu\mu}G_{\nu\nu}-G_{\mu\nu}^2\right) g(G_\mu, G_\nu) (\mu \,G_\mu+\nu \,G_\nu-G)\right)
\eea
Variation of this action with respect to $G$ produces the following equation:
\bea
G_{\mu\mu}G_{\nu\nu}-G_{\mu\nu}^2=\frac{f(\mu,\nu)}{g(G_\mu, G_\nu)}
\eea
The equation (\ref{Ricciflat}) is a particular case, when $f= \tilde{a}\,\mu$ and $g=e^{G_\mu+G_\nu}$.

\vspace{0.3 cm}
\noindent
\textbf{Remark.} The variational problem above may be related to one of optimal transport theory \cite{Evans}. In the latter setup the relevant problem is to maximize the functional
\bea\label{action2}
\mathcal{S}=\int\,d\mu\,d\nu\,f(\mu, \nu)\,\tilde{G}(\mu, \nu)+\int \,ds\,dt\,g(s,t)\,\tilde{K}(s, t),
\eea
with $f>0, g>0$, subject to the condition
\bea
\tilde{G}(\mu, \nu)+\tilde{K}(s, t)\leq (\mu-s)^2+(\nu-t)^2
\eea
Changing variables to
\bea
\tilde{G}(\mu, \nu)=\mu^2+\nu^2-G(\mu, \nu),\quad \tilde{K}(s, t)=s^2+t^2-K(s,t),
\eea
one is to minimize 
\bea
\tilde{\mathcal{S}}[G, K]=\int\,d\mu\,d\nu\,f(\mu, \nu)\,G(\mu, \nu)+\int \,ds\,dt\,g(s,t)\,K(s, t)
\eea
subject to
\bea
G(\mu, \nu)+K(s, t)\geq \mu s+\nu t
\eea
It clearly follows that $G(\mu, \nu)\geq \underset{(s, t)}{\mathrm{max}}(\mu s+\nu t-K(s, t)):=K^\vee(\mu, \nu)$ and $K(s, t)\geq \underset{(\mu, \nu)}{\mathrm{max}}(\mu s+\nu t-G(\mu, \nu)):=G^\vee(s, t)$. Therefore $\tilde{\mathcal{S}}[G, K] \geq \tilde{\mathcal{S}}[G, K=G^\vee]$, hence in the minimizing configuration $K=G^\vee$, meaning that $K$ and $G$ are Legendre dual.

\section{Killing-Yano forms: the definition}\label{KYapp}

First of all, a Killing-Yano form is a 2-form $\omega_{jk}$ on $\mathcal{M}$ satisfying the equation $\nabla_i \omega_{jk}+\nabla_j \omega_{ik}=0$. By definition, a conformal Killing-Yano form (CKYF) is a 2-form $\omega_{jk}$ on $\mathcal{M}$ satisfying an equation of the form
\bea
\mathscr{D}\omega=0,
\eea
where $\mathscr{D}\omega$ is a 3-tensor, which is a linear combination of covariant derivatives $\nabla_{i}\omega_{jk}$, symmetric w.r.t. the first pair of indices and fully traceless. Being symmetric w.r.t. $i\leftrightarrow j$, we can write it as follows:
\bea
(\mathscr{D}\omega)_{ijk}=\nabla_{i}\omega_{jk}+\nabla_{j}\omega_{ik}+a\,g_{ij}\,g^{mn} \nabla_m \omega_{nk}+b\,(g_{ik}\,g^{mn} \nabla_m \omega_{nj}+g_{jk} g^{mn} \nabla_m \omega_{ni})
\eea
Requiring this 3-tensor to be completely traceless, i.e. $(\mathscr{D}\omega)^i_{\;ik}=(\mathscr{D}\omega)^i_{\;ji}=0$, we get
\bea
b+1+{aD\over 2}=0,\quad\quad -1+a+b\,(D+1)=0\quad\Rightarrow \quad b={1\over D-1}, \quad a=-{2\over D-1}\,.
\eea
Therefore the CKYF condition takes the form
\bea\label{ckyt1}
(\mathscr{D}\omega)_{ijk}=\nabla_{i}\omega_{jk}+\nabla_{j}\omega_{ik}+{1\over D-1}\left(g_{ik}\,g^{mn} \nabla_m \omega_{nj}+g_{jk} g^{mn} \nabla_m \omega_{ni}-2\,g_{ij}\,g^{mn} \nabla_m \omega_{nk} \right)=0
\eea
We can give an equivalent definition by requiring that $(D\omega)_{ijk}$ is skew-symmetric w.r.t. the interchange $j \leftrightarrow k$, i.e.
\bear
&& (\widetilde{\mathscr{D}}\omega)_{ijk}:={1\over 3} ((\mathscr{D}\omega)_{ijk}-(\mathscr{D}\omega)_{ikj})=\\ \label{ckyt2}&&=\nabla_i\omega_{jk}-{1\over 3} T_{ijk}+{1\over D-1}\,\left(g_{ik}\,g^{mn}\nabla_m \omega_{nj}-g_{ij} g^{mn} \nabla_m \omega_{nk}\right)=0\\ \label{Tdef}&&
\textrm{where}\quad\quad T_{ijk}=\nabla_i \omega_{jk}+\nabla_k \omega_{ij}-\nabla_j \omega_{ik}
\eear
The two conditions (\ref{ckyt1}), (\ref{ckyt2}) are equivalent. The tensor $T$ here, which is anti-symmetric in all pairs of indices, is proportional to the exterior derivative of $\omega$, i.e. $T\;\propto\; d\omega$.

\section{Non-holomorphic Killing vector fields\newline on Calabi-Yau twofolds}\label{TNUT}

In Proposition\;\ref{CYholvec} we showed that on a Calabi-Yau threefold without parallel vector fields every Killing vector is holomorphic. In complex dimension two, i.e. for a Calabi-Yau 2-fold, the situation is different. In that case we have a parallel holomorphic 2-form $\Omega_{ij}$, and the dualization of (\ref{parallelform}) gives
\bea\label{hconst}
\nabla_\mu\,h=0,\quad\quad\textrm{where}\quad\quad h=\widetilde{\Omega}^{ij}\,F_{ij}\,.
\eea
In particular, in this case $h$ is a function, and the above equation implies
\bea
h=h_0=\mathrm{const.}\eea
One can see how this scenario is realized in practice. The relevant example is the Taub-NUT space (see \cite{LeBrun} for a detailed discussion of the K\"ahler structure of this space), which has the metric
\bear\label{NUT}
&&ds^2=V\,(dx^2+dy^2+dz^2)+V^{-1}\,(dt+A)^2\, ,\\ \nonumber &&V=a+\frac{1}{r},\quad\quad dA=\ast\, dV\,,\quad\quad a>0\,.
\eear
One can define an integrable complex structure as a map
\bea
\mathcal{J}:\quad\quad dx \to V^{-1}\,(dt+A),\quad\quad dy \to dz\,.
\eea
Two Killing vectors ${\dd\over \dd t}$ and $y\,{\dd\over \dd z}-z\,{\dd\over \dd y}$, generating translations along $t$ and rotations in the $(y, z)$-plane, are holomorphic. On the other hand, the metric (\ref{NUT}) has isometry $SO(3)\times U(1)$, where $U(1)$ is the group of translations along the periodic direction $t$, and $SO(3)$ is generated by rotations in the $(x, y, z)$-space\footnote{The gauge field $A$ is in general not invariant under such transformations, but rather shifts by $A\to A+d\Phi$, where $\Phi$ is a function of $(x, y, z)$. Therefore we also need to appropriately compensate by shifts of the $t$-variable, $t\to t-\Phi$.)}.

\vspace{0.3cm}\noindent
The Killing vectors that lie in $\mathfrak{so}(3)\setminus \mathfrak{u}(1)$ -- the complement to the subgroup $\mathfrak{u}(1)$ of rotations in the $(y, z)$ plane, are \emph{not} holomorphic. Their action may be characterized by using the fact that the Taub-NUT is a hyper-K\"ahler manifold, with three symplectic forms $\varpi=\varpi_1, \varpi_2, \varpi_3$, each of which is Hermitian w.r.t. its own complex structure, $I, J$ or $K$. The vector fields generating the $\mathfrak{so}(3)$ rotate the three K\"ahler forms, i.e.
\bea
\mathfrak{L}_v\,(\varpi_1, \varpi_2, \varpi_3)=\,a_v\circ (\varpi_1, \varpi_2, \varpi_3)\,\quad\quad a_v\in\mathfrak{so}(3)\,.
\eea
For the form $\varpi=\varpi_1$ this implies
\bea
\mathfrak{L}_v\,\varpi=\alpha\, \varpi_2+\beta\,\varpi_3\,,\quad\quad \alpha, \beta = \mathrm{const.}
\eea
Here we assume that $v$ is non-holomorphic, in which case $\alpha$ and $\beta$ are not simultaneously zero. On the other hand, it is known that on a hyper-K\"ahler manifold, the complex two-form $\Omega:=\varpi_2+i\,\varpi_3$ is of type $(2, 0)$ w.r.t. the complex structure $I$ -- the one, in which $\varpi$ is Hermitian. The formula above may now be recast in the form
\bea
\mathfrak{L}_v\,\varpi={\alpha-i\,\beta\over 2}\,\Omega+\mathrm{c.c.}
\eea
Comparing with (\ref{killF}), we find
\bea
F\sim \Omega\,,
\eea
with a constant proportionality factor. Finally, it is easily seen that $\Omega$ is the Calabi-Yau two-form, as $\Omega \wedge \widebar{\Omega}\sim \mathrm{vol.}$ To check this, one might recall that $\omega_1, \omega_2, \omega_3$ are the three K\"ahler forms for the same metric, therefore $\omega_1\wedge \omega_1=\omega_2\wedge\omega_2=\omega_3\wedge \omega_3={1\over 2}\, \Omega \wedge \widebar{\Omega}$. As a result,
\bea
\widetilde{\Omega}^{ij}\,F_{ij}=\mathrm{const.}\,,
\eea
as required by (\ref{hconst}).

\renewcommand\refname{\begin{center} \centering\normalfont\scshape  References\end{center}}
\bibliography{refsdelpezzoarxiv}

\begin{thebibliography}{10}

\bibitem{EH}
T.~Eguchi and A.~J. Hanson, ``{Asymptotically Flat Selfdual Solutions to
  Euclidean Gravity},'' {\em Phys.Lett.}, vol.~B74, p.~249, 1978.

\bibitem{GH}
G.~Gibbons and S.~Hawking, ``{Gravitational Multi - Instantons},'' {\em
  Phys.Lett.}, vol.~B78, p.~430, 1978.

\bibitem{CdO}
P.~Candelas and X.~C. de~la Ossa, ``{Comments on Conifolds},'' {\em
  Nucl.Phys.}, vol.~B342, pp.~246--268, 1990.

\bibitem{Gompf}
R.~E. {Gompf} and A.~I. {Stipsicz}, {\em {4-Manifolds and Kirby Calculus}}.
\newblock Graduate Studies in Mathematics, Vol. 20, American Mathematical
  Society, 1999.

\bibitem{LuPope1}
W.~Chen, H.~Lu, and C.~Pope, ``{Kerr-de Sitter black holes with NUT charges},''
  {\em Nucl.Phys.}, vol.~B762, pp.~38--54, 2007.

\bibitem{Gauduchon}
V.~{Apostolov}, D.~M. {Calderbank}, and P.~{Gauduchon}, ``{Hamiltonian 2-forms
  in K\"ahler geometry. I: General theory.},'' {\em {J. Differ. Geom.}},
  vol.~73, no.~3, pp.~359--412, 2006.

\bibitem{Moroianu}
A.~Moroianu and U.~Semmelmann, ``Twistor forms on {K}\"ahler manifolds,'' {\em
  Ann. Sc. Norm. Super. Pisa Cl. Sci. (5)}, vol.~2, no.~4, pp.~823--845, 2003.

\bibitem{Calabi}
E.~Calabi, ``On {K}\"ahler manifolds with vanishing canonical class,'' in {\em
  Algebraic geometry and topology. {A} symposium in honor of {S}. {L}efschetz},
  pp.~78--89, Princeton University Press, Princeton, N. J., 1957.

\bibitem{Coevering}
C.~van Coevering, ``Regularity of asymptotically conical ricci-flat k{\"a}hler
  metrics.'' arXiv:0912.3946.

\bibitem{Goto}
R.~Goto, ``Calabi-{Y}au structures and {E}instein-{S}asakian structures on
  crepant resolutions of isolated singularities,'' {\em J. Math. Soc. Japan},
  vol.~64, no.~3, pp.~1005--1052, 2012.

\bibitem{Joyce}
D.~Joyce, ``Asymptotically locally {E}uclidean metrics with holonomy {${\rm
  SU}(m)$},'' {\em Ann. Global Anal. Geom.}, vol.~19, no.~1, pp.~55--73, 2001.

\bibitem{Keldysh}
M.~V. Keldysh, ``On certain cases of degeneration of equations of elliptic type
  on the boundry of a domain,'' {\em Doklady Akad. Nauk SSSR (N.S.)}, vol.~77,
  pp.~181--183, 1951.

\bibitem{Guillemin}
V.~{Guillemin}, ``{Kaehler structures on toric varieties.},'' {\em {J. Differ.
  Geom.}}, vol.~40, no.~2, pp.~285--309, 1994.

\bibitem{Delzant}
T.~Delzant, ``Hamiltoniens p\'eriodiques et images convexes de l'application
  moment,'' {\em Bull. Soc. Math. France}, vol.~116, no.~3, pp.~315--339, 1988.

\bibitem{Greene}
C.~Beasley, B.~R. Greene, C.~I. Lazaroiu, and M.~R. Plesser, ``{D3-branes on
  partial resolutions of Abelian quotient singularities of Calabi-Yau
  threefolds},'' {\em Nucl. Phys.}, vol.~B566, pp.~599--640, 2000.

\bibitem{Coeveringcompactcoh}
C.~van Coevering, ``Ricci-flat {K}\"ahler metrics on crepant resolutions of
  {K}\"ahler cones,'' {\em Math. Ann.}, vol.~347, no.~3, pp.~581--611, 2010.

\bibitem{Sparks}
J.~Sparks, ``Sasaki-{E}instein manifolds,'' in {\em Surveys in differential
  geometry. {V}olume {XVI}. {G}eometry of special holonomy and related topics},
  vol.~16 of {\em Surv. Differ. Geom.}, pp.~265--324, Int. Press, Somerville,
  MA, 2011.

\bibitem{BottTu}
R.~Bott and L.~W. Tu, {\em Differential forms in algebraic topology}, vol.~82
  of {\em Graduate Texts in Mathematics}.
\newblock Springer-Verlag, New York-Berlin, 1982.

\bibitem{PZTmain}
L.~A. Pando~Zayas and A.~A. Tseytlin, ``{3-branes on spaces with $R \times S^2
  \times S^3$ topology},'' {\em Phys.Rev.}, vol.~D63, p.~086006, 2001.

\bibitem{Romans}
L.~J. Romans, ``{New Compactifications of Chiral $N=2 d=10$ Supergravity},''
  {\em Phys. Lett.}, vol.~153B, pp.~392--396, 1985.

\bibitem{Prokhorov}
T.~Kishimoto, Y.~Prokhorov, and M.~Zaidenberg, ``Group actions on affine
  cones,'' in {\em Affine algebraic geometry}, vol.~54 of {\em CRM Proc.
  Lecture Notes}, pp.~123--163, Amer. Math. Soc., Providence, RI, 2011.

\bibitem{Pedersen}
H.~{Pedersen} and Y.~{Poon}, ``{Hamiltonian constructions of K\"ahler-Einstein
  metrics and K\"ahler metrics of constant scalar curvature.},'' {\em {Commun.
  Math. Phys.}}, vol.~136, no.~2, pp.~309--326, 1991.

\bibitem{Bykov}
D.~V. Bykov, ``{The differential geometry of blow-ups},'' {\em Theor. Math.
  Phys.}, vol.~185, no.~2, pp.~1636--1648, 2015.
\newblock [Teor. Mat. Fiz. 185, no.2, 313 (2015)].

\bibitem{GMSW}
J.~P. Gauntlett, D.~Martelli, J.~Sparks, and D.~Waldram, ``{Sasaki-Einstein
  metrics on $S^2 \times S^3$},'' {\em Adv. Theor. Math. Phys.}, vol.~8, no.~4,
  pp.~711--734, 2004.

\bibitem{Conti}
D.~Conti, ``Cohomogeneity one {E}instein-{S}asaki 5-manifolds,'' {\em Comm.
  Math. Phys.}, vol.~274, no.~3, pp.~751--774, 2007.

\bibitem{Svartholm}
N.~{Svartholm}, ``{Die L\"osung der Fuchsschen Differentialgleichung zweiter
  Ordnung durch hypergeometrische Polynome.},'' {\em {Math. Ann.}}, vol.~116,
  pp.~413--421, 1939.

\bibitem{Slavyanov}
S.~Y. {Slavyanov} and W.~{Lay}, {\em {Special functions. A unified theory based
  on singularities. With a foreword by Alfred Seeger.}}
\newblock Oxford: Oxford University Press, 2000.

\bibitem{Szego}
G.~{Szeg\"o}, ``{Orthogonal polynomials. 4th ed.}.'' {American Mathematical
  Society (AMS), 432 p.}, 1975.

\bibitem{Whittaker}
E.~{Whittaker} and G.~{Watson}, ``{A course of modern analysis. An introduction
  to the general theory on infinite processes and of analytic functions; with
  an account of the principal transcendental functions. 4th ed., reprinted.}.''
  {Cambridge: At the University Press. 608 p. (1962).}, 1962.

\bibitem{Urbas}
J.~I.~E. Urbas, ``The equation of prescribed {G}auss curvature without boundary
  conditions,'' {\em J. Differential Geom.}, vol.~20, no.~2, pp.~311--327,
  1984.

\bibitem{MS}
D.~Martelli and J.~Sparks, ``{Resolutions of non-regular Ricci-flat Kahler
  cones},'' {\em J.Geom.Phys.}, vol.~59, pp.~1175--1195, 2009.

\bibitem{Chervonyi}
Y.~Chervonyi and O.~Lunin, ``{Killing(-Yano) Tensors in String Theory},'' {\em
  JHEP}, vol.~09, p.~182, 2015.

\bibitem{Santillan}
O.~P. Santillan, ``{Hidden symmetries and supergravity solutions},'' {\em J.
  Math. Phys.}, vol.~53, p.~043509, 2012.

\bibitem{Morozov}
V.~V. Morozov and K.~S. En, ``On imprimitive groups of the three-dimensional
  complex space,'' {\em Kazan. Gos. Univ. Uchen. Zap.}, vol.~115:14,
  pp.~69--85, 1955.

\bibitem{Evans}
L.~C. Evans, ``{Partial differential equations and Monge-Kantorovich mass
  transfer},'' {\em Current Developments in Mathematics, International Press},
  1997.

\bibitem{LeBrun}
C.~LeBrun, ``Complete {R}icci-flat {K}\"ahler metrics on {${\bf C}^n$} need not
  be flat,'' in {\em Several complex variables and complex geometry, {P}art 2
  ({S}anta {C}ruz, {CA}, 1989)}, vol.~52 of {\em Proc. Sympos. Pure Math.},
  pp.~297--304, Amer. Math. Soc., Providence, RI, 1991.

\end{thebibliography}
\bibliographystyle{ieeetr}

\end{document}